\documentclass[11pt]{article}

\usepackage[margin=1in]{geometry}
\usepackage{amsmath,amssymb,amsthm,mathtools,bm,bbm}
\usepackage{graphicx}
\graphicspath{{figures/}}
\usepackage{enumitem}%
\usepackage{algorithm}
\usepackage{algpseudocode}
\usepackage[colorlinks=true,linkcolor=blue,citecolor=blue,urlcolor=blue]{hyperref}
\usepackage{cite}
\usepackage{xcolor}
\usepackage{cleveref}
\usepackage{caption}
\theoremstyle{plain}
\newtheorem{theorem}{Theorem}[section]
\newtheorem{lemma}[theorem]{Lemma}
\newtheorem{proposition}[theorem]{Proposition}

\theoremstyle{definition}
\newtheorem{remark}[theorem]{Remark}
\newtheorem{definition}[theorem]{Definition}

\newtheorem{assumption}{Assumption}
\newcommand{\R}{\mathbb{R}}
\newcommand{\Dorig}{D_0}
\newcommand{\piorig}{\pi_0}
\newcommand{\Dnew}{D}
\newcommand{\pinew}{\pi}

\newcommand{\E}{\mathbb{E}}
\newcommand{\Exp}{\mathrm{Exp}}
\newcommand{\Pp}{\mathbb{P}}
\newcommand{\1}{\mathbbm{1}}

\newcommand{\iid}{\stackrel{\mathrm{i.i.d.}}{\sim}}

\newcommand{\CV}{{\mathrm{CV}}}

\newcommand{\gradnorm}{{m_1}}

\newcommand{\la}{\lambda}
\newcommand{\loc}{{\mathrm{loc}}}
\newcommand{\N}{\mathcal N}
\newcommand{\psorm}{p_{\mathrm{SORM}}}

\newcommand{\tail}{{\mathrm{tail}}}

\newcommand{\Var}{\mathrm{Var}}
\newcommand{\tr}{\mathrm{Tr}}
\newcommand{\us}{{\mathcal U}}
\numberwithin{equation}{section}

\title{Geometry-informed importance sampling for rare events}

\author{
Qinyi Wang$^{1}$, Anya Katsevich$^{1}$, and Shanyin Tong$^{2}$ \\[0.5em]
$^{1}$Department of Statistical Science, Duke University \\
\texttt{\{qinyi.wang,anya.katsevich\}@duke.edu} \\[0.25em]
$^{2}$Department of Mathematics, University of Pennsylvania \\
\texttt{tong3@sas.upenn.edu}
}

\date{\today}

\begin{document}
\maketitle
\begin{abstract}
We propose and rigorously analyze \textbf{GeoIS}, a new importance sampling method for estimating rare event probabilities under a Gaussian distribution that exploits local event geometry. In rotated coordinates, the proposal is a mixture of the laws of $(X,\hat\phi(X)+Y)$ and $(X,\hat\phi(X)-Y)$, where $X$ is Gaussian in coordinates tangent to the event boundary at the instanton, $\hat\phi$ is the quadratic approximation to the boundary there, and $Y$ is exponentially distributed. The upper component shifts toward the event interior and is an asymptotic approximation to the optimal zero-variance proposal, while the lower component covers event mass below the approximate boundary, preserving unbiasedness. We prove a fully explicit, non-asymptotic upper bound on the relative mean-squared error of GeoIS that holds on a high-probability event. The bound quantifies the effect of boundary geometry, rarity, and dimension. It remains bounded as rarity increases, and depends as a mild polynomial on dimension. To our knowledge, our theoretical analysis is the first to address either the high-dimensional or pre-asymptotic setting in which direct asymptotic approximations to the rare probability are unreliable. We implement our method numerically in several examples, including a steady-state diffusion problem. The numerical experiments confirm the qualitative dependence on key parameters predicted by the theory, and show that GeoIS markedly outperforms benchmark importance sampling methods.
\end{abstract}

\section{Introduction}
Rare events such as financial crashes, epidemic outbreaks, and earthquakes can have substantial real-world impacts. Accurately estimating their probabilities can help us better understand such events and make more informed decisions when preparing for or responding to them. In this work, we focus on estimating rare probabilities $p\ll1$ under Gaussian distributions. Specifically, we consider $p=\piorig(\Dorig)$, where $\pi_0:=\mathcal N(0_{d+1},\la_0^{-1}I_{d+1})$ is the Gaussian \emph{reference} distribution and $\Dorig\subset\R^{d+1}$ is the rare event. This formulation includes a variety of rarity regimes, such as the case where $\la_0\gg1$, or $\Dorig$ is far from the origin, or both. It also includes non-Gaussian rare events, since if $U\sim \mathcal N(0_{d+1},\la_0^{-1}I_{d+1})$, then $p=\Pp(U\in \Dorig)=\Pp(F(U)\in F(\Dorig))$ for any bijective function $F$. We are particularly interested in high-dimensional settings, where the combination of rarity, high dimensionality, and high cost of evaluating event membership (i.e. the indicator $\1\{u\in\Dorig\}$ for a given $u\in\R^{d+1}$) makes accurate probability estimation especially challenging.

A natural approach to rare-event estimation is to use asymptotic methods, since rarity provides useful asymptotic structure. The key idea is that if $u^*=\arg\min_{u\in\Dorig}\|u\|$ is unique, then most of the rare event mass stems from a neighborhood of this single most likely point under the Gaussian density.  Thus it suffices to understand the rare-event geometry near this point, called the \emph{instanton}. In particular, replacing the log Gaussian density and event boundary by Taylor expansions of suitable order at $u^*$ gives the leading-order asymptotic approximation to $p$. Namely, in fixed $d$, it is known~\cite[Theorem~5]{breitung1996geometric},~\cite[Section~8.3]{bleistein1975asymptotic} that
\begin{equation}\label{intro-sorm}
\begin{gathered}
p = \psorm ( 1+\mathcal O(\la_0^{-1})), \quad\la_0\to\infty, \\
\psorm:=\frac{e^{-{\la_0\|u^*\|^2}/2}}{\sqrt{2\pi\la_0\|u^*\|^2\det H}}.
\end{gathered}\end{equation}
Up to a rescaling factor, $H$ is the difference between the Hessians of local graph representations of the boundary of $\Dorig$ and the Gaussian level set $\{\|u\|=\|u^*\|\}$ it is tangent to. Generically, $H\succ 0$, which corresponds to the constrained minimum at $u^*$ being nondegenerate. %
SORM is the acronym for second-order reliability method, the name given to the leading expansion~\eqref{intro-sorm} in the reliability engineering literature~\cite{hu2021second}. ``Second-order" refers to the fact that the boundary is approximated by a quadratic at $u^*$. 

In the important, high-dimensional regime,~\eqref{intro-sorm} provides no meaningful guarantee because the $\mathcal O(\la_0^{-1})$ error does not make explicit the dependence on dimension itself and other $d$-dependent quantities such as instanton length and boundary curvature, which can completely change the order of magnitude of $\psorm$'s approximation accuracy. Recent work~\cite{katsevich2025asymptotic} by the second author quantifies these high-dimensional effects for the first time, deriving an explicit high-dimensional bound on the SORM error. The work also presents matching lower bounds, confirming that the error bound is tight. Therefore,~\cite{katsevich2025asymptotic} establishes the right notion of high-$d$ asymptotic regime: the regime in which the high-$d$ SORM error bound is vanishingly small.

The specification of the high-dimensional rare-event asymptotic regime also delineates a corresponding, high-$d$ \emph{pre-asymptotic} regime, in which the SORM error bound is not negligible. This new pre-asymptotic regime is the one we take up in the present work. We develop a rigorously justified numerical procedure for accurate high-dimensional rare-event estimation in the pre-asymptotic regime where SORM alone is insufficiently accurate. A key focus of our work is to give not a limiting asymptotic description but a \emph{quantitative error bound for our numerical procedure}. Indeed, \emph{if asymptotics apply, then SORM is already accurate}, and there is no need for numerics in the first place.

\paragraph{Importance sampling for accurate pre-asymptotic rare-event estimation.}
The natural way to accurately approximate rare probabilities $p$ in the pre-asymptotic regime is to use an \emph{exact} numerical method, meaning, a method which can be made arbitrarily accurate by expending sufficient computational effort. For example, the simplest approach is to use the standard Monte Carlo estimator, which counts the proportion of samples drawn from $\piorig$ that landed in $\Dorig$. This estimator is inefficient, however. It requires a number of samples proportional to \(1/p\) to achieve fixed relative accuracy, making it impractical for very rare events, especially when event membership is expensive to evaluate. More efficient approaches applicable to problems in our setting include directional simulation~\cite{ditlevsen1990general}, subset simulation~\cite{au2001estimation}, and importance sampling~\cite{bucklew2007introduction}. As we now explain, importance sampling (IS) is particularly well-suited for exploiting asymptotic structure to improve efficiency, even in the pre-asymptotic regime.

The idea of IS is to draw $N$ samples $U_i$ from a \emph{proposal} distribution $\hat\pi\neq\piorig$ that places more probability mass in $\Dorig$ than does $\piorig$, and correct for the change of distribution through importance weights. The resulting estimator is 
\begin{equation}\label{IS-def}\hat p_N= \frac1N\sum_{i=1}^N\frac{\piorig(U_i)}{\hat\pi(U_i)}\1\{U_i\in \Dorig\},\qquad U_i\iid\hat\pi.\end{equation}We refer to the full summands in~\eqref{IS-def}, including the event indicator, as the importance weights. 

The bias $\E[\hat p_N]-p$ of this estimator is nonzero if and only if $\Dorig$ is not fully contained in the support of $\hat\pi$, and this bias does not vanish with increasing $N$. Thus for IS to be exact (as defined above), the support of $\hat\pi$ must contain $\Dorig$. For such a proposal, the relative mean-squared error (MSE) of $\hat p_N$ is given by
\begin{equation}\label{rel-var}\E\left[(\hat p_N/p-1)^2\right]=\chi^2(\piorig\vert_{\Dorig}\,\vert\!\vert\,\hat\pi)/N.\end{equation}
Here, $\piorig\vert_{\Dorig}$ is the reference distribution restricted to the rare event and normalized to be a probability distribution, while $\chi^2(\cdot\,\vert\!\vert\,\cdot)$ is the chi-squared divergence between two densities. Thus the theoretically optimal choice of proposal is $\hat\pi=\piorig\vert_{\Dorig}$ itself, but it cannot be used because its density depends on the unknown $p$, and the density of $\hat\pi$ is needed to implement~\eqref{IS-def}.

For other unbiased proposals, $\hat p_N$ still satisfies our desired exactness property: it can be made arbitrarily accurate by taking $N$ large enough. To make the estimator not just exact but computationally feasible, $\hat\pi$ must be chosen to ensure $\chi^2(\piorig\vert_{\Dorig}\,\vert\!\vert\,\hat\pi)$ is not too large. Two classical benchmarks in rare-event simulation are logarithmic efficiency and bounded relative error, which are choices of proposal that ensure $\log\chi^2(\piorig\vert_{\Dorig}\,\vert\!\vert\,\hat\pi)=o(\la_0)$ or $\mathcal O(1)$ as $\la_0\to\infty$, respectively~\cite{asmussen2007rare, juneja2006rare}. 

A natural strategy for designing a proposal $\hat\pi$ is to use a tractable asymptotic approximation to $\piorig\vert_{\Dorig}$. This principle has extensive precedent in other rare-event settings; see e.g.~\cite{blanchet_efficient_2008,blanchet_liu_2012,broniatowski_caron_2014,liu_xu_2014,lecuyer_etal_2011,reijsbergen_etal_2018}. We now review IS methods developed for our specific Gaussian rare-event setting and the available guarantees for their accuracy.

\paragraph{Prior work in our setting.} 
In our setting, the asymptotic approximation to $\piorig\vert_{\Dorig}$ has the form 
\begin{equation}\label{piD-approx}
\hat\pi=\mathrm{Law}(X,\psi(X)+Y),
\end{equation}
after an orthogonal transformation taking $u^*$ to $(0_d,\|u^*\|)$, where $0_d$ is the $d$-dimensional origin. Here, $\psi:\R^d\to\R$ describes the event boundary, $X$ has a $d$-dimensional Gaussian distribution and $Y$ is exponentially distributed. That $\hat\pi$ approximates $\piorig\vert_{\Dorig}$ was shown by~\cite{lapinski2019multivariate} in the case of fixed $d$ and a flat boundary, in which case $\psi(x)\equiv \|u^*\|$. More recently, the aforementioned work~\cite{katsevich2025asymptotic} derived a non-asymptotic bound on the total variation distance $\mathrm{TV}(\hat\pi, \piorig\vert_{\Dorig})$ for a general curved event boundary and $d\gg1$. This work implies $\psi$ can be taken either to be the exact boundary of $\Dorig$ or its quadratic Taylor approximation at $(0_d,\|u^*\|)$. 

Both choices of $\psi$ present obstacles to implementation in IS: the exact event boundary is typically inaccessible, whereas a quadratic approximation induces bias: since $\hat\pi$ is supported only above the graph $(x,\psi(x))$, and $\psi$ is not the true boundary, part of $D$ could lie outside the support of $\hat\pi$. 

The first work proposing to use~\eqref{piD-approx} as an IS proposal is~\cite{maes1993asymptotic}, motivated by heuristic asymptotic arguments. The authors assume that $\psi$ is either the true boundary or a further improvement on the quadratic Taylor approximation. However, no rigorous theory on the accuracy of the resulting estimator is given. 

Following this early work, extensive recent literature on IS for Gaussian rare events uses Gaussian proposals, rather than proposals based on the Gaussian–exponential approximation~\eqref{piD-approx}; see e.g.~\cite{peherstorfer_multifidelity_2018, geyer2019cross, uribe2021cross, tong2021extreme, el_masri_improvement_2021, tong2023large}. The work~\cite{tong2021extreme} studies the choice $\hat\pi=\mathcal N(u^*, \la_0^{-1}I_{d+1})$ (called LSIS), and the other works numerically find a proposal within a chosen Gaussian family by approximately minimizing the KL divergence from the optimal conditional distribution $\piorig\vert_{\Dorig}$ to the proposal. This is called the cross entropy (CE) method in the rare-event literature, which more generally denotes minimizing the KL objective among some family of distributions~\cite{de2005tutorial,geyer2019cross}. The focus on Gaussian proposals is natural: they are convenient to sample from, tractable for optimizing the KL divergence objective, and have full support in $\R^{d+1}$, avoiding the support-induced bias associated with~\eqref{piD-approx}. They can also achieve logarithmic efficiency: the informal asymptotic analysis in~\cite{tong2021extreme} establishes this for LSIS in fixed $d$. 

Much of this literature tackles the numerical challenges of fitting Gaussian proposals in high dimensions. The resulting algorithms are complex, making rigorous analysis difficult. When theory is available, as in~\cite{tong2021extreme}, it concerns asymptotic performance and the fixed-$d$ case. Thus, the literature has yet to meet the two main objectives of this work:
to derive a rigorous guarantee on an IS estimator's accuracy in 1) the high-dimensional regime, and 2) the pre-asymptotic regime in which IS is actually needed, rather than in the asymptotic regime in which SORM is accurate. A third natural objective is also yet to be achieved, even in fixed dimension: to design an implementable IS proposal that provably has bounded relative error, not just logarithmic efficiency.

\paragraph{Concurrent work.} The concurrent work~\cite{li2026beyond} achieves the third objective: the authors design a new proposal distribution and prove the resulting IS estimator achieves bounded relative error (in fact, even vanishing relative error, with a suitable parameter scaling) in fixed $d$. Their proposal distribution is essentially the same one derived concurrently and independently in the present work, and described below. But the authors' theoretical analysis holds in the fixed $d$, large $\la_0$ asymptotic regime, so our first two objectives are not addressed. 

\paragraph{Our method and theoretical guarantees.} 
We develop GeoIS, an IS method based on the Gaussian--exponential approximation to $\piorig\vert_{\Dorig}$, and derive an explicit non-asymptotic bound on its relative error in the high $d$ case, on a high-probability event $A_N$; see~\eqref{intro:CV} below. Our theoretical analysis includes fixed $d$ as a special case, and in the fixed $d$, large $\la_0$ asymptotic regime, GeoIS has bounded relative error on $A_N$. Thus we achieve all three of our objectives in this work. 

To describe the method, we continue to use the rotated coordinates described below~\eqref{piD-approx}. Let $\psi$ be the quadratic Taylor approximation to the event boundary at the instanton $(0_d,\|u^*\|)$, and $\hat\pi_+=\mathrm{Law}((X,\psi(X)+Y))$. As discussed above, this proposal can omit part of $D$ and thereby introduce bias. To resolve this issue, we use a mixture of two densities as our proposal: $\hat\pi = w\hat\pi_+ + (1-w)\hat\pi_-$ for a fixed $w\in(0,1)$, with $\hat\pi_\pm = \mathrm{Law}((X, \psi(X)\pm Y))$.

Adding this second mixture component ensures $\hat\pi$ is supported on all of $\R^{d+1}$, so it leads to an unbiased estimator $\hat p_N$. That the two mixture components are mirror reflections across $\psi$ is convenient but not itself the motivation for the lower mixture component. The true motivation for the choice of $-Y$ in $\hat\pi_-$ is to ensure enough mass is placed in the strip between the true boundary and its quadratic approximation, in order to control the denominators of the importance weights in~\eqref{IS-def}. The optimal vertical distribution depends on the strip's height, which is unknown because it depends on the exact boundary. Our chosen distribution for $-Y$ is designed to be robust to a broad range of strip heights.

GeoIS implementation is straightforward. Most of the computation is confined to a one-time, pre-sampling stage, in which the instanton $u^*$, and the second derivative of the event boundary at this point, must be obtained. These are the same quantities needed to implement the SORM formula~\eqref{intro-sorm}, so it is reasonable to assume they are accessible. Once these quantities are computed, generating a proposal sample $U_i$ requires only Gaussian and exponential draws, evaluation of the quadratic boundary approximation, and a single membership test $\1\{U_i\in\Dorig\}$.

We analyze the performance of the GeoIS estimator on an event $A_N=\cap_{i=1}^N\{U_i\in\Omega\}$, where the $U_i\iid\hat\pi$, and $\Omega$ is a local region such that $U_i\in\Omega$ with high probability. We prove that 
\begin{equation}\label{intro:CV}
\begin{gathered}
\E\left[\left(\frac{\hat p_N}{p}-1\right)^2\1_{A_N}\right] = \frac1N\left(w^{-1}-1 + \frac{\epsilon_+}{w} + \frac{\epsilon_-}{1-w}\right) + \mathcal E_{\exp}, \\
 |\mathcal E_{\exp}|\lesssim \exp(-\la\kappa^2/48), \qquad \Pp(A_N^c) \leq N\exp(-\la\kappa^2/4),
\end{gathered}
\end{equation} and give explicit bounds on $\epsilon_\pm$ in terms of $d$, $\Dorig$, and $\la=\la_0\|u^*\|^2$, where the latter is an aggregate rarity parameter. The choice of $\kappa$ depends on the geometry of $\Dorig$ and in some cases, can be as large as constant with respect to $\la$. In any case, our conditions guarantee that $\la\kappa^2\to\infty$ as $\la\to\infty$. As a first consequence of~\eqref{intro:CV}, we conclude that for absolute $c,c'$, we can freely increase $N$ up to $\exp(c\lambda\kappa^2)$ while keeping the probability of $A_N^c$ below $\exp(-c'\lambda\kappa^2)$ and, at the same time, decreasing the relative MSE on $A_N$ at the expected $1/N$ rate. This is because $|\mathcal E_{\exp}|\lesssim 1/N$ for the specified range of $N$.

The term $(w^{-1}-1)/N$ can be understood as the relative MSE in the limit where both the conditional-density and boundary approximations are exact: $\hat\pi_+=\piorig\vert_{\Dorig}$, while $\hat\pi_-$ is supported in $\Dorig^c$. This term therefore reflects the cost of allocating a fraction $1-w$ of samples to the lower component when it is unneeded. 

The terms $\epsilon_\pm$ quantify the departure from this limit. Our explicit bounds on these terms combine dimension, rarity, and local boundary derivatives. For example, the generically dominant term in the bound on both $\epsilon_+$ and $\epsilon_-$ is $\tau_3d/\sqrt{\la}$. Here, $\tau_3$ is a weighted operator norm of the tensor of third-order boundary derivatives at the instanton, in certain rescaled coordinates.  %
Interestingly, both $\tau_3d/\sqrt{\la}$ and the other quantities in our upper bounds on $\epsilon_\pm$ are precisely the same ones that appear in the high-dimensional upper bound on $|p/\psorm - 1|$ from~\cite{katsevich2025asymptotic}. This makes the delineation between the SORM asymptotic regime introduced by~\cite{katsevich2025asymptotic}, and the GeoIS pre-asymptotic regime, especially clear. Namely, when the upper bounds on $\epsilon_\pm$ are only bounded by constants, accuracy of $\psorm$ is not guaranteed but accuracy of GeoIS is, by drawing a sufficient (but controlled) number $N$ of proposal samples. 

Of course, it would be natural to compare GeoIS not only to the asymptotics-based SORM, but also to other IS methods, for which the relevant comparison is between the pre-factors in front of $1/N$. However, as described above, there are no comparable high-dimensional error bounds for other IS methods. For a comparison between methods in the fixed $d$, large $\lambda_0$ asymptotic regime, see the aforementioned concurrent work~\cite{li2026beyond}.

\subsection{Theoretical significance}
As already described, our explicit and non-asymptotic bounds are unique in the literature in their ability to give informative guarantees in the high-dimensional regime. Another significant and powerful feature of our result is that it applies essentially to any event $\tilde D$ that is rare 
under a Gaussian $\mathcal N(\mu, \Sigma)$, for which rarity stems from $\tilde D$ being far from the mean $\mu$ on the scale set by $\Sigma$, and for which mass is concentrated near the single instanton, provided the boundary of $\tilde D$ is locally smooth and curving away from the Gaussian level set it touches. Indeed, any such rare probability $p=\Pp(\mathcal N(\mu, \Sigma)\in\tilde D)$ can be written in the form studied here: $p=\Pp(\mathcal N(0_{d+1}, I_{d+1})\in\Dorig)$, setting $\la_0=1$. Standardizing is convenient because it forces rarity to be encoded only in the event, rather than both in the reference distribution and the event. Furthermore, one can rotate $\Dorig$ without loss of generality, so that the instanton $u^*$ is $u^*=(0_d,\sqrt\la)$ for some $\la\gg1$. One can now fully characterize $\Dorig$ by its local shape, $t\geq \phi_0(x)+\sqrt{\la}$. Our bound on relative MSE quantifes the dependence on dimension $d$, distance $\sqrt{\lambda}$, and local boundary shape $\phi_0$, \emph{without prescribing how they vary together as the event becomes rarer}. As two special cases of this general framework, in Section~\ref{sec:asymp}, we work out the form of our bounds for a family of dilated events $\Dorig$ and a family of translated events $\Dorig$, allowing $d$ to grow in both regimes. In contrast, the asymptotic approach of~\cite{li2026beyond} locks in the specific rarity regime $p=\Pp(\mathcal N(0_{d+1}, I_{d+1})\in\sqrt{\la_0}\Dorig)$ for fixed $\Dorig$ and $\la_0\to\infty$.

Next, the fact that our bounds are valid in the pre-asymptotic regime sets our work apart from other rigorous theoretical analyses of IS in the literature. Those works characterize algorithmic performance in the limit in which direct asymptotic approximations themselves become accurate. Our guarantee goes beyond this regime, quantifying the performance of IS when it is needed most: when asymptotic approximations alone are not precise enough. Importantly, our explicit local relative MSE bound is not just valid in the pre-asymptotic regime but also depends favorably on the key parameters. The growth with $d$ is a mild polynomial, and increasing $\la_0$ only improves the bound. Indeed, the range of sample sizes $N$ for which $A_N$ is high-probability expands, and the terms $\epsilon_\pm$ vanish, so that the local relative MSE remains bounded with respect to $\la_0$.  The latter feature is notable given that logarithmic efficiency alone remains a desirable target for rare-event IS methods~\cite{bai2022rare,deo2025achieving}, even though it allows the relative MSE to deteriorate polynomially in $\la_0$.

A related important distinction between our results and previous theoretical guarantees on IS is that previous bounds are inexplicit in ways that hinder understanding of the algorithm's performance, whereas our bounds give much more complete information. For example, the relative MSE bound of~\cite{li2026beyond} appears to decay at the expected $1/N$ rate. However, the prefactor depends in an unquantified way on an auxiliary parameter which must be chosen to depend on $N$. Thus the true dependence on sample size may have a worse rate than $1/N$. In contrast, our bounds are fully explicit except for suppressed dependence on absolute constants. 

Next, we describe a key technical challenge in the theoretical analysis of IS performance, how we handle it, and how overall, our theoretical approach allows us to derive these significantly better (flexible, pre-asymptotic, explicit, high-dimensional) results than are available in the literature.
\paragraph{The challenge of importance weight blow-up.}
A well-known and long-studied problem with importance sampling is that even a well-chosen proposal density can produce extreme importance weights in the tails. When such weights occur frequently enough to destabilize the estimator, one can truncate them~\cite{ionides2008truncated} or smooth them~\cite{vehtari2024pareto}. In other settings, including ours, weight blow-up occurs only on a low-probability event and causes no issues in practice. Nevertheless, 
these rare draws must be accounted for in a rigorous analysis of IS performance. One approach is to study the expected relative absolute error $\E|\hat p_N/p-1|$, which is less sensitive than relative MSE to extreme weights. This is done in~\cite{chatterjee2018sample}, but the error has the suboptimal rate $N^{-1/4}$ rather than the expected $N^{-1/2}$. Another approach, used in~\cite{li2026beyond} and here, is to bound the relative MSE on the event that all $N$ draws lie in a region $\Omega$ where the weights do not blow up, and to show that this event has high probability.

The difficulty with this localization approach is that the size of $\Omega$ enters both the containment probability and the local relative MSE bound. In~\cite{li2026beyond}, the auxiliary parameter mentioned above is an $R$ such that $\Omega$ has radius $R/\sqrt\la$, in our notation. The authors' local relative MSE bound scales as $1/N$ for fixed $R$, but to keep all $N$ draws in the local region with high probability as $N$ grows, one must increase $R$ with $N$. Thus as discussed, the overall rate in $N$ of the local relative MSE is unknown and may be worse than $1/N$. The work~\cite{schillings2020convergence} also uses localization in a related IS context, but considers $N$ fixed, so the tradeoff as $N$ grows does not arise.

\paragraph{Our analysis.}
In our own localization analysis, we explicitly track how the local relative MSE depends on the radius $\kappa$. This allows us to obtain an explicit rate in $N$, even when $\kappa$ is chosen as a function of $N$. Our $\kappa$ is also larger than the corresponding radius $R/\sqrt\la$ of~\cite{li2026beyond}, since $\la\kappa^2$ at least diverges with $\la$, and $\kappa$ may even be of constant order in $\la$. This is convenient because the larger the region, the higher the probability of samples falling in it, and hence the more samples $N$ that can be used. 

The main challenge in our proof is to control the second moment of the importance weight in this large region $\Omega$, sharply enough to give informative bounds in the high-dimensional, pre-asymptotic regime. The first step is to take advantage of the shared, simple Gaussian--exponential structure of $\hat\pi_+$ and $\hat\pi_-$. This allows us to obtain explicit preliminary bounds on the second moment, in terms of expectations of exponentiated boundary and density approximation errors on $\Omega$. We give separate estimates for these expectations inside an inner ball of radius $\sim\sqrt{d/\la}$, and the surrounding annulus extending to the full radius $\kappa$. In the inner ball, we have more precise control of the remainders via Taylor estimates. We combine these with Gaussian concentration to derive precise, dimension- and geometry-dependent bounds. In the annulus, the boundary Taylor remainder in the exponent is larger and must be handled differently. We resolve the issue by showing that the quadratic decay of the Gaussian density overwhelms the Taylor remainder. 

The second moment estimate in the large annulus is precisely what motivates choosing $Y\sim\Exp(\la)$ in the lower mixture component $\hat\pi_-=\mathrm{Law}(X,\psi(X)-Y)$. This is discussed in Section~\ref{subsec:discuss}.

\subsection{Numerical performance}
 We study the numerical performance of GeoIS in four rare-event examples. The first two specify $\Dorig$ geometrically as the epigraph of a parabola and the upper exterior of an ellipsoid. The third concerns a rare portfolio loss, and the fourth is a rare endpoint exceedance event in a one-dimensional steady-state diffusion problem with random permeability field. For the mixture weight $w$, we use the default choices $w=0.5$ and $w=0.9$, except for the first example with an exact quadratic boundary, for which we also study $w=1$. In principle, approximate knowledge of $\epsilon_-$ can be used to inform the choice of $w$. In particular, one should choose $w$ close to 1 when $|\epsilon_-|\ll1$,  as can be seen from~\eqref{intro:CV}. 
We leave the possibility of adaptation to future work, though taking $w=1$ for the quadratic boundary is a simple example of this adaptation, since $\epsilon_-=0$ in this case.

For each example, we compare GeoIS to the IS methods of~\cite{tong2021extreme,tong2023large} discussed above. The methods in these works are direct competitors to GeoIS: they likewise use instanton-based asymptotic information to guide the construction of IS proposals, and are designed for high-dimensional Gaussian rare events, with $\Dorig$ potentially defined in terms of an expensive numerical model. The first of these methods is LSIS~\cite{tong2021extreme}, which uses $\hat\pi=\mathcal N(u^*,\la_0^{-1}I_{d+1})$. The second method, introduced in~\cite{tong2023large}, selects $\hat\pi$ by applying the CE method to the Gaussian family, with the KL divergence optimization initialized at LSIS. We call this combination LSIS+CE. Finally, due to intrinsic low-dimensional structure in the diffusion problem (see Section~\ref{subsec:PDE} for more details), it is natural to compare GeoIS to LAIS in this example. LAIS is another algorithm from the same work~\cite{tong2023large}. It also uses the CE method with a Gaussian family, but exploits low-dimensional structure to improve the scalability of Gaussian proposal fitting. 

We do not include the Gaussian--exponential method of~\cite{maes1993asymptotic} in the numerical comparison, because it requires locating points on the rare-event boundary for each sample. These repeated boundary searches can be costly when the event is defined through a complex numerical model. Both GeoIS and the above benchmark methods avoid this additional computation.

In our four examples, we study the performance of GeoIS and the competing methods as sample size $N$, dimension $d$, the Gaussian inverse variance scale $\la_0$, and the instanton distance $\|u^*\|$ are varied. Across these different regimes, the empirical relative variance of GeoIS is substantially smaller than the benchmarks, with improvements of several orders of magnitude in some cases. It is also noticeably more robust to increasing rarity. In genuinely high-dimensional problems without intrinsic low-dimensionality, dimension reduction via LAIS does not make sense, and LSIS+CE deteriorates severely as fitting the Gaussian covariance becomes increasingly difficult. Thus among the competing methods, only LSIS is viable, but its empirical relative variance remains several orders of magnitude higher than that of GeoIS as dimension increases. Finally, the observed dependence of the GeoIS relative variance on rarity, dimension, and geometry are consistent with our theoretical bounds. Our numerical results demonstrate the efficiency, robustness and wide applicability of GeoIS for estimating rare-event probabilities in high-dimensional problems, including those involving complex models.

\subsection*{Organization}
The remainder of the paper is organized as follows. Section~\ref{sec:main} introduces the rare-event setting, GeoIS algorithm, assumptions, and our theoretical guarantee on its performance. Section~\ref{sec:asymp} specializes the performance guarantee to the context of two rarity regimes: dilation and translation of the rare event. Section~\ref{sec:numerical} presents numerical experiments for the parabolic, ellipsoidal, portfolio, and diffusion PDE examples. Section~\ref{sec:proof-kappa} proves the performance guarantee, and we make concluding remarks in Section~\ref{sec:conclude}. The appendices provide quantitative verification of the geometric assumptions, implementation details for the numerical examples, and technical lemmas used in the proof of the main result.

\subsection*{Notation}
All probability measures in this work are absolutely continuous with respect to Lebesgue measure; we abuse notation by using the same letter to denote both the measure and its density with respect to Lebesgue. 

For a symmetric positive definite matrix $H\succ0$ and a sufficiently smooth function $F$, the $H$-weighted operator norm of the $k$th derivative tensor of $F$ at $x$ is given by
\begin{equation}\label{F-H}
\|\nabla^kF(x)\|_H:= \sup_{\|H^{1/2}x'\|=1}\left|\frac{d^k}{da^k}F(x+ax')\big\vert_{a=0}\right|.
\end{equation} We write $\|\nabla^kF(x)\|$ as shorthand for $\|\nabla^kF(x)\|_{I_d}$. For (not necessarily square) matrices $A$, the quantity $\|A\|$ is the usual matrix operator norm, and for vectors $v$, the quantity $\|v\|$ is the usual Euclidean norm.

For nonnegative \(a,b\), we write \(a\lesssim b\) if \(a\leq Cb\); similarly, \(a=\mathcal O(b)\) means \(|a|\leq Cb\). Unless otherwise stated, the constant \(C\) is absolute. Subscripts indicate permitted dependence on specified parameters, as in \(\lesssim_\rho\) and \(\mathcal O_\rho\). We write \(a\asymp b\) when both \(a\lesssim b\) and \(b\lesssim a\).

\section{Main results}\label{sec:main}
In Section~\ref{sec:setup}, we introduce the rare-event setting and a change of coordinates. In Section~\ref{sec:sampler} we present and motivate the GeoIS proposal. Section~\ref{subsec:theory} presents our theoretical guarantee and some initial intuition, and we conclude with further discussion on the choice of lower mixture component in Section~\ref{subsec:discuss}.

\subsection{Set-up}\label{sec:setup}
Let $\piorig=\mathcal N(0_{d+1},\la_0^{-1}I_{d+1})$. For a rare-event set $\Dorig\subset\R^{d+1}$ not containing the origin, our goal is to estimate $p=\piorig(\Dorig)$. Assume that $\Dorig$ has a unique closest point $u^*$ to the origin and let $\ell$ and $n$ denote the length and corresponding unit vector for $u^*$:
\begin{equation}\label{eq:instanton}
u^*=\arg\min_{u\in\Dorig}\|u\|^2,\qquad \ell=\|u^*\|,\qquad n=\frac{u^*}{\ell}.
\end{equation}
The point $u^*$ is called the \emph{instanton}, and it must lie on the boundary $\partial\Dorig$ of $\Dorig$. In reliability engineering, $u^*$ is referred to as the design point~\cite{zhao_general_1999, breitung2024return}.
We now describe the geometry of $\Dorig$ and its boundary near $u^*$. To do so, let $Q\in\R^{(d+1)\times d}$ have orthonormal columns spanning the orthogonal complement of $n$, so that
\begin{equation}\label{eq:Q}
Q^\top Q=I_d,\qquad Q^\top n=0_d.
\end{equation}
\begin{assumption}[Smooth one-sided boundary]\label{assume:bdy} There exist an open neighborhood $V_0\subset\R^d$ of $0_d$, a function $\phi_0\in C^4(V_0)$, and a height $\mu_0>0$, such that
\begin{align}
\phi_0(0_d)=0,\qquad H:=I_d+\ell\nabla^2\phi_0(0_d)\succ0,\label{H-phi0}\\[0.5em]
\{u^*+Qv+sn\,:\, v\in V_0,\;\; \phi_0(v)\leq s\leq\phi_0(v)+\mu_0\}&\subseteq D_0, \tag{strip}\\
\{u^*+Qv+sn\,:\, v\in V_0,\;\; s<\phi_0(v)\}&\subseteq D_0^c.\tag{one-sidedness}
\end{align}
\end{assumption} 
This assumption states that near $u^*$, the boundary $\partial\Dorig$ can locally be represented as the graph $s=\phi_0(v)$ in the coordinates $u=u^*+Qv+sn$. Furthermore, $\Dorig$ must contain a strip of height $\mu_0$ above the boundary over $V_0$ (``strip") and it must contain no points below the boundary over this same neighborhood (``one-sidedness"). 

The fact that the boundary of $\Dorig$ can be described by $\phi_0$, and that $u^*$ is the closest point in $\Dorig$ to the origin, imply $v\mapsto \|u^*+Qv+\phi_0(v)n\|^2=(\ell+\phi_0(v))^2 + \|v\|^2$ has an unrestricted local minimizer at $v=0_d$. This immediately yields the following two additional properties of $\phi_0$:
$$\nabla\phi_0(0_d)=0_d, \qquad H=I_d+\ell\nabla^2\phi_0(0_d)\succeq0.$$
Thus the requirement $H\succ0$ in~\eqref{H-phi0} can be understood as  nondegeneracy of the instanton.
\begin{remark}
    The smoothness condition $\phi_0\in C^4(V_0)$ in Assumption~\ref{assume:bdy} is needed for our theoretical guarantee, but the GeoIS algorithm itself uses only the second-order derivative of $\phi_0$ at $0_d$.
\end{remark}

Now, we apply a sequence of bijective transformations both to the event $\Dorig$ and to the random variable $U\sim\piorig=\mathcal N(0_{d+1},\la_0^{-1}I_{d+1})$. First, take $u\mapsto (Q,n)^\top u$. This moves the instanton from $u^*$ to the point $(0_d,\ell)$ on the $(d+1)$st axis, and leaves $\piorig$ unchanged. Next, apply $u\mapsto \frac1\ell u$. This compresses the event and concentrates the Gaussian to have new variance $(\ell^2\la_0)^{-1}$. Finally, apply $u\mapsto u-(0_d,1)$, which shifts the event and Gaussian both down by $1$ on the $(d+1)$st axis. Together, these transformations give
\begin{equation}\label{D0toD}\begin{gathered}
         p=\piorig(\Dorig)=\Pp(U\in\Dorig) = \Pp(\tau(U) \in \tau(\Dorig))=\pi(D),\\
    D = \tau(\Dorig),\quad\pi=  \mathcal N\bigl((0_d,-1),\la^{-1}I_{d+1}\bigr),\quad \la=\ell^2\la_0,
\end{gathered}
\end{equation} where $\tau(u) = \tfrac1\ell(Q,n)^\top u - (0_d,1)$. Thus the Gaussian center is now at $(0_d,-1)$, the new instanton is at the origin, and the distance between the two is 1 rather than $\ell$. We can think of $\la=\ell^2\la_0$ as an aggregate rarity parameter. The relationship between $\Dorig$ and $D$ implies the following relationship between $\phi_0$ and the boundary $\phi$ of $D$:
$$\phi(x) = \phi_0(\ell x)/\ell, \qquad H=I_d+\nabla^2\phi(0_d).$$ 
Locally, $D$ can be described by $\{(x,t)\,:\,t\geq\phi(x)\}$.

\subsection{Importance sampling preliminaries and GeoIS algorithm}\label{sec:sampler}
We motivate and describe GeoIS in the new coordinates introduced at the end of the previous section. Algorithm~\ref{alg:geoIS} gives the implementation in original coordinates. 

The general importance-sampling estimator of
$p=\pinew(D)$ based on a proposal density $\hat\pi$ is
\begin{equation}\label{IS}
\hat p_N
=
\frac1N\sum_{i=1}^N
\1\{(X_i,T_i)\in D\}
\frac{\pinew(X_i,T_i)}{\hat\pi(X_i,T_i)},
\qquad
(X_i,T_i)\iid\hat\pi.
\end{equation}
Here $\pinew$ is the density of $\mathcal N((0_d,-1),\la^{-1}I_{d+1})$. Let $\pi_D(x,t):=\frac1p\pinew(x,t)\1\{(x,t)\in D\}$
denote the density $\pi$ conditioned on the rare event. Then the relative bias
of~\eqref{IS} is
$$
\E[\hat p_N/p]-1
=
-\pi_D(D\setminus\operatorname{supp}(\hat\pi)).
$$
This bias cannot be reduced by drawing more samples from the same proposal,
so we seek a proposal whose support contains $D$. For such proposals,
$\hat p_N$ is unbiased and
\begin{equation}\label{var-chi2}
\CV^2(\hat p_N)
=
\Var(\hat p_N/p)
=
\frac1N\chi^2(\pi_D\Vert\hat\pi),
\end{equation}
where $\chi^2(\rho\Vert\mu)=\int\left(\frac{d\rho}{d\mu}\right)^2\,d\mu-1$ is the chi-squared divergence between $\rho$ and $\mu$. This motivates the following design principle:
$$
\text{construct a tractable $\hat\pi$ that approximates $\pi_D$
and whose support contains $D$.}
$$
To construct a tractable approximation to $\pi_D$, we start with a tractable approximation to the event $D$, which requires only the boundary's second derivative at the instanton.  Namely, let $\hat\phi(x)=\tfrac12x^\top \nabla^2\phi(0_d)x$ be the quadratic Taylor approximation to $\phi$, noting that $\phi(0_d)=0$ and $\nabla\phi(0_d)=0_d$, since this is true for $\phi_0$. We then use $D\approx \{(x,t)\,:\,t\geq\hat\phi(x)\}$. Next, we study the density of $\pi=\mathcal N((0_d,-1),\la^{-1}I_{d+1})$  in $(x,y)$ coordinates, where $y=t-\hat\phi(x)$ measures vertical displacement from the approximate boundary. We have the following key identity, whose short proof can be found in Section~\ref{subsec:MSE}:
\begin{align}
  \pinew(x,\hat\phi(x)+y) &= \psorm\, \varrho(x) \la e^{-\la y} e^{-\frac\la2(\hat\phi(x)+y)^2},\label{key-identity}\\ 
  \psorm &:=
\frac{e^{-\la/2}}{\sqrt{2\pi\la \det H}}.\label{psorm}
\end{align}
Here, $\psorm$ is precisely the SORM approximation to $p$, and $\varrho$ is the density of $\mathcal N(0_d, (\la H)^{-1})$. The identity~\eqref{key-identity} is valid for every $(x,y)\in\R^d\times\R$. From here, the density of $\pi_D$ is obtained by dividing~\eqref{key-identity} by $p=\pi(D)$ and restricting the support to $(x,y)$ such that $(x,\hat\phi(x)+y)\in D$.
To motivate the proposal, consider the regime $\la\to\infty$ with $d$ and the normalized event $D$ fixed. The following approximations describe the density near the instanton $(0_d,0)$:
\begin{equation}\label{key-approx}\begin{split}
\pi_D(x,\hat\phi(x)+y) &= \frac{\psorm}{p} \varrho(x) \la e^{-\la y} e^{-\frac\la2(\hat\phi(x)+y)^2}\1\{(x,\hat\phi(x)+y)\in D\}  \\
&\approx \varrho(x) \la e^{-\la y} e^{-\frac\la2(\hat\phi(x)+y)^2}\1\{y\geq 0\}\\
&\approx \varrho(x) \la e^{-\la y}\1\{y\geq 0\}.
\end{split}
\end{equation}
The first approximation replaces the event by its quadratic approximation and uses $p_{\mathrm{SORM}}/p\approx1$ (as is well-known for fixed $d,D$, and $\la_0\to\infty$). To justify the second approximation, recall $\varrho$ is the density of $X\sim\mathcal N(0_d,(\la H)^{-1})$ and note $\la e^{-\la y}\1\{y\geq0\}$ is the density of $Y\sim\Exp(\la)$. Since $\sqrt{\la}X$ and $\la Y$ have distributions independent of $\la$, and $\hat\phi$ is quadratic, $A:=\la(\hat\phi(X)+Y)$ also has a distribution independent of $\la$. Consequently, $e^{-\frac\la2(\hat\phi(X)+Y)^2}=e^{-A^2/2\la}$ converges to 1 in probability, as $\la\to\infty$. Thus dropping this factor is justified for typical draws of the Gaussian--exponential product. The above argument leads to the approximation $\pi_D \approx\mathrm{Law}(X,\hat\phi(X)+Y)$, where
\begin{equation}\label{XYdef}
 X\sim \mathcal N(0_d,(\la H)^{-1}), \;Y\sim\Exp(\la),\; X\perp Y.   
\end{equation}
The accuracy of this approximation is quantified in total variation distance in~\cite{katsevich2025asymptotic}. 

Next, we address the second aspect of the design principle: the support of the proposal distribution should contain $D$. The above law is only supported above the quadratic approximation to the boundary and may miss mass in the true event $D$. For this reason, we add a second mixture component to our proposal, supported in the region $\{(x,\hat\phi(x)-y)\,:\,x\in\R^d,\,y\ge0\}$. 
Substituting $-y$ for $y$ in~\eqref{key-identity} preserves the Gaussian factor $\varrho(x)$, motivating the same choice $X\sim\mathcal N(0_d,(\la H)^{-1})$ for the lower component. We therefore take this component to be the law of $(X,\hat\phi(X)-Y)$, where $Y\geq0$ is independent of $X$. We again choose $Y\sim\Exp(\la)$. The justification for this choice is deferred to Section~\ref{subsec:discuss}.

The above rationale leads us to our choice of proposal distribution, $\hat\pi = w\hat\pi_++(1-w)\hat\pi_-$. Here $w\in (0,1)$, $\hat\pi_\pm$ is the density of $(X,\hat\phi(X)\pm Y)$, and $X,Y$ are as in~\eqref{XYdef}. We substitute this $\hat\pi$ into the general IS estimator~\eqref{IS}, and draw the random vectors $(X_i, T_i)$ i.i.d. from $\hat\pi$ by setting
\begin{equation}\label{XTYJ}
    (X_i, T_i)=\begin{cases}
    (X_i,\,\hat\phi(X_i)+Y_i), &J_i=1,\\
    (X_i, \,\hat\phi(X_i)-Y_i), &J_i=0.
\end{cases} 
\end{equation}
Here,
$J_i$ are drawn i.i.d. from $\mathrm{Bernoulli}(w)$ and $(X_i, Y_i)$ are drawn i.i.d. from~\eqref{XYdef}, and independently of the $J_i$. Using the identity~\eqref{key-identity} to simplify~\eqref{IS}, the resulting estimator $\hat p_N$ can be written in the following form:
\begin{equation}\label{hatpgeo}
\hat p_N=\left(\frac1N\sum_{i=1}^N W_i\right)\psorm,
\end{equation} where
\begin{align}
W_i&=\1\{(X_i,T_i)\in D\}\,e^{-\la T_i^2/2}\,
\begin{cases}
\,1/w,&J_i=1,\\
\,e^{2\la Y_i}/(1-w),&J_i=0.
\end{cases}\label{tildWi-components}\end{align}
The connection between~\eqref{hatpgeo} and~\eqref{IS} is that $W_i\,\psorm=(\pi/\hat\pi)(X_i,T_i)\1\{(X_i,T_i)\in D\}$. %

Algorithm~\ref{alg:geoIS} summarizes the sampling and weight calculations in the original coordinates.
	\begin{algorithm}[H]
		\captionsetup{labelfont={bf}}
		\caption{GeoIS for estimating $p=\pi_0(D_0)$}\label{alg:geoIS}
		\algrenewcommand{\algorithmicrequire}{\textbf{Input:}}
		\begin{algorithmic}[1]
        \setlength{\itemsep}{0.25em}
        \Statex
\hspace*{\dimexpr-\labelwidth-\labelsep\relax}%
\parbox[t]{\dimexpr\linewidth+\labelwidth+\labelsep\relax}{\textbf{Input:} distribution $\pi_0=\mathcal N(0_{d+1},\la_0^{-1}I_{d+1})$, rare-event set $D_0$ with unique instanton and locally twice-differentiable boundary,  mixture weight $w\in(0,1)$, and sample size $N$.}
			\State \textbf{Prestage:} obtain instanton and geometric information:
			\Statex \hspace{\algorithmicindent}Solve for instanton $u^*$ from \eqref{eq:instanton}, set $\ell=\|u^*\|$ and $n=u^*/\ell$. 
			\Statex \hspace{\algorithmicindent}\parbox[t]{\dimexpr\linewidth-\algorithmicindent\relax}{Build matrix $Q\in \R^{(d+1)\times d}$ following \eqref{eq:Q}; represent $D_0$ boundary as
$u=u^*+Qv+\phi_0(v)n$ near $v=0_d$.} 
\Statex \hspace{\algorithmicindent}Compute boundary Hessian $\nabla^2\phi_0(0_d)$.
			\Statex \hspace{\algorithmicindent}Set $H= I_d+\ell\nabla^2\phi_0(0_d)$, $\hat\phi(x)=\tfrac12x^\top (\ell\nabla^2\phi_0(0_d))x$, and $\la=\ell^2\la_0$.
            \Statex \hspace{\algorithmicindent}\textbf{Require} $H\succ0$.
            \Statex \hspace{\algorithmicindent}Compute $p_{\mathrm{SORM}}$ as in~\eqref{psorm}.
			\State \textbf{Sampling:} Draw proposal samples and compute weights:
			\Statex For {$i = 1,\ldots,N$}
	       \Statex\hspace{\algorithmicindent} 
            \parbox[t]{\dimexpr\linewidth-\algorithmicindent\relax}{%
Draw $X_i\sim\mathcal N(0_d,(\la H)^{-1})$,
$Y_i\sim\operatorname{Exp}(\la)$,
$J_i\sim\operatorname{Bernoulli}(w)$, independently of one another
and of all previous draws $j<i$.}\Statex\hspace{\algorithmicindent} Set $T_i=\hat\phi(X_i)+Y_i$ if $J_i=1$ and $T_i=\hat\phi(X_i)-Y_i$ if $J_i=0$.
			\Statex \hspace{\algorithmicindent} Map to the original coordinates: $U_i= u^*+\ell(QX_i+T_i n)$.
			\Statex \hspace{\algorithmicindent} Compute \begin{equation}\label{tildW}
			     W_i= 
			\1\{U_i\in D_0\}e^{-\frac\la2T_i^2}\begin{cases}
			    1/w, &J_i=1,\\
                e^{2\la Y_i}/(1-w), &J_i=0.
			\end{cases}.\end{equation}
			\State \Return estimated probability $\hat p_N=(\frac1N\sum_{i=1}^N W_i)\,\psorm$
as in~\eqref{hatpgeo}.
		\end{algorithmic}
	\end{algorithm}
Note that $(X_i, T_i)=\tau(U_i)$, where $\tau$ is defined below~\eqref{D0toD}. Thus $U_i\in D_0$ is equivalent to $(X_i,T_i)\in D$, so~\eqref{tildW} is the same as~\eqref{tildWi-components}.

\begin{figure}%
    \centering
     \includegraphics[width=0.6\textwidth]{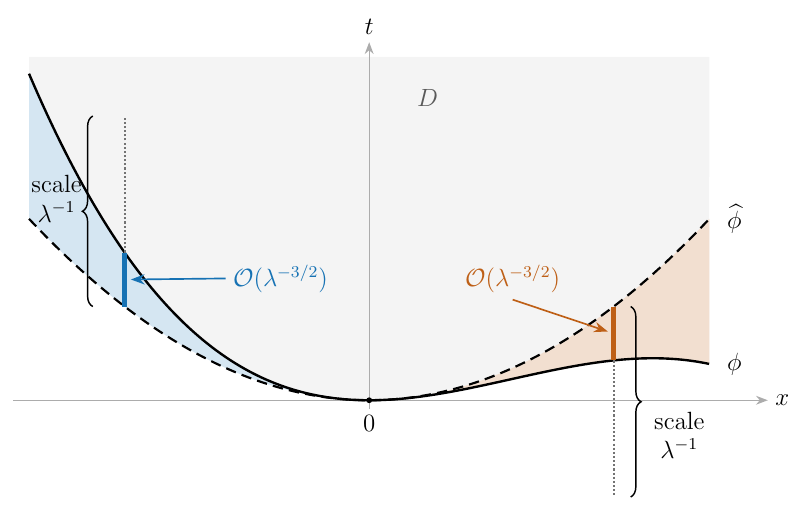}
    \caption{Samples $(X_i,\hat\phi(X_i)+Y_i)$ from the upper mixture component fall outside of $D$ when they land in the lefthand, light-blue region, in which $\phi$ lies above $\hat\phi$. For this to happen, the sampled $Y_i$ must be no larger  than the height of the strip, which scales as $(\phi-\hat\phi)(X_i)\sim \|X_i\|^3=\mathcal O(\la^{-3/2})$ for typical values of $X_i$. But $Y_i\sim\Exp(\la)$, so typical draws of $\hat\phi(X_i)+Y_i$ (see the dotted vertical line on the left) range between $\hat\phi(X_i)$ and $\hat\phi(X_i)+c/\la$. Therefore, drawing from the strip is unlikely. Thus samples from the upper mixture component are highly likely to fall in $D$. Symmetrically, samples $(X_i,\hat\phi(X_i)-Y_i)$ from the lower mixture component fall inside $D$ when they land in the righthand, orange region, in which $\phi$ lies below $\hat\phi$. Typical draws of $\hat\phi(X_i)-Y_i$ (see the dotted vertical line on the right) range between $\hat\phi(X_i)$ and $\hat\phi(X_i)-c/\la$, so drawing from the strip of height $\mathcal O(\la^{-3/2})$ is unlikely. Thus samples from the lower mixture component are highly likely to fall outside of $D$.}
    \label{fig:W}
\end{figure}

Since $\hat p_N$ is unbiased by construction, we see from~\eqref{hatpgeo} that the role of the weights $ W_i$ is to, on average, correct the approximation error due to SORM: $\E[\frac1N\sum_{i=1}^N\ W_i] =  p/\psorm$. 
Meanwhile, the variance of the $W_i$ determines the variance of $\hat p_N$. To understand the behavior of the  $W_i$, assume again for simplicity that $D,d$ are fixed. We claim that \begin{equation}\label{law-Wi}
    \mathrm{Law}( W_i) \Rightarrow \frac1w\mathrm{Bernoulli}(w), \qquad \la\to\infty.
\end{equation}
Indeed, $W_i$ is nonzero only when $(X_i,T_i)\in D$ and as explained in Figure~\ref{fig:W}, samples from the upper and lower mixture components lie inside $D$ with probability tending to one and zero, respectively. The latter limit implies $\mathrm{Law}(W_i\mid J_i=0)\Rightarrow\delta_0$.  Moreover, conditional on $J_i=1$, we have $\la T_i^2\to0$ in probability, as explained below~\eqref{key-approx}. The weight formula~\eqref{tildWi-components} therefore gives $\mathrm{Law}(W_i\mid J_i=1)\Rightarrow\delta_{1/w}$. Combining these limits with $J_i\sim\mathrm{Bernoulli}(w)$ yields~\eqref{law-Wi}.

The variance of the limiting law in~\eqref{law-Wi} is $w^{-1}-1$, suggesting $\Var(\frac1N\sum_{i=1}^NW_i)\approx \frac1N(w^{-1}-1)$. Recalling~\eqref{hatpgeo}, this predicts $\Var(\hat p_N/p)\approx \frac1N(w^{-1}-1)$ as well, using $\psorm\approx p$. In Proposition~\ref{prop:kappa}, we confirm this heuristic argument, proving that $\frac1N(w^{-1}-1)$ is indeed the dominant contribution to the localized relative MSE of $\hat p_N$. Note that bounding the MSE requires second-moment control of $W_i$, whereas the above informal argument only justifies control of $W_i$ in  probability.

The purpose of $\hat\pi_-$ is to cover the part of $D$ that is missed by $\hat\pi_+$. Since the argument in Figure~\ref{fig:W} shows samples from $\hat\pi_-$ rarely fall inside $D$, our choice $Y\sim\Exp(\la)$ for the downward sampling may seem wasteful. The reason to do this is to safeguard against less likely, larger draws of $X_i$, for which the strip above the true boundary and below the approximate quadratic boundary is very wide. Samples drawn toward the bottom of this wide strip receive large importance weights if $Y$ is overly-concentrated near the top of the strip. We give a more quantitative explanation for this choice in Section~\ref{subsec:discuss}.

\paragraph*{GeoIS for level-set events}
A common rare-event setting is the  case in which a response function $F:\R^{d+1}\to\R$ maps the random parameter to an observable, and $\Dorig:=\{u\in\R^{d+1}:F(u)\geq z\}$ for a given threshold $z$. If $F$ is twice continuously differentiable near the instanton $u^*$ and $\nabla F(u^*)\neq0$, then first-order optimality gives $n=\nabla F(u^*)/\|\nabla F(u^*)\|$. Implicit differentiation of the boundary yields
\begin{equation}\label{eq:implicit-BH}
\begin{split}
H &= Q^\top\left(I_{d+1}-\frac{\ell}{\|\nabla F(u^*)\|}\nabla^2F(u^*)\right)Q,
\end{split}
\end{equation} where $\ell=\|u^*\|$ as usual. Thus GeoIS for level-set events involves the instanton $u^*$ and the first and second derivatives of $F$ at this point.

\subsection{Assumptions and error guarantee}\label{subsec:theory}
To state our guarantee for $\hat p_N$, we need one more assumption and a few definitions. 
\begin{definition}[$H$-weighted tangential neighborhoods]
For $r>0$, define
$$
\us_r:=\{x\in\R^d:\|H^{1/2}x\|\leq r\}.
$$
\end{definition}

\begin{assumption}[Local cap and no return]\label{assume:no-return}
Let $V_0$ be as in Assumption~\ref{assume:bdy}. There exists $r_1>0$ such that 
$\ell\us_{r_1}\subseteq V_0$ 
and $\gamma>0$ such that
\begin{align}
    \phi(x)-\frac12x^\top \nabla^2\phi(0_d)x&\geq-\frac13x^\top Hx\qquad \forall\; x\in\us_{r_1},\tag{local cap}\\
\|(x,t)-(0_d,-1)\|^2&\geq1+\gamma,\qquad \forall\; (x,t)\in\Dnew\cap (\us_{r_1}^c\times\R).\tag{no return}
\end{align}
\end{assumption}
The local cap condition prevents the true boundary $\phi(x)$ from falling too far below its quadratic approximation. The no-return condition says that, after leaving the local neighborhood $\us_{r_1}$, every point in $D$ has Gaussian cost uniformly larger than the cost at the instanton. 
\begin{remark}[Local vs global conditions]
    The strip part of Assumption~\ref{assume:bdy} and local cap part of Assumption~\ref{assume:no-return} pertain to local properties of $\Dorig$ (equivalently, of $\Dnew$) and its boundary near the instanton. In contrast, the one-sidedness part of Assumption~\ref{assume:bdy} and no-return part of Assumption~\ref{assume:no-return} require global information about $\Dorig$.
\end{remark}
\begin{definition}[$H$-weighted derivative norms]\label{def:norms}
For $j=2,3,4$ and any $\kappa\geq0$ such that $\ell\us_\kappa\subseteq V_0$, define
$$
\tau_j(\kappa):=\sup_{x\in\us_\kappa}\|\nabla^j\phi(x)\|_H,
$$ where $\|\cdot\|_H$ is as in~\eqref{F-H}.
\end{definition}
\noindent In particular,
$$
\tau_2(0)=\|\nabla^2\phi(0_d)\|_H=\|I_d-H^{-1}\|=\max_{i=1,\dots,d}\frac{|b_i|}{1+b_i},
$$
where $b_1,\dots,b_d$ are the eigenvalues of $\nabla^2\phi(0_d)$. The derivative quantities can also be expressed in the original coordinates. Since $\phi(x)=\phi_0(\ell x)/\ell$,
\begin{equation}\label{eq:weighted-rescaling}
\tau_j(\kappa)=\ell^{j-1}\sup_{v\in\ell\us_\kappa}\|\nabla^j\phi_0(v)\|_H,\qquad j=2,3,4,
\end{equation}

Next, we specify a suitable local region $\us_\kappa$ used in the statement and proof of our main result. Fix $0<\rho<1/2$.
\begin{definition}[$\rho$-admissible localization radius]\label{def:kappa}
A radius $\kappa>0$ is $\rho$-admissible if
\begin{equation}\label{kap:admit}
\frac{12}{\lambda}
\left(3d\log\frac{\lambda}{d}+2(1\vee\log\det H)\right)
\leq\kappa^2\leq\min(r_1^2,3\gamma,4\mu_0/\ell),\qquad \kappa\tau_3(\kappa)\leq\rho,
\end{equation} where $\mu_0$ is as in Assumption~\ref{assume:bdy} and $r_1,\gamma$ are as in Assumption~\ref{assume:no-return}.
\end{definition}
The lower bound on $\kappa$ ensures the probability of leaving $\us_\kappa$ is sufficiently small, and controls the relative-error contributions arising from localizing the estimator; see~\eqref{hatploc1} below. In particular, the term $\log \det H$ in~\eqref{kap:admit} appears because relative error normalizes by $p$, which under our assumptions is no less than $\psorm/2$. In turn, $\psorm$ involves $\det H$, recalling~\eqref{psorm}. The upper bounds ensure that the local geometric assumptions hold within $\us_\kappa$. Finally, the bound $\kappa\tau_3(\kappa)\leq\rho$ controls the variation of the boundary curvature inside $\us_\kappa$.

We will measure the performance of $\hat p_N$ by comparing it to a coupled localized estimator $\hat p_{\loc}$, which never enters a lower-tail region:
\begin{equation}\label{hatploc1}
    \begin{split}
        \hat p_{\loc}&:
=p_{\mathrm{SORM}}\frac{1}{N}\sum_{i=1}^N
 W_i\,\1\{J_i=1\ \text{or}\ X_i\in\mathcal U_\kappa\},
    \end{split}
\end{equation} where the $ W_i$ are as in~\eqref{tildWi-components}. This localized estimator sets weights to zero whenever $(X_i,T_i)$ is drawn from the lower component ($J_i=0$) and $X_i\in\us_\kappa^c$. 
\begin{proposition}\label{prop:kappa}
Suppose Assumptions~\ref{assume:bdy} and~\ref{assume:no-return} hold, $\la\geq2(d+1)$, and $\kappa$ is $\rho$-admissible. Let
\begin{equation}\label{def:epsilon-234}\begin{gathered}
\epsilon_2:=\tau_2(0)\frac d{\sqrt\la},\qquad \epsilon_3:=\tau_3(0)\frac d{\sqrt\la},\qquad \epsilon_4^2:=\tau_4(\kappa)\frac{d^2}{\la}, \\
 \mathcal R = (\epsilon_2\vee\epsilon_3\vee\epsilon_4)^2 + \epsilon_4^4/d.
\end{gathered}\end{equation}Then there is an absolute constant $C>0$ and a constant $C_\rho$ depending only on $\rho$ such that
\begin{align}
\Pp(\hat p_N=\hat p_{\loc})&\geq1-Ne^{-\la\kappa^2/4},\label{eq:kappa-prob-clean}\\
\CV_\loc^2:=\E\left[\left(\frac{\hat p_{\loc}}{p}-1\right)^2\right]&=\frac1N\left\{w^{-1}-1+\frac{\epsilon_+}{w}+\frac{\epsilon_-}{1-w}\right\}+\mathcal E_{\exp},\label{MSE-decomp}
\end{align}
where 
\begin{align}
|\epsilon_+|&\lesssim_\rho \,e^{C_\rho\mathcal R}\,\left[\epsilon_3+ \epsilon_2^2+\epsilon_4^2 +\la^{-1}\right],\label{MSE-err+}\\[4pt]
0\leq\epsilon_-&\lesssim_\rho \,e^{C_\rho\mathcal R}\,\left[\epsilon_3+\epsilon_4^2\right],\label{MSE-err-}\\[4pt]
0\leq\mathcal E_{\exp}&\lesssim e^{C\mathcal R}\,\left[e^{-\la\kappa^2/24}+N^{-1} e^{-\la\kappa^2/48}\right].
\label{Eexpbd}
\end{align}
In particular, if $\epsilon_2\vee\epsilon_3\vee\epsilon_4 \leq \eta$ for any $\eta$, then
\begin{equation}\label{MSE-2}
\begin{gathered}
|\epsilon_+|\lesssim_{\eta,\rho} \,\epsilon_3 + \epsilon_2^2+\epsilon_4^2+\la^{-1}, \qquad 0\leq\epsilon_-\lesssim_{\eta,\rho}\, \epsilon_3+\epsilon_4^2, \\[4pt]
 0\leq\mathcal E_{\exp}\lesssim_{\eta} e^{-\la\kappa^2/24}+N^{-1} e^{-\la\kappa^2/48}.
\end{gathered}\end{equation}
\end{proposition}
The proof is given in \Cref{sec:proof-kappa}. By taking e.g. $\rho=1/4$ in~\eqref{kap:admit}, the bounds in~\eqref{MSE-err+}-\eqref{Eexpbd} are indeed explicit up to absolute constants, as claimed in the introduction. We discuss sufficient conditions for verifying the assumptions of the proposition in Appendix~\ref{app:asymp}, in the case where $\Dorig$ is an epigraph or a super level set of a function on $\R^{d+1}$.
\begin{remark}\label{rk:cv}
    Note that $\CV^2_\loc$ is the relative MSE of $\hat p_\loc$, and thus satisfies the bias-variance decomposition, in relative form: $\CV^2_\loc=(\E[\hat p_\loc/p]-1)^2 + \Var(\hat p_\loc/p)$. The second term is known as the squared coefficient of variation of $\hat p_\loc$, or $\CV^2(\hat p_\loc)$. Thus $\CV^2_\loc$ is equal to the sum of the squared relative bias term (which turns out to be exponentially small) and $\CV^2(\hat p_\loc)$.  
\end{remark}
\paragraph*{Interpretation.}
The leading term $(w^{-1}-1)/N$ in~\eqref{MSE-decomp} is the contribution predicted by the approximation $\mathrm{Law}( W_i)\approx \mathrm{Bernoulli}(w)/w$, as discussed below Algorithm~\ref{alg:geoIS}. The terms $\epsilon_\pm$ are corrections to this limiting description. Specifically, $\epsilon_+$ measures how much $\pi_D$ deviates from $\hat\pi_+$ in the upper region, and $\epsilon_-$ is a second moment of $\pi_D/\hat\pi_-$ in the lower, local region. Only draws in the strip above $\phi$ and below $\hat\phi$ contribute to $\epsilon_-$; see the green triangle in Figure~\ref{fig:region}. Thus $\epsilon_-$ is small when the strip is narrow and $\pi_D/\hat\pi_-$  is controlled there. We discuss how to choose the lower proposal $\hat\pi_-$ to ensure this in Section~\ref{subsec:discuss}. The exponentially small remainder $\mathcal E_{\exp}$ in the relative MSE decomposition of $\hat p_\loc$ arises from the bias term in the bias-variance decomposition. Note that $\hat p_\loc$ is biased because its support omits mass in $D$. 

As a first, narrow reading of the proposition, consider fixed $d$ and $D$. Then the generically dominant term in the bounds on $\epsilon_\pm$ is $\epsilon_3=\tau_3(0)d/\sqrt{\lambda}\sim\la^{-1/2}$. More broadly, the proposition quantifies how dimension, rarity, and boundary geometry jointly determine the size of the corrections $\epsilon_\pm$. The result is extremely flexible and can be used in a broad range of regimes, because it holds without any assumptions on the magnitude of the terms $\epsilon_2,\epsilon_3,\epsilon_4$ appearing in the bounds. The only nontrivial assumption is that $\la\geq 2(d+1)$ and that the lower bound on $\kappa^2$ in the first inequality in~\eqref{kap:admit} is smaller than the upper bound on $\kappa^2$.

\paragraph*{Comparison to SORM.}
The relationship between GeoIS and SORM is clear from~\eqref{hatpgeo}: the GeoIS estimator multiplies $p_{\mathrm{SORM}}$ by the correction factor $\frac1N\sum_{i=1}^N W_i$, which is a random unbiased estimate of $p/\psorm$. Thus SORM has good relative accuracy if $p/\psorm$ is close to 1 (by definition), while GeoIS has good relative accuracy if the random \emph{estimator} of $p/\psorm$ has low variance. Interestingly, both relative accuracies depend on the same geometric quantities $\epsilon_2,\epsilon_3,\epsilon_4$. Indeed, for GeoIS, these terms appear in our bounds~\eqref{MSE-err+}-\eqref{MSE-err-} on $\epsilon_\pm$. For high-dimensional SORM,~\cite{katsevich2025asymptotic} shows $|p/\psorm-1|\ll1$ if $\epsilon_j\ll1$, $j=2,3,4$. But for GeoIS, the $\epsilon_j$ need \emph{not} be small, since additional sampling (larger $N$) can be used to compensate for large constant values of $\epsilon_j$, thereby improving the accuracy of GeoIS to the needed tolerance.

\paragraph*{Role of localization.}
    The estimator $\hat p_\loc$ is only a theoretical device. We can control the relative error of $\hat p_N$ another way which does not explicitly depend on $\hat p_\loc$. Namely, Markov's inequality gives that for any $\delta>0$,
\begin{equation}\label{in-prob}
\Pp\left(\left|\frac{\hat p_N}{p}-1\right|\geq \delta\right) \leq \frac{\CV^2_\loc}{\delta^2}+ \Pp(\hat p_N\neq\hat p_\loc),
\end{equation} and the two quantities on the righthand side can be bounded using~\eqref{eq:kappa-prob-clean} -\eqref{Eexpbd}. We have chosen to formulate our results as in Proposition~\ref{prop:kappa} because together, the second moment bound~\eqref{MSE-decomp} and high-probability bound~\eqref{eq:kappa-prob-clean} are stronger than~\eqref{in-prob}. Furthermore,~\eqref{eq:kappa-prob-clean}-\eqref{MSE-decomp} give insight into the algorithm behavior: the proposal samples never enter a certain tail region with high probability, and on this event, the relative MSE of $\hat p_N=\hat p_\loc$ is controlled by~\eqref{MSE-decomp}.

\paragraph*{Choosing $N$ and $\kappa$.}
Combining~\eqref{eq:kappa-prob-clean} with~\eqref{MSE-decomp} and~\eqref{MSE-2} we get that for all $N\leq e^{\la\kappa^2/24}$, \begin{equation}\label{N-kappa}
       \Pp(\hat p_N=\hat p_\loc)\geq1-e^{-5\la\kappa^2/24},\qquad \CV_\loc^2\lesssim_{\rho,\eta}\frac{1}{Nw(1-w)}.
   \end{equation} To get the latter bound, we simply substitute the coarse bounds $|\epsilon_\pm|\leq C(\rho,\eta)$, $\mathcal E_{\exp}\lesssim_\eta 1/N$ into~\eqref{MSE-decomp}. Thus we are free to take $N$ as large as needed, up to  $N=e^{\la\kappa^2/24}$, to drive down $\CV^2_\loc$ while keeping $\Pp(\hat p_N=\hat p_\loc)$ close to 1. Recall that $\kappa$ itself is a theoretical instrument. To obtain a rigorous guarantee on the performance of the method for some desired large value of $N$, one can choose a larger $\kappa$, provided it continues to satisfy the admissibility condition~\eqref{kap:admit}. %

\subsection{Choice of lower mixture component}\label{subsec:discuss}
In Section~\ref{sec:sampler}, we have already motivated the form $\hat\pi_- = \mathrm{Law}(X,\hat\phi(X)-Y)$, where $X\sim\mathcal N(0_d, (\la H)^{-1})$ and $Y\geq0$ is independent of $X$. Thus it remains to motivate the particular choice of distribution $Y\sim\Exp(\la)$ used in $\hat\pi_-$. We do so in the regime of fixed $d$ and $D$, fixed mixture weight $w\in(0,1)$, and $\la\to\infty$.

The lower component enters the MSE decomposition~\eqref{MSE-decomp} through $\epsilon_-$. As a minimal criterion, the distribution of $Y$ should be chosen to ensure $\epsilon_-=o(1)$, to avoid increasing the leading MSE term $(w^{-1}-1)/N$.  
Now, we show in Lemma~\ref{lma:loc-MSE} that $\epsilon_- =\int(\pi_D/\hat\pi_-)^2(x,t)\1_{\us_\kappa}(x)\hat\pi_-(x,t)dxdt$. Let $\varrho_Y$ for the density of $Y$ to be chosen, so that $\hat\pi_-(x,\hat\phi(x)-y)=\varrho(x)\varrho_Y(y)$. Using the formula for $\epsilon_-$ and the identity in the first line of~\eqref{key-approx}, we obtain
 \begin{equation}\label{ER}
     0\leq \epsilon_-\leq \left(\frac{p_{\mathrm{SORM}}}{p}\right)^2 \E\left[\1\{X\in\mathcal U_\kappa\}\int_0^{r(X)_+}\frac{\lambda^2e^{2\lambda y}}{\varrho_Y(y)}\,dy\right],\qquad r(x) = \hat\phi(x)-\phi(x).
 \end{equation} Here, we bounded the factor $e^{-\la(\hat\phi(X)-y)^2}$ from~\eqref{key-approx} by 1. See Section~\ref{subsec:MSE} for more details on the above calculation. The factor $(\psorm/p)^2$ is not affected by the choice of $\varrho_Y$, so we focus only on  the second term, the expectation. The inner integral in this expectation motivates the form $Y\sim\Exp(m)$, for some $m>0$ to be chosen. Indeed, this family is a natural choice given the form of the numerator, and allows us to take the integral in $y$ explicitly. Further, it provides sufficient flexibility to optimize the resulting bound with respect to $m$, which we do now. Substituting $\varrho_Y(y)=me^{-my}$ above, taking the integral in $y$, and using $e^{b} - 1\leq be^{b}$ for $b\geq0$ gives
 \begin{equation}\label{E-exp}\begin{split}
\E\left[\1\{X\in\mathcal U_\kappa\}\int_0^{r(X)_+}\frac{\lambda^2e^{2\lambda y}}{\varrho_Y(y)}\,dy\right]&\leq \frac{\la}{m}\E\left[(\la r(X))_+e^{(2\la+m) r(X)_+}\1_{\us_\kappa}(X)\right]\\
& \lesssim \frac{\sqrt{\la}}m\left(\E\left[e^{C(\la+m)\|X\|^3}\1\{\|X\|\leq C\kappa\}\right]\right)^{1/2}.
 \end{split}
 \end{equation}
 The second inequality can be formally proved using Cauchy--Schwarz and a Taylor remainder bound. Informally, it follows by noting $(\la r(X))_+ \lesssim \la^{-1/2}$, since $X\sim\la^{-1/2}$ and $r(x)$ is approximately cubic. To get the indicator in the second line, we assume $H$ is bounded below away from zero. %
The bound in the second line reveals a tradeoff in choosing $m$: increasing $m$ decreases the prefactor but increases the exponential expectation. We choose $m$ to ensure that the prefactor is as small as possible subject to (1) the exponential expectation remaining bounded by a constant, and (2) the chosen $m$ not overly restricting $\kappa$. 

Roughly speaking, the criterion ensuring (1) is that for all $\|x\|\lesssim \kappa$, the cubic $(\la+m)\|x\|^3$ should not overwhelm the negative quadratic $-\la x^\top Hx/2 \asymp -\la\|x\|^2$ in the exponent of the Gaussian density (assuming $H\asymp I_d$), when taking the integral that gives the Gaussian expectation. This is satisfied if $\kappa\leq c\la/(\la+m)$ for a sufficiently small but constant $c$. Thus ensuring (1) enforces an upper bound on the allowable $\kappa$ which depends on the chosen $m$. This explains the need for (2). The reason it is desirable to allow $\kappa$ to be large in the first place is that it extends the range of sample sizes $N$ covered by our performance guarantee, as discussed below~\eqref{N-kappa}. This can be interpreted as a sign of robustness of the algorithm.  
 
Now, consider the condition $\kappa\leq c\la/(\la+m)$. Even for $m=0$, this imposes the requirement that $\kappa$ be smaller than a fixed constant. We can freely increase $m$ up to  $\mathcal O(\la)$ without changing this requirement on $\kappa$ by more than a constant factor.  Subject to $m=\mathcal O(\la)$, choosing $m\asymp\la$ gives the best decay order in the prefactor $\sqrt{\la}/m$. This explains our choice $Y\sim\Exp(\la)$. 

Alternatively and more directly, consider the second line of~\eqref{E-exp}. The presence of $\la$ in the exponent shows there is no harm in increasing $m$ up to $\mathcal O(\la)$. The reasoning about $\kappa$ shows there \emph{is} harm in taking $m$ larger, and since $\epsilon_-$ is already negligible compared to the fixed $w^{-1}-1$, no additional benefit. We do not claim that this argument is appropriate in all settings. For example, when choosing $w$ adaptively, it can be advantageous to obtain a smaller bound on $\epsilon_-$ by increasing $m$, at the cost of requiring $\kappa$ to be smaller.

\section{Interpreting the guarantee in two representative regimes}\label{sec:asymp}
In this section, we study our performance guarantee in two common settings of increasing rarity. The goal is to simplify both the assumptions and conclusion of Proposition~\ref{prop:kappa} in these settings. We consider the rare probability from the perspective of a standard Gaussian:
$$p = \Pp(\mathcal N(0_{d+1}, I_{d+1})\in\sqrt{\la_0}\Dorig).$$ The two regimes we consider are dilation and translation. Dilation refers to letting $\la_0\to\infty$ and fixing $\Dorig$. Translation refers to $\la_0=1$ and $\Dorig = (0_d,\ell)+D_{00}$, with $\ell\to\infty$ and $D_{00}$ fixed.

\paragraph*{Dilation regime.} 
Consider a sequence of problems with $\Dorig$ remaining ``fixed" as $\lambda_0\to\infty$. If $d$ grows together with $\la_0$, then $\Dorig$ cannot literally remain fixed. However, the intuition can be formalized by supposing Assumptions~\ref{assume:bdy} and~\ref{assume:no-return} hold with positive constants $r_1,\gamma,\mu_0$ that are uniform along the sequence, that $\ell \asymp 1$, and that for a fixed $r_0\in(0,r_1]$, the quantities $\|H\|$, $\tau_2(0)$, $\tau_3(r_0)$, and $\tau_4(r_0)$ are uniformly bounded. Since all of these conditions pertain to $\Dorig$ only, uniformity of the geometric constants is natural in the ``fixed" $\Dorig$ regime.

In addition, assume $d/\lambda_0\to0$. Then there exists a constant $c>0$ such that the following bounds hold for all sufficiently large $\lambda_0$:
\begin{equation}\label{P-CV-asymp}
\begin{split}
    \Pp(\hat p_N=\hat p_\loc)&\geq 1-Ne^{-c\lambda_0},\\
    \CV_\loc^2&=\frac{1}{N}\bigg(w^{-1}-1+\frac{\mathcal O\bigl(e^{Cd^2/\la_0}\,d/\sqrt{\lambda_0}\bigr)}{w(1-w)}\bigg)+\mathcal O(e^{-c\lambda_0}).
\end{split}
\end{equation} The constant $c$ and the implicit constants in the big-$\mathcal O$'s are independent of $d,\lambda_0,w$, and $N$. Equation~\eqref{P-CV-asymp} follows by choosing a small but fixed $\kappa>0$, which is $(1/4)$-admissible for all $\lambda_0$ large enough. This is possible since the lower bound in~\eqref{kap:admit} goes to zero as $d/\lambda_0\to0$. We also use, in the exponent $\mathcal R$ defined in~\eqref{def:epsilon-234}, that $\epsilon_4^2 + \epsilon_4^4/d = \epsilon_4^2(1 + \tau_4(\kappa)d/\la)\leq Cd^2/\la_0$. This follows by boundedness of $\tau_4(r_0)$ and the assumption $d/\la_0\to0$.

This analysis gives different performance guarantees in three regimes. In the first, $d/\la_0\to0$ but $d^2/\la_0\to\infty$. Then~\eqref{P-CV-asymp} applies, but guarantees only that the relative error can be brought below a fixed tolerance with an exponentially large number of samples $N$. The second regime is the most fruitful pre-asymptotic regime: $d^2/\lambda_0\leq C$ for an arbitrarily large but constant $C$. In this regime, GeoIS is guaranteed to bring down the relative error below a fixed tolerance with a possibly large but fixed number of samples $N$. Finally, the third regime is the truly asymptotic one: $d^2/\la_0\to0$. Then we get the precise scaling $\CV^2_\loc = (w^{-1}-1 + o(1))/N + \mathcal O(e^{-c\la_0)}$.

\paragraph*{Translation regime.}
Next, fix $\la_0=1$ and let $D_0 = (0_d,\ell) + D_{00}$, with $\ell\to\infty$ and ``fixed" $D_{00}$. In this setting, we have nontrivial interaction between the rarity parameter $\la=\ell^2$ and the $\ell$-rescaled derivatives $\tau_j$, $j=2,3,4$. To understand this interaction without being weighed down by technicalities, assume $D_{00}$ is the exact epigraph $\{(x,t)\in\R^d\times\R\,:\, t\geq\phi_0(x)\}$. We consider $D_{00}$ fixed in the sense that $\phi_0$ should depend on $\ell$ through $d$, which is allowed to grow with $\ell$.

Now, recall that $H=I_d + \ell\nabla^2\phi_0(0_d)$ must be positive definite. To ensure this holds as $\ell\to\infty$, we see that $\nabla^2\phi_0(0_d)$ cannot have negative eigenvalues. This is a nontrivial distinction between this translation regime and the previous dilation regime, in which there is no such restriction on $\nabla^2\phi_0(0_d)$. Geometrically, the reason for this condition is that the Gaussian level set $\{\|u\|=\ell\}$  is increasingly flat as $\ell\to\infty$. 

To simplify the discussion further, we assume not only that $\nabla^2\phi_0(0_d)\succeq 0$ but that $\phi_0$ is globally $m$-strongly convex. Furthermore, assume $\phi_0$ is $C^4$ in $\{\|v\|\leq R_0\}$, with
$$
\|\nabla^2\phi_0(0_d)\|\leq C_2,\qquad \sup_{\|v\|\leq R_0}\|\nabla^j\phi_0(v)\|\leq C_j,\quad j=3,4,
$$
with constants $m,R_0>0$ and $C_2,C_3,C_4\geq0$ uniform along the sequence. Suppose also that $d(\log\ell)/\ell\to0$. Under these assumptions, we prove the following result at the end of Appendix~\ref{app:explicit}:
\begin{equation}\label{P-CV-asymp-ell}
\begin{split}
   \Pp(\hat p_N=\hat p_\loc)&\geq 1-Ne^{-c\ell}, \\
    \CV_\loc^2&=\frac{1}{N}\bigg(w^{-1}-1+\frac{\mathcal O\bigl(e^{Cd^2/\ell}\,d/\sqrt{\ell}\bigr)}{w(1-w)}\bigg)+\mathcal O(e^{-c\ell}). 
\end{split}
\end{equation} Here, $c,C$ and all implicit constants do not depend on $d,\ell,w$, or $N$.

To explain where the $d/\sqrt\ell$ scaling stems from, consider the term $\epsilon_3=\tau_3(0)d/\ell$ (recalling $\la_0=1$), which is generically the dominant term in the bounds ~\eqref{MSE-err+} and~\eqref{MSE-err-} on $\epsilon_\pm$. We now show $\tau_3(0)=\mathcal O( \sqrt\ell)$, yielding $d/\sqrt\ell$ in total. Under our assumptions, $H=I_d + \ell\nabla^2\phi_0(0_d) \succeq  m\ell I_d$. Thus, a third-order $H$-weighted operator norm is a factor $\ell^{3/2}$ smaller than its unweighted counterpart. Recalling~\eqref{eq:weighted-rescaling}, we then have $\tau_3(0) = \ell^2\|\nabla^3\phi_0(0)\|_H \lesssim_{C_3,m} \ell^2/\ell^{3/2}=\ell^{1/2}$, as claimed.

As in the previous asymptotic setup, there are three distinct regimes: when $d(\log\ell)/\ell\to0$ but $d^2/\ell\to\infty$, the guarantee for $\CV^2_\loc$ is valid, but the factor multiplying $N^{-1}$ blows up exponentially. When $d^2/\ell = \mathcal O(1)$, the factor is bounded above, and when $d^2/\ell\to0$, the factor is precisely $w^{-1}-1 + o(1)$.

\section{Theory and numerical simulations in examples}\label{sec:numerical}
In this section, we study four examples numerically \footnote{
Code and results for the numerical experiments are available at
\url{https://github.com/singularity0310/GeoIS-for-high-dim-rare-event}.}, and for the first three, we also provide rigorous theoretical guarantees.
In the first example, the rare event has an exact quadratic boundary. 
The second is the upper half of the exterior of an ellipsoid. 
The third is a portfolio-loss example with a nonquadratic boundary. 
The fourth is a diffusion PDE example with an implicitly defined rare-event boundary.  These four examples demonstrate how GeoIS compares to other methods  and how its performance depends on sample size, dimension, and rarity.

\paragraph{Methods.}
We compare GeoIS with several benchmark importance sampling methods: LSIS~\cite{tong2021extreme} and CE initialized with LSIS~\cite{tong2023large}. For the diffusion PDE example, we also include LAIS-dm~\cite{tong2023large} in the comparison. For CE, we treat LSIS as round-0. Each round of CE uses \(N\) samples both to evaluate the estimator and to fit the next proposal. %
Thus, CE round~\(r\) requires a total of \((r+1)N\) samples. See Appendix~\ref{app:baselines} for implementation details. 

As we demonstrate and explain in the quadratic example, CE breaks down at large $d$. We therefore omit CE from high-$d$ simulations for the other three examples, to focus on the better-performing methods.

For GeoIS, we use $w=0.5$ for the mixture weight in all four examples. We also consider a larger $w$. For
the exact quadratic boundary, the upper component covers the entire
event, so we can take $w=1$. In the other three examples, we take $w=0.9$.  Adaptively choosing $w$ based on the size of $\epsilon_-$ can improve $\CV^2_\loc$, as seen from the decomposition~\eqref{MSE-decomp}. We leave the possibility of choosing $w$ adaptively to future work. 

\paragraph{Reported quantities.}For each method and parameter setting, we perform $R=100$ independent
repetitions. We report two quantities: the empirical $\CV^2$ of the probability estimator and $\log_{10}\bar p_N$,
where $\bar p_N$ is the mean of that method's $R$ probability estimates.

To explain how we compute the empirical $\CV^2$, let
$W_{ij}$ be the importance weight of sample $j=1,\dots, N$ in repetition
$i=1,\dots,R$. For GeoIS, these are the weights
in~\eqref{tildWi-components}. For LSIS and CE, if $U_{ij}$ is drawn
from proposal $q$, its weight is
\[
  W_{ij}
  =\1\{U_{ij}\in\Dorig\}\frac{\piorig(U_{ij})}{q(U_{ij})}.
\]  For each method, we compute the
sample mean $\overline W$ and sample variance $S_{W}^2$ of all
$RN$ weights, and report
\[
  \widehat{\CV}^2(\hat p_{N})
  =S_{W}^2/(N\overline W^{2}).
\]
This is an estimator of $\CV^2(\hat p_N) = \Var(\hat p_N/p)$. 
 
For the diffusion example in \Cref{subsec:PDE}, since LAIS-dm reuses and reweights samples, samples across iterations may be dependent. We therefore compute the empirical $\CV^2$ for all four methods from the independent repetitions. Writing
$\hat p_{N}^{(i)}$ for the estimate from repetition $i$ and
$\bar p_{N}$ for the mean of these $R$ estimates, we report
\[
\widehat{\mathrm{CV}}^2(\hat p_N) =\frac{1}{R-1}\sum_{i=1}^{R}\left(\widehat p_N^{(i)}-\bar p_N\right)^2/\bar p_N^2.\]

\paragraph{Comparison with theory.}
It is well-known that the empirical diagnostic $\widehat{\mathrm{CV}}^2(\hat p_N)$ can severely underestimate the global quantity $\CV^2(\hat p_N)=\Var(\hat p_N/p)$ when rare, large weights are absent from the sample. This is not an issue for us because we do not target the global $\CV^2$ in the first place. We instead view $\widehat{\mathrm{CV}}^2(\hat p_N)$ as an estimator of $\CV^2_\loc$, the relative MSE of $\hat p_\loc$, which appears in our theoretical bounds. Estimating this quantity reliably does not require one to explore rare regions of state space. To see why $\widehat{\mathrm{CV}}^2(\hat p_N)$ estimates $\CV^2_\loc$, consider the high-probability event $A_N$ that all of the lower-component samples $(X_i,T_i)$ satisfy $X_i\in\us_\kappa$. On $A_N$, we have $\hat p_N = \hat p_\loc$, and $\widehat{\mathrm{CV}}^2(\hat p_N)$ estimates $\CV^2(\hat p_\loc\mid A_N)$. In turn this is close to $\CV^2_\loc$, up to an exponentially-small term.

Some of our parameter sweeps extend beyond the ranges specified by the sufficient conditions in our propositions. Even outside these ranges, the observed trends in the empirical \(\CV^2\) of GeoIS are often qualitatively consistent with the theory.

\subsection{Exact quadratic boundary}\label{sub:quad}

We first consider a rare-event set with quadratic boundary,
\begin{equation}\label{eq:D-parabolic}
\Dorig=\left\{(x,t)\in\R^d\times\R:t\geq\ell+\frac12\sum_{i=1}^da_ix_i^2\right\}.
\end{equation}
The case $t\geq\ell+\frac12x^\top Ax$ for a general symmetric $A$ can be brought into the diagonal form above by a rotation of the coordinates $x$. We estimate $p=\piorig(\Dorig)$ with $\piorig=\mathcal N(0_{d+1},\la_0^{-1}I_{d+1})$. The boundary $\phi$ of the normalized event $\Dnew$ is also exactly quadratic, so $\hat\phi=\phi$.
We can then bound $\CV^2(\hat p_N)$ directly without localization. We prove the following result in Appendix~\ref{app:quadratic-details}.
\begin{proposition}\label{prop:quadratic}
Consider the parabolic event~\eqref{eq:D-parabolic}. Assume $1+\ell a_i>0$ for $i=1,\ldots,d$. Define
$$
\tau_2(0)=\max_{1\leq i\leq d}\frac{|\ell a_i|}{1+\ell a_i}.
$$
Then for all $N\geq1$ and $w\in(0,1]$,
\begin{equation}\label{eq:parabolic-cv2}\begin{split}
\E[(\hat p_N/p-1)^2] &= \frac1N\left(w^{-1}-1 + \frac{\epsilon_+}{w}\right),\\
0\leq\epsilon_+&\leq \exp\left(\frac{3\tau_2(0)^2 d^2}{4\la_0\ell^2} + \frac{2}{\la_0\ell^2}\right) - 1. 
\end{split}
\end{equation}
Here, $w=1$ denotes the upper-only proposal, which remains unbiased.
\end{proposition}
\begin{figure}[htbp]
    \centering
    \includegraphics[width=0.6\textwidth]{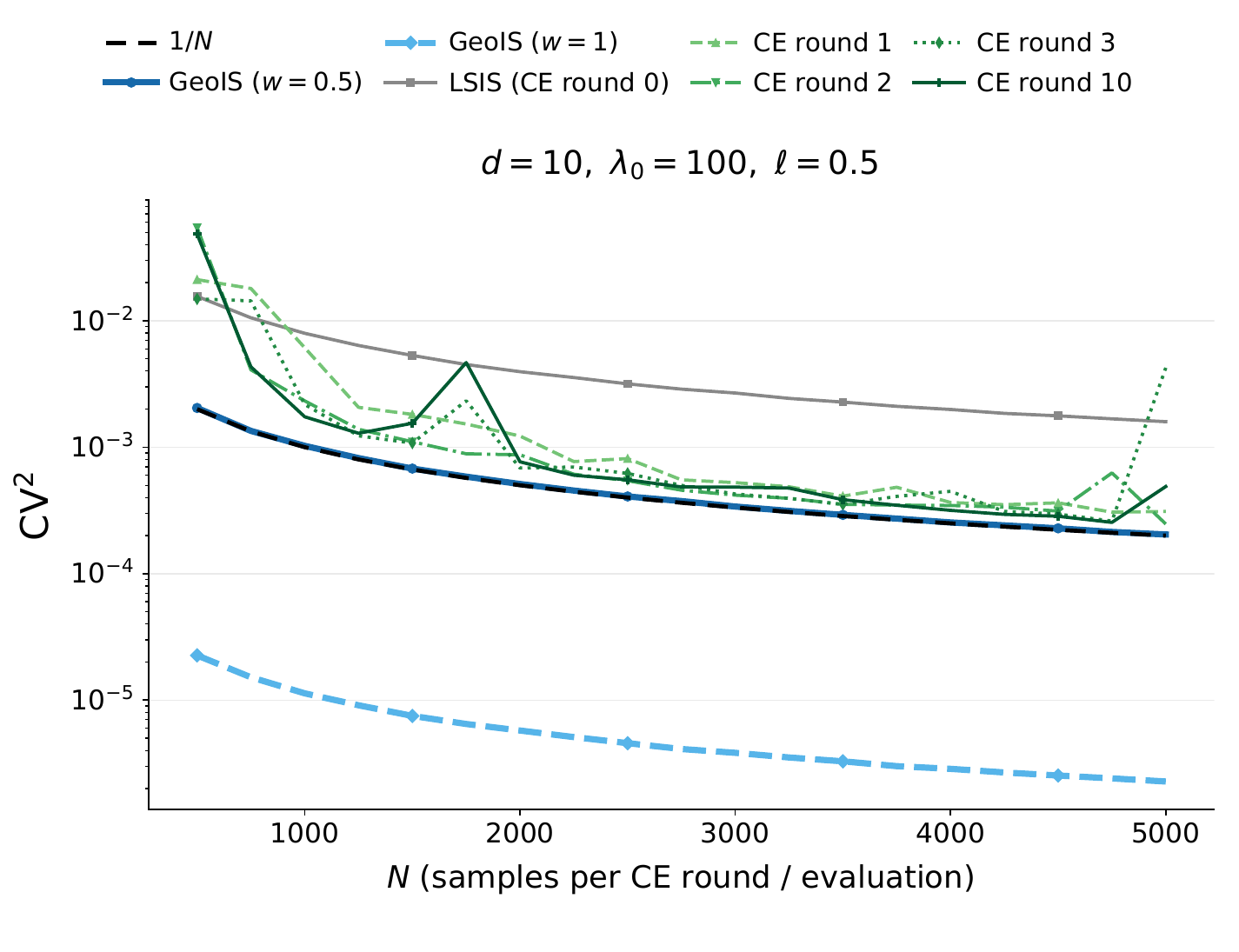}
    \caption{Comparison of empirical $\mathrm{CV}^2$ for different methods under different sample size \(N\), for the parabolic example.}
    \label{fig:quadratic_N}
\end{figure}
\begin{figure}[htbp]
    \centering
    \includegraphics[width=0.85\textwidth]{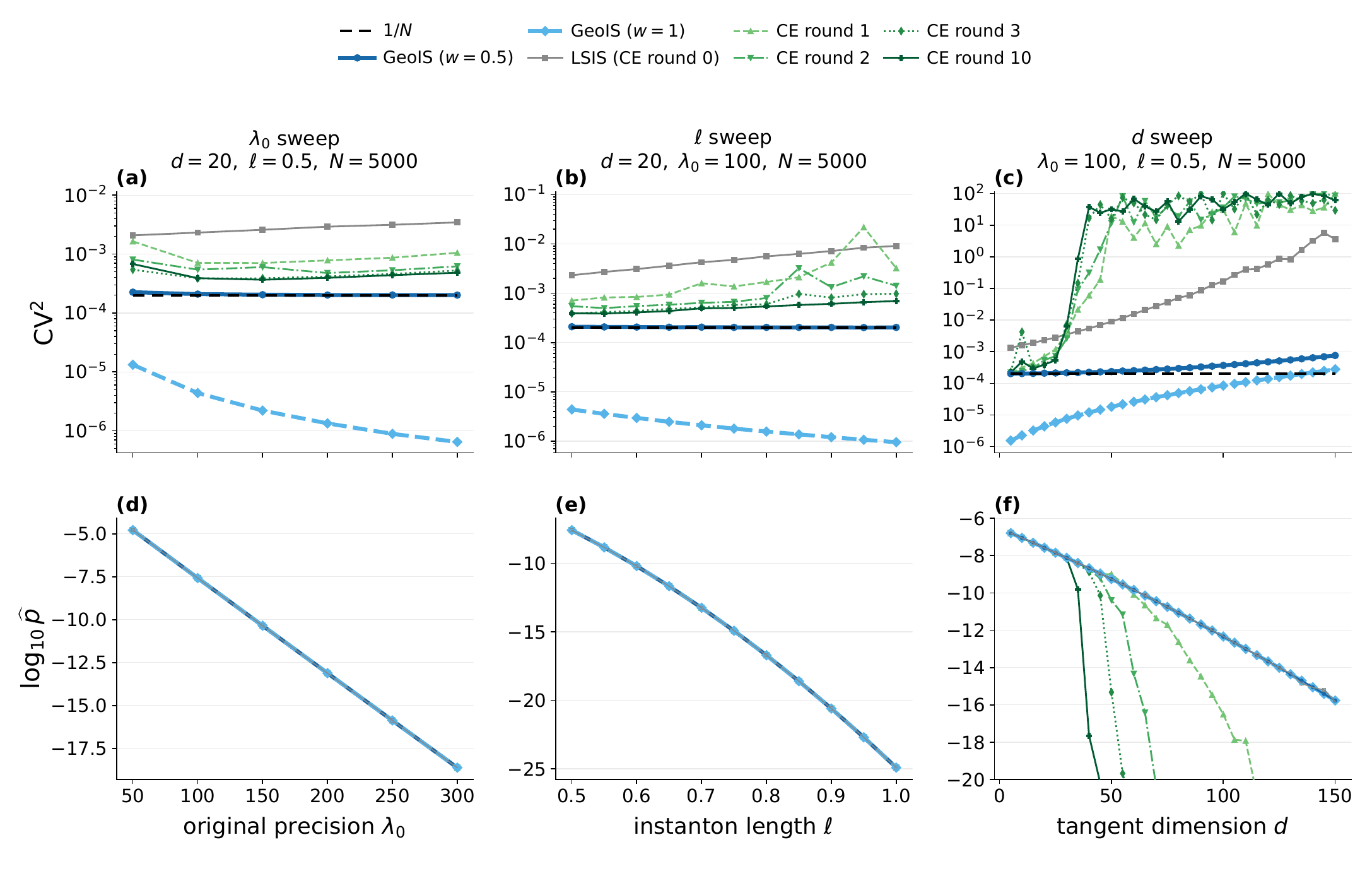}
    \caption{Comparison of empirical $\mathrm{CV}^2$ for different methods under different dimensions $d$ and rareness parameters $\lambda_0$ and $\ell$ with $N=5000$, for the parabolic example. Panels (a)--(c) show the empirical $\mathrm{CV}^2$ under the \(\la_0\), \(\ell\), and \(d\) sweeps, respectively, and panels (d)--(f) show the corresponding log-probability estimates.}
    \label{fig:quadratic_sweeps}
\end{figure}
For numerical experiments, we take $a_i=1/2$ for all $i$. We study the performance of GeoIS with $w=1/2$ and with $w=1$. Since $\hat\phi=\phi$, the upper component $\hat\pi_+$ covers the entire rare-event set, so the $w=1$ proposal $\hat\pi=\hat\pi_+$ remains unbiased.

\Cref{fig:quadratic_N} compares the empirical $\mathrm{CV}^2$ as $N$ varies for fixed $d,\la_0,\ell$. We see that GeoIS with $w=0.5$ improves on LSIS by about a factor of 10, while GeoIS with $w=1$ achieves a further dramatic reduction in $\CV^2$ over the $w=0.5$ case. Although the CE proposals have similar $\CV^2$ to GeoIS with $w=0.5$, they require a greater number of samples: recall that CE round $r$ actually uses $(r+1)N$ samples, rather than only $N$ samples. Note also that the $\CV^2$ curve for GeoIS with $w=0.5$ closely follows the $1/N$ reference line, consistent with the leading contribution \(N^{-1}(w^{-1}-1)=1/N\) to $\CV^2_\loc$ in our theoretical result; see~\eqref{eq:parabolic-cv2}.

We next study the performance of these methods under different rarity levels \(\la_0\), instanton lengths \(\ell\), and dimensions \(d\) in Figure~\ref{fig:quadratic_sweeps}. In the \(\la_0\) and \(\ell\) sweeps, both GeoIS proposals achieve substantially smaller \(\CV^2\) than LSIS and CE, especially when accounting for the additional samples that CE needs. Furthermore, GeoIS $\CV^2$ stays constant (for $w=0.5$) or even decreases (for $w=1$) with increasing rarity, i.e. larger $\la_0$ or $\ell$. Both GeoIS proposals are also significantly more robust than LSIS and CE to increasing $d$. In particular, CE completely breaks down for large $d$. This is because each full Gaussian CE update estimates a \((d+1)\) dimensional mean and a \((d+1)\times (d+1)\) covariance matrix from the \(N\) samples. For a fixed $N$, these estimates become unreliable as $d$ increases, so the KL optimization to find a good Gaussian proposal fails. This explains the large $\CV^2$ for the CE proposals in panel (c) of \Cref{fig:quadratic_sweeps}, and the severe underestimate of $p$ in panel (f). The need for many samples to reliably update the Gaussian proposal is also evident in \Cref{fig:quadratic_N}, where CE performs poorly for $N<2000$, even at $d=10$.

For GeoIS, the behavior of the empirical $\CV^2$ when varying $\la_0$, $\ell$, and $d$ is consistent with the theoretical predictions in~\eqref{eq:parabolic-cv2}.  For $w=0.5$, we have $\CV^2(\hat p_N)=(1+2\epsilon_+)/N$, which is close to the $1/N$ reference level when $\epsilon_+$ is small. This is consistent
with the nearly constant empirical $\CV^2$ in the $\la_0$ and $\ell$ sweeps and at small $d$ in the dimension sweep.
For $d\gtrsim50$ in Figure~\ref{fig:quadratic_sweeps} panel~(c),
the empirical $\CV^2$ rises above this reference level,
consistent with an increasingly significant $\epsilon_+$.
For $w=1$, the leading contribution vanishes and
$\CV^2(\hat p_N)=\epsilon_+/N$, making the dependence of the $\CV^2$ on the model parameters more directly visible, via $\epsilon_+$. For our choice $a_i=1/2$ for all $i$, we have $\tau_2(0)=\ell/(2+\ell)$, so the exponent in the bound on $\epsilon_+$ is
\[
\frac{3\tau_2(0)^2d^2}{4\la_0\ell^2}+\frac{2}{\la_0\ell^2}=
\frac{3d^2}{4\la_0(2+\ell)^2}+\frac{2}{\la_0\ell^2}.
\]
This bound decreases
with $\la_0$ and $\ell$ and increases with $d$, and we observe the same behavior in panels (a), (b), (c).

\subsection{Ellipsoid level set}\label{ex:ellipsoid}
\begin{figure}[H]
    \centering
    \includegraphics[width=0.5\textwidth]{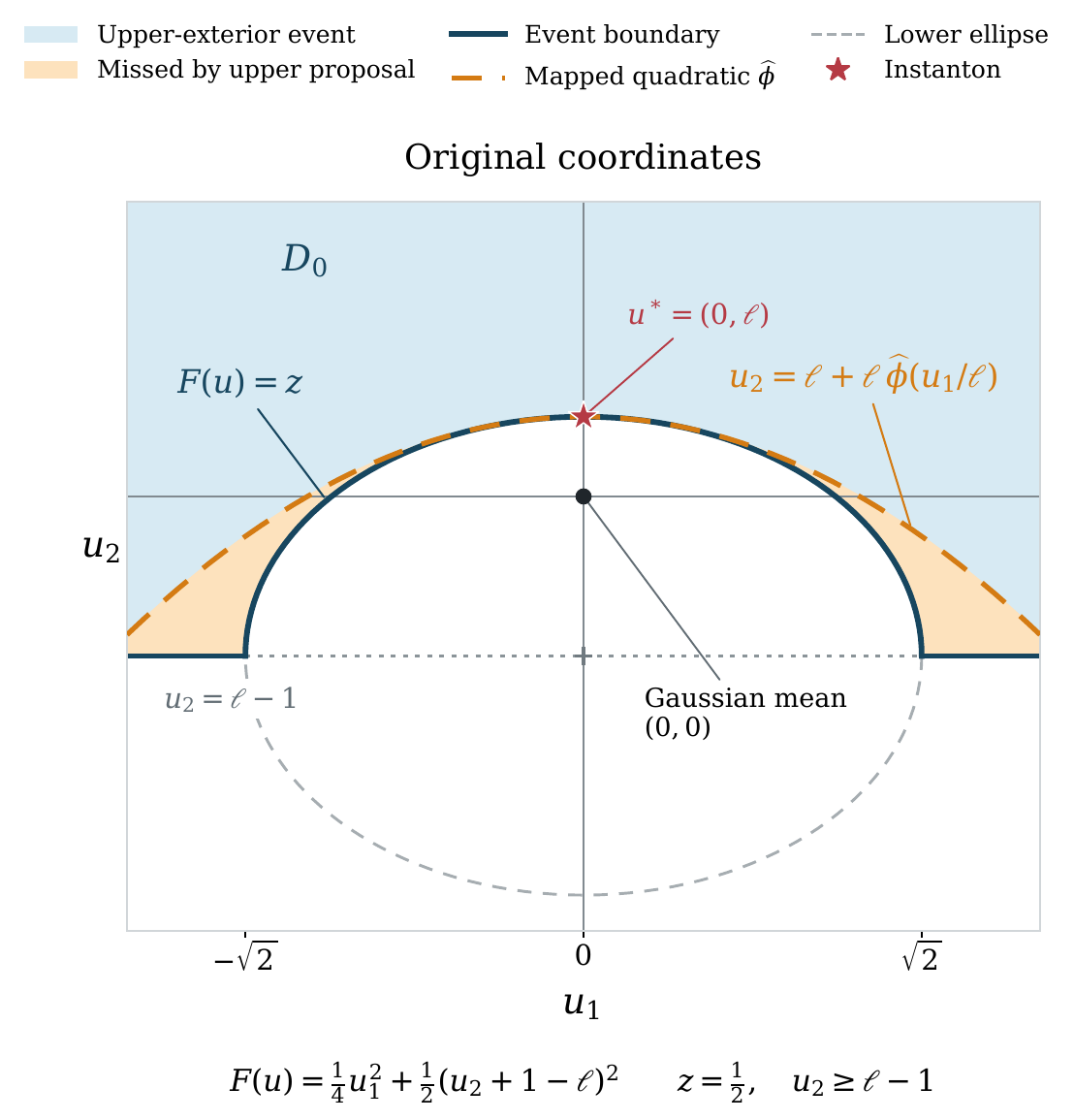}
    \caption{Geometry of the upper-exterior ellipsoid for \(d=1\), with \(a_1=1/2\) and \(\ell=1/3\).}
    \label{fig:ellipsoid_geometry}
\end{figure}

Let $a_1,\ldots,a_{d}>0$ and $0<\ell<1$. Define 
\begin{equation}\label{F-ellipse}
F(u)=\frac12\sum_{i=1}^da_iu_i^2 + \frac{1}{2}(u_{d+1}+1-\ell)^2.
\end{equation} 
The level set $F(u)=1/2$ is an ellipsoid centered at $(\ell-1)e_{d+1}$. We consider the upper exterior event outside this ellipsoid,  
\[\Dorig=\{u\in\R^{d+1}:F(u)\geq \tfrac12,\;u_{d+1}\geq\ell-1\}, \quad u\sim\pi_0 = \mathcal N(0_{d+1},\la_0^{-1}I_{d+1}).\]
Figure~\ref{fig:ellipsoid_geometry} illustrates the event geometry and the quadratic boundary approximation.

We prove the following error estimate in Appendix~\ref{app:ellipsoid-details}.
\begin{proposition}\label{prop:ellipsoid}
Consider the ellipsoidal event $\Dorig$ described above. Let $a=\max_{1\leq i\leq d}a_i$.
There exist absolute constants $k,\eta,c,C>0$ such that, for every $a_1,\dots,a_d>0$, and $d,\ell,\la_0$ satisfying
\begin{equation}\label{h-ell-la-cond}
h:=1-\ell a > 0,\qquad 
\la_0\ell^2\geq \frac{\eta d}{h^3}\log\frac{e}{h},
\end{equation}
it holds
\begin{align}
  \Pp(\hat p_N=\hat p_{\loc})
  &\geq 1-Ne^{-c\la_0\ell^2 h^3}, \label{eq:ellipsoid-localization}\\
  \left|\CV_{\loc}^2-\frac{w^{-1}-1}{N}\right|
  &\lesssim \frac{d^2/[\la_0(\ell h)^2]}{Nw(1-w)}\exp\left(\frac{Cd^2}{\la_0(\ell h)^2}\right) +e^{-c\la_0\ell^2 h^3},
  \label{eq:ellipsoid-cv2}
\end{align}
where $\kappa$ defining $\hat p_{\loc}$ is given by $\kappa=kh^{3/2}$. %
\end{proposition}
The usually leading-order term $\epsilon_3=\tau_3(0)d/\sqrt\la$ in the bounds~\eqref{MSE-err+} and~\eqref{MSE-err-} on $\epsilon_\pm$ vanishes for the ellipsoid. Thus the dominant terms are $\epsilon_4^2$ and $\epsilon_2^2$, which scale as $\frac{d^2}{\la_0(\ell h) ^2}$ in the ellipse setting. 

In the numerical experiments, we take \(a_i=1/2\) for \(i=1,\ldots,d\). Unless \(\ell\) itself is varied, we set \(\ell=1/3\). Here and in the remaining examples, we study the performance of GeoIS with $w=0.5$ and $w=0.9$, because the support of the $w=1$ upper-only proposal may not cover the entire event. 

\begin{figure}[H]
    \centering
    \includegraphics[width=0.6\textwidth]{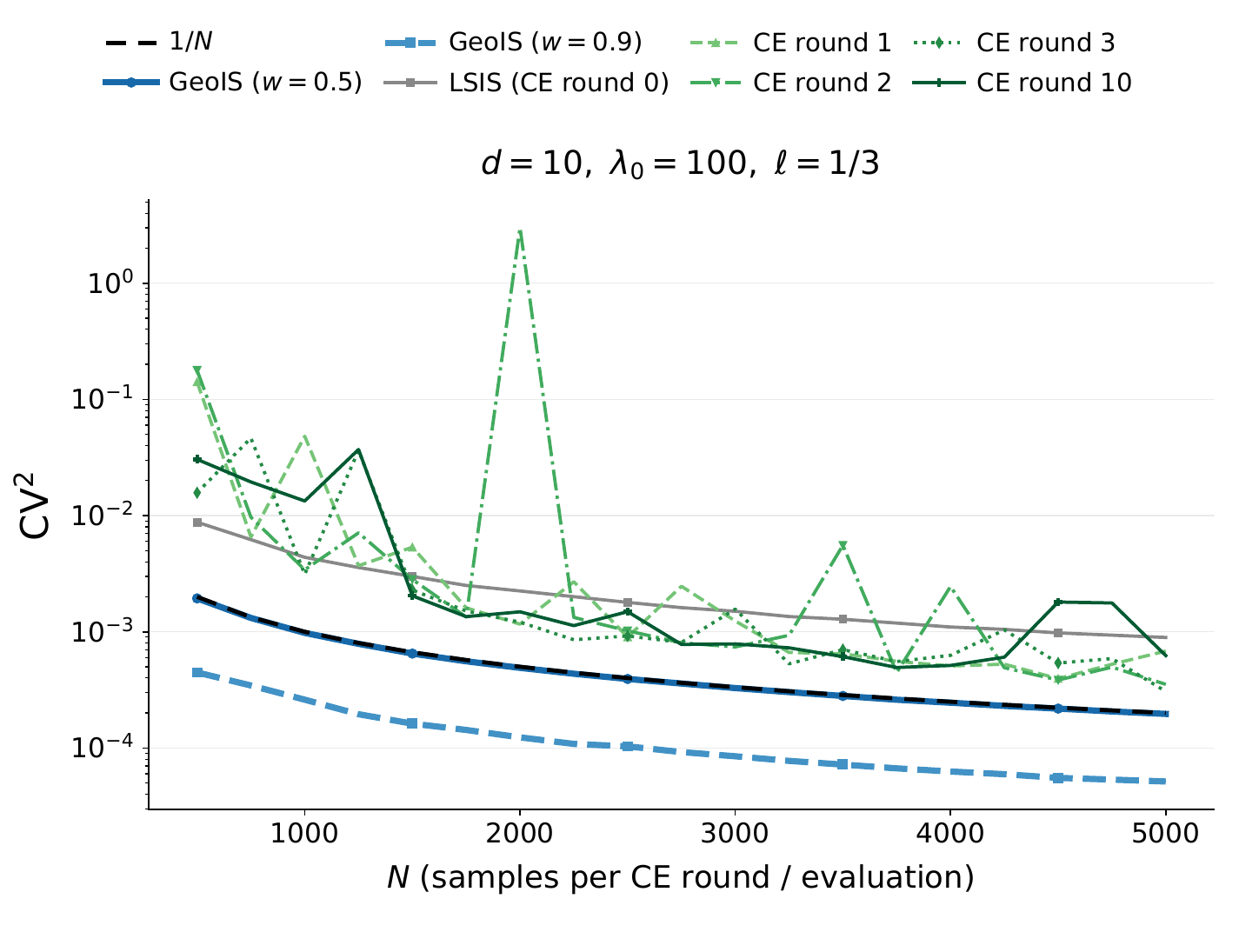}
    \caption{Comparison of empirical $\mathrm{CV}^2$ for GeoIS, LSIS, and CE as the sample size $N$ varies, for the ellipsoidal event.}
    \label{fig:ellipsoid_N}
\end{figure}

In \Cref{fig:ellipsoid_N}, we compare GeoIS with LSIS and CE rounds as $N$ varies, fixing $d,\la_0,\ell$. Across the \(N\) sweep, GeoIS reduces the empirical \(\CV^2\) relative to LSIS by about a factor of 5 with \(w=0.5\) and a factor of 20 with \(w=0.9\).
GeoIS with \(w=0.9\) also has approximately ten times lower empirical \(\CV^2\) than the best CE round.
The \(w=0.5\) curve closely follows the \(1/N\) reference line, and
the lower empirical \(\CV^2\) at \(w=0.9\) is qualitatively consistent
with the decrease in the baseline term from \(1/N\) to \(1/(9N)\)
in~\eqref{eq:ellipsoid-cv2}.
\begin{figure}[H]
    \centering
    \includegraphics[width=0.85\textwidth]{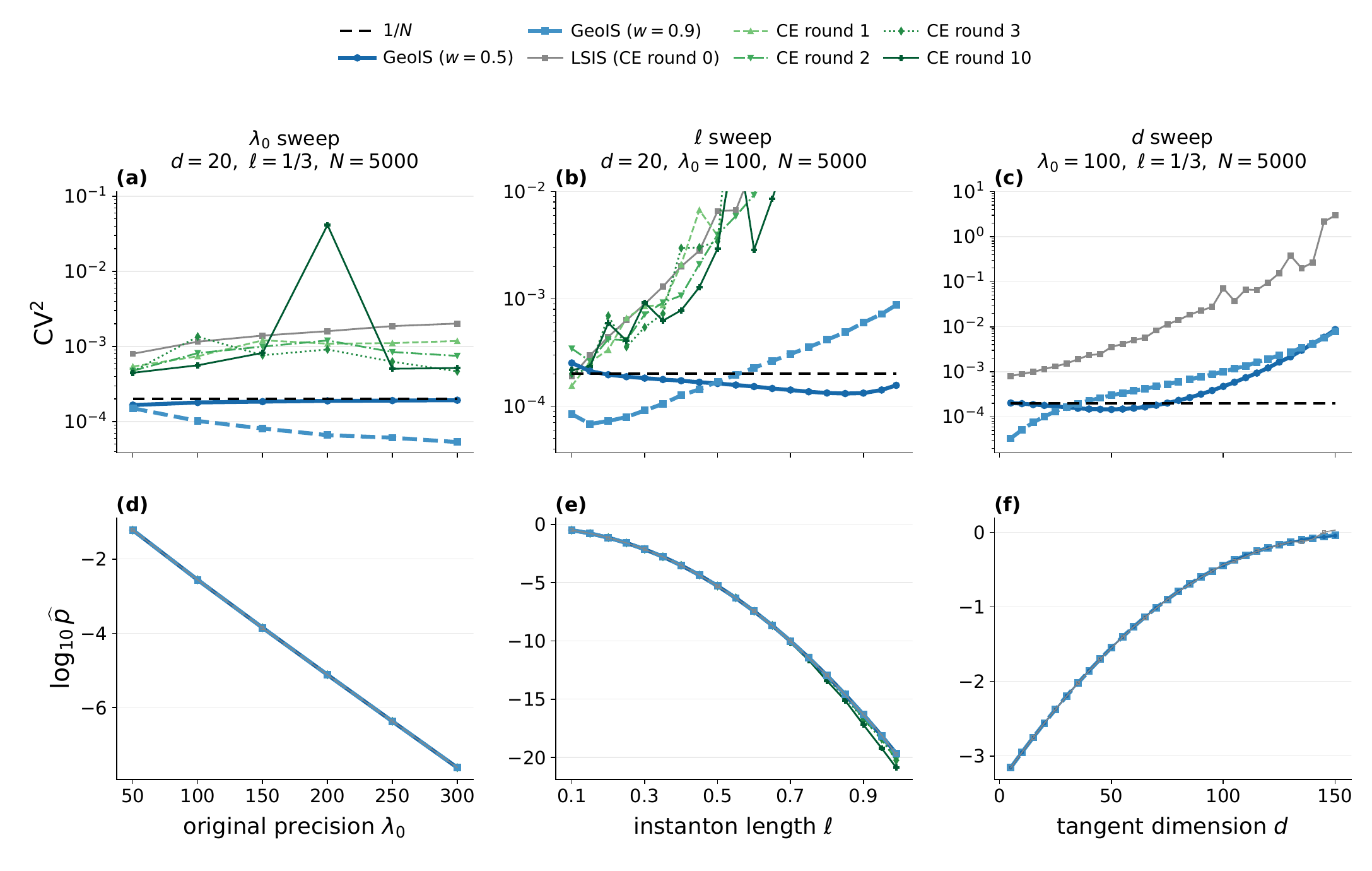}
    \caption{Comparison of performance for different methods under different dimensions \(d\) and rarity parameters \(\lambda_0\) and \(\ell\) with \(N=5000\), for the ellipsoid example. Panels (a)--(c) show the estimator \(\mathrm{CV}^2\) under the \(\lambda_0\), \(\ell\), and \(d\) sweeps, respectively. Panels (d)--(f) show the corresponding log-probability estimates.}
    \label{fig:ellipsoid_sweeps}
\end{figure}

We next vary \(\la_0\), \(\ell\), and \(d\) in \Cref{fig:ellipsoid_sweeps}.
Across the three parameter sweeps, both GeoIS proposals achieve substantially lower empirical \(\CV^2\) than LSIS and CE.
Increasing \(w\) from \(0.5\) to \(0.9\) lowers the baseline term from \(1/N\) to \(1/(9N)\), so departures from the baseline are more visible on the logarithmic scale for the $w=0.9$ case. In panel (a), we see that the empirical $\CV^2$ moves closer to its baseline, which is $1/(9N)\approx 2.2\times 10^{-5}$, as $\la_0$ increases. Panel (c) shows it moves farther away from the baseline as $d$ increases. Both trends are consistent with our predictions in~\eqref{eq:ellipsoid-cv2}.

Varying \(\ell\) changes both the rarity of the event and the geometry. As \(\ell\) increases, the instanton moves farther from the Gaussian center and the event becomes rarer, so we expect GeoIS $\CV^2$ to be closer to the baseline. But for large enough $\ell$, the boundaries of the  ellipse and Gaussian level set approach one another, i.e. \(H=(1-\ell/2)I_d\) degenerates. This moves us away from the asymptotic baseline. The nonmonotone behavior in $\ell$ is captured in our theoretical bound~\eqref{eq:ellipsoid-cv2}, via the term $(\ell h)^2 = (\ell[1-\ell a])^2$ in the denominator. The nonmonotonicity with respect to $\ell$ is also evident in panel (b). To zoom in on the GeoIS behavior, we truncate the y-axis in panel (b), which is why the LSIS and CE $\CV^2$ curves are cut off.

\subsection{Portfolio example}\label{sec:portfolio}
We next consider a financial application adapted from the portfolio model in \cite[Section~6.1]{tong2022optimization}.
Consider a portfolio containing $d+1$ equally weighted stocks, with unit initial values. We assume the stock prices $V_i$, $i=1,\dots,d+1$ are independent and follow a log-normal model:
\begin{equation}
V_i(t)=\exp\left[\left(\theta_i-\frac{\sigma_i^2}{2}\right)t+\sqrt{t}\,\xi_i\right],\quad \xi_i\stackrel{\text{ind.}}{\sim} \mathcal N(0,\sigma_i^2).
\end{equation}
Thus the $i$th stock is driven by an independent random rate of return \(\xi_i\sim\mathcal N(0,\sigma_i^2)\).
We study the probability that $\frac{1}{d+1}\sum_{i=1}^{d+1}V_i(\la_0^{-1}) \leq z$ for some $0<z<1$. This is the event that in the short time $t=\la_0^{-1}$, the portfolio drops to a fraction $z$ of its initial value. 

For simplicity, set \(\sigma_i=1\), \(\theta_i=1/2\). Then $V_i(\la_0^{-1}) = e^{U_i}$, where $U_i=\la_0^{-1/2}\xi_i\iid\mathcal N(0,\la_0^{-1})$. The probability of interest can then be written as
\begin{equation}\label{eq:D-portfolio}
    p=\Pp\left(\frac1{d+1}\sum_{i=1}^{d+1}e^{U_i}\leq z\right), \quad
    (U_1,\dots, U_{d+1})\sim\mathcal N(0_{d+1},\lambda_0^{-1}I_{d+1}).
\end{equation}
For fixed $d$, the normalized portfolio value $\frac1{d+1}\sum_{i=1}^{d+1}e^{U_i}$ converges to $1$ in probability as $\la_0\to\infty$, so the event that the normalized portfolio value is below some threshold $z<1$ is indeed rare. Below, we assume that $z\leq e^{-a_0}$ for some fixed $a_0>0$, or equivalently, that \(\lvert\log z\rvert\geq a_0\), to keep the event rare. 

See Appendix~\ref{app:portfolio-details} for the proof of the following error estimate.
\begin{proposition}\label{prop:portfolio}
Consider the rare portfolio event described above. Fix $a_0>0$. There exist constants
$k_{a_0},\eta_{a_0},c_{a_0},C_{a_0}>0$, depending only on $a_0$,
such that, for every $d,\la_0, z$ satisfying
\begin{equation}\label{port-assume}
   0<z\leq e^{-a_0}, \qquad \la_0 \geq \eta_{a_0}d(1+\log d + \log(1+|\log z|)),  
\end{equation}
the following bounds hold:
\begin{align}
\Pp(\hat p_N=\hat p_{\loc})
&\geq 1-Ne^{-c_{a_0}\la_0},
\label{eq:portfolio-localization}\\
\left|\CV_{\loc}^2-\frac{w^{-1}-1}{N}\right|
&\lesssim_{a_0}
\frac{d/\sqrt{|\log z|\la_0}}{Nw(1-w)}
\exp\left(C_{a_0}\frac{d^2}{|\log z|\la_0}\right)
+e^{-c_{a_0}\la_0}.
\label{eq:portfolio-cv2}
\end{align}
Here, $\kappa$ defining $\hat p_{\loc}$ is given by $\kappa=k_{a_0}/(|\log z|\sqrt d)$.
\end{proposition} 

\begin{figure}[htbp]
    \centering
    \includegraphics[width=0.6\textwidth]{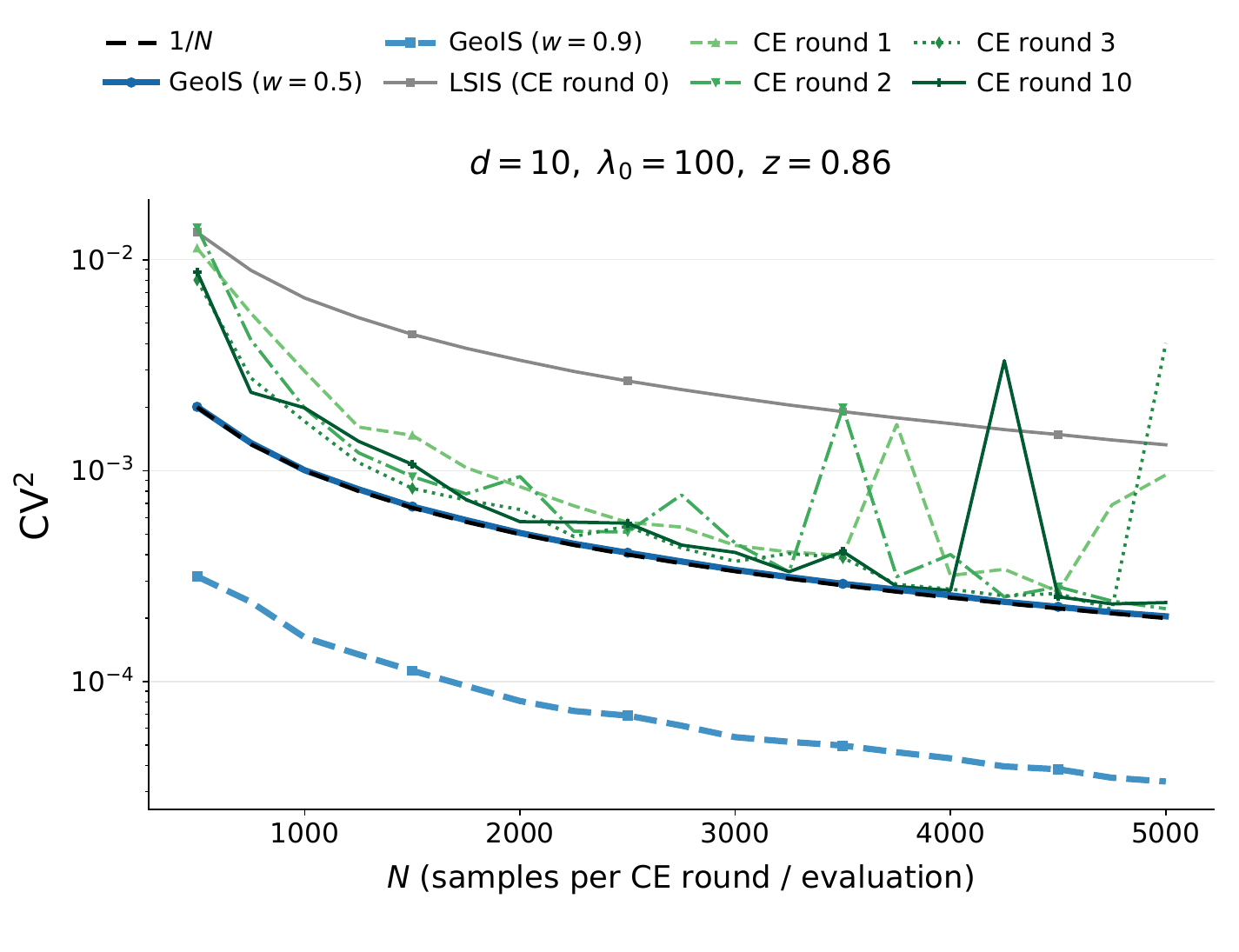}
    \caption{Comparison of empirical $\mathrm{CV}^2$ for GeoIS, LSIS, and CE as the sample size $N$ varies, for the portfolio example.}
    \label{fig:portfolio_N}
\end{figure}
\begin{figure}[htbp]
    \centering
    \includegraphics[width=0.85\textwidth]{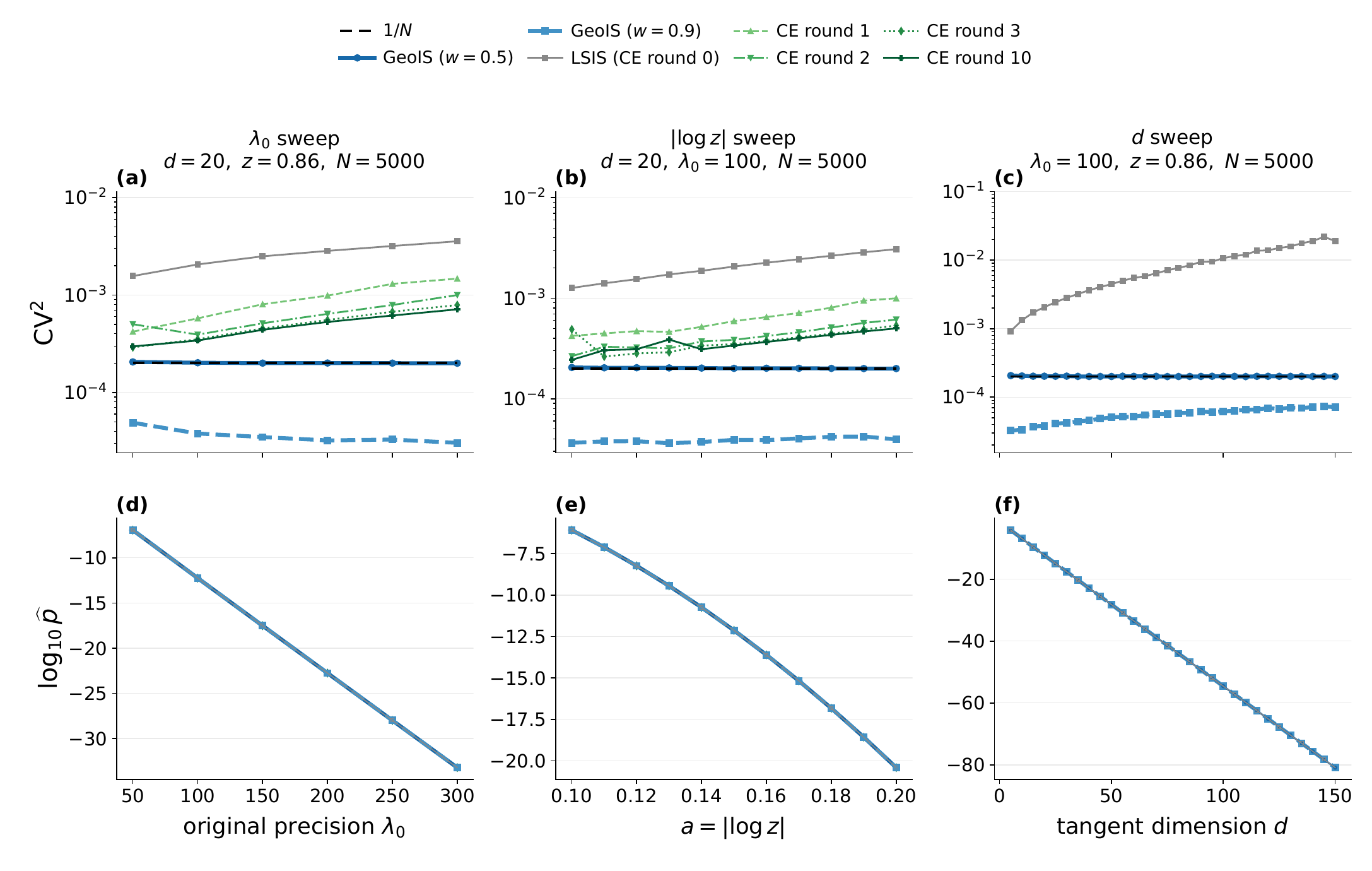}
    \caption{Comparison of estimator performance across the \(\la_0\),
\(\lvert\log z\rvert\), and \(d\) sweeps for the portfolio example
with \(N=5000\). Panels (a)--(c) show the empirical \(\mathrm{CV}^2\)
for the three sweeps, respectively. Panels (d)--(f) show the
corresponding log-probability estimates.}
    \label{fig:portfolio_sweeps}
\end{figure}

In \Cref{fig:portfolio_N}, we compare GeoIS with LSIS and CE rounds as \(N\) varies, fixing $d,\la_0,z$. 
Across the tested range of \(N\), GeoIS with \(w=0.5\) reduces the empirical \(\CV^2\) by about a factor of 10 relative to LSIS, and its curve closely follows the \(1/N\) reference line. The advantage over LSIS is even more pronounced for \(w=0.9\), with a roughly fortyfold reduction. GeoIS with \(w=0.9\) achieves a much lower empirical \(\CV^2\) than CE does. %

Next, we vary \(\la_0\), \(\lvert\log z\rvert\), and \(d\) in \Cref{fig:portfolio_sweeps}, the three independent parameters appearing in the bound~\eqref{eq:portfolio-cv2}.
Across all three sweeps, both GeoIS proposals achieve lower empirical \(\CV^2\) than LSIS and CE. The \(w=0.5\) curve stays close to \(1/N\). GeoIS with \(w=0.9\) achieves the lowest empirical \(\CV^2\) at every tested point. Its empirical \(\CV^2\) decreases overall with \(\la_0\), increases modestly with \(d\), and varies little across the $|\log z|$ sweep.
The empirical \(\CV^2\) of LSIS generally increases with rarity and dimension.

\subsection{Diffusion problem with PDE}\label{subsec:PDE}
We finally consider a one-dimensional steady-state diffusion problem from \cite{tong2023large}, with random diffusion coefficient $\alpha$.
The state $\nu$ satisfies
\begin{equation}\label{eq:diffusion-eq}
-\frac{\partial}{\partial x}\left(\alpha(x;u)\frac{\partial}{\partial x}\nu(x;u)\right)=1,\; x\in[0,1],  \text{ with boundary condition }\nu(0;u)=0,\; \frac{\partial \nu}{\partial x}(1;u)=0.
\end{equation}
The rare event of interest is $\nu(1;u)\geq z$.
	
We model $\alpha$ as a log-normal random field.
The reference Gaussian log-field has mean $\mu_Z=-\sigma_Z^2/2$ and covariance $C(x,x')=\sigma_Z^2\exp(-|x-x'|/L_c)$, with $\sigma_Z^2=\log(1.01)$ and $L_c=0.01$. Let $\{\gamma_m,e_m\}_{m=1}^{d+1}$ denote the first $d+1$ eigenpairs of this covariance kernel. The diffusion coefficient is then defined by the following Karhunen--Lo\`eve expansion
\begin{equation}
	\alpha(x;u)=\exp\left(\mu_Z+\sum_{m=1}^{d+1}\sqrt{\gamma_m}\,e_m(x)u_m\right),\qquad
	u\sim\piorig=\mathcal N(0_{d+1},\la_0^{-1}I_{d+1}).
\end{equation}

Thus, the rare event probability can be written as the following problem, where $F$ involves a PDE solve to \eqref{eq:diffusion-eq},
\begin{equation}
	p=\piorig(\Dorig),\qquad \Dorig=\{u\in\R^{d+1}:F(u)\geq z\}, \quad F(u):=\nu(1;u).
\end{equation}
	
Increasing \(\la_0\) suppresses the random fluctuations of the coefficient field. As \(\la_0\to\infty\),  \(\alpha(x; u)\to e^{\mu_Z}\), so the diffusion problem approaches its deterministic limit. For a fixed threshold \(z\) above the deterministic endpoint response, the event \(F(u)\geq z\) therefore becomes increasingly rare as \(\la_0\) increases.

We discretize the diffusion equation using piecewise linear finite elements on $512$ equidistant spatial elements. The instanton $u^*$, defined in \Cref{sec:setup}, is computed numerically by solving the first-order optimality conditions, with gradients evaluated using a discrete adjoint solver. Since the event is specified implicitly through the PDE response function $F$, we compute the matrix $H$ using~\eqref{eq:implicit-BH}. We obtain the full tangent curvature spectrum from Hessian-vector products approximated by forward finite differences of the adjoint gradient. %

For this PDE-related example, we include an additional benchmark method: LAIS-dm from \cite{tong2023large}, which exploits the problem's intrinsic low-dimensional structure by updating the Gaussian mean and covariance only within an LDT-informed subspace (selected using curvature information at the instanton). Let $r$ denote the dimension of this subspace, and $J$ denote the number of adaptive IS iterations. For a total sample budget \(N\), the LAIS uses \(N_{\mathrm{ce}}=N/J\) samples per iteration. In the final iteration, LAIS uses deterministic-mixture MIS weights to reweight all $N$ evaluations from the adaptive process. In our experiments, we use $J=5$ and compare subspace dimensions $r=3$ and $r=5$.

All four methods share an optimization stage to compute the instanton $u^*$, whose cost is independent of the sample size $N$. 
GeoIS and LAIS additionally require Hessian information: GeoIS uses the full tangent Hessian, whereas LAIS constructs an $r$-dimensional LDT-informed subspace from the normal direction and $r-1$ dominant curvature directions.
Both constructions use Hessian-vector products, and each product requires one additional forward PDE solve and one adjoint PDE solve. Assembling the full tangent Hessian requires $d$ such products, while the number required by LAIS depends on the retained subspace dimension $r$ and the eigensolver. Since these setup costs are small relative to the sampling cost in the tested regimes, we exclude them from the figures below and report only the number of samples used in the final estimator by each method.

\begin{figure}[htbp]
\centering
\includegraphics[width=0.9\textwidth]{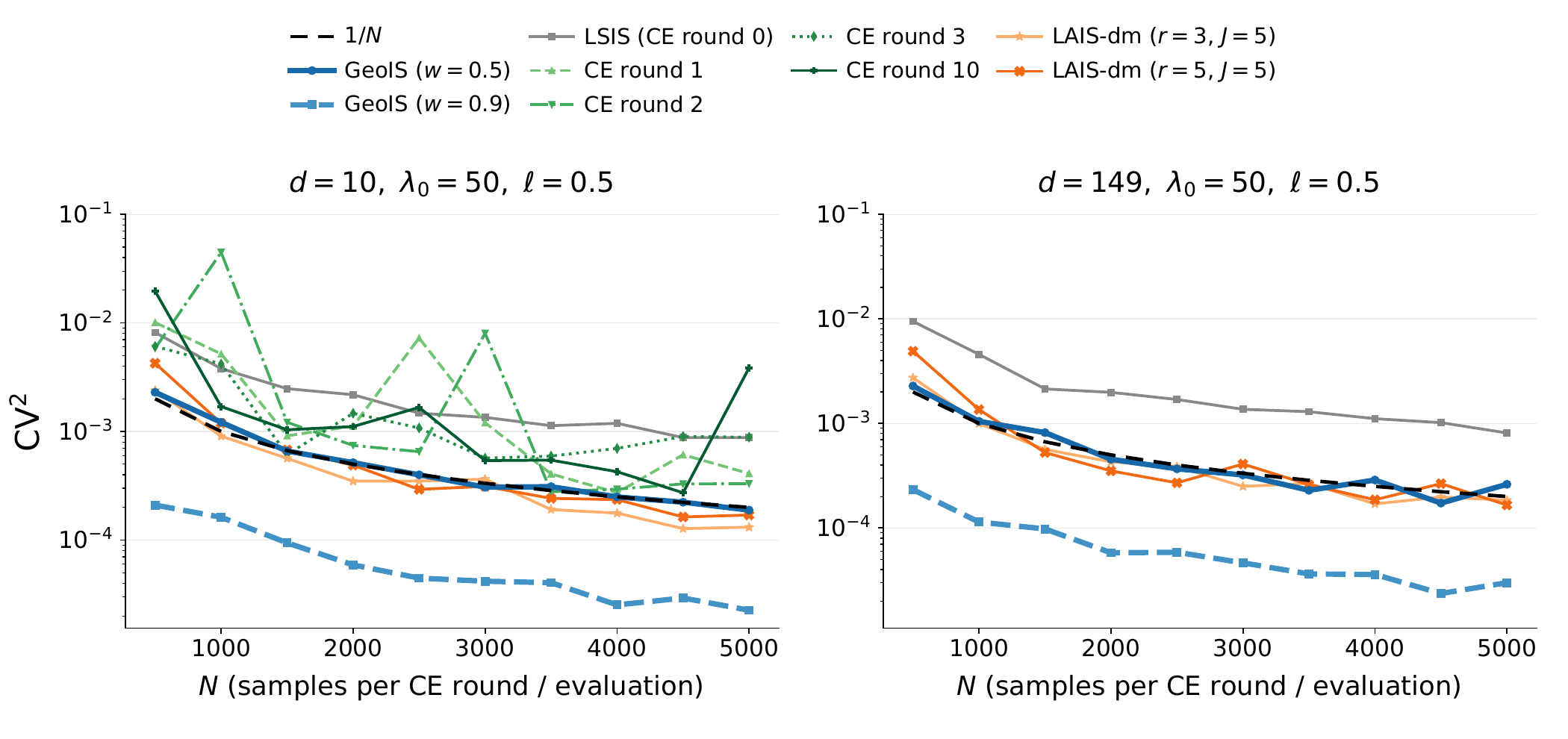}
\caption{Empirical \(\mathrm{CV}^2\) under the \(N\)-sweep for the diffusion PDE example. The left and right panels correspond to \(d=10\) and \(d=149\), with \(z=0.506535\) and \(z=0.506581\). The total number of samples (including those used in previous adaptive rounds) is $N$ for GeoIS and LAIS and $(i+1)N$ for CE round $i$.}
\label{fig:diffusion_N}
\end{figure}
\begin{figure}[htbp]
	\centering
	\includegraphics[width=0.8\textwidth]{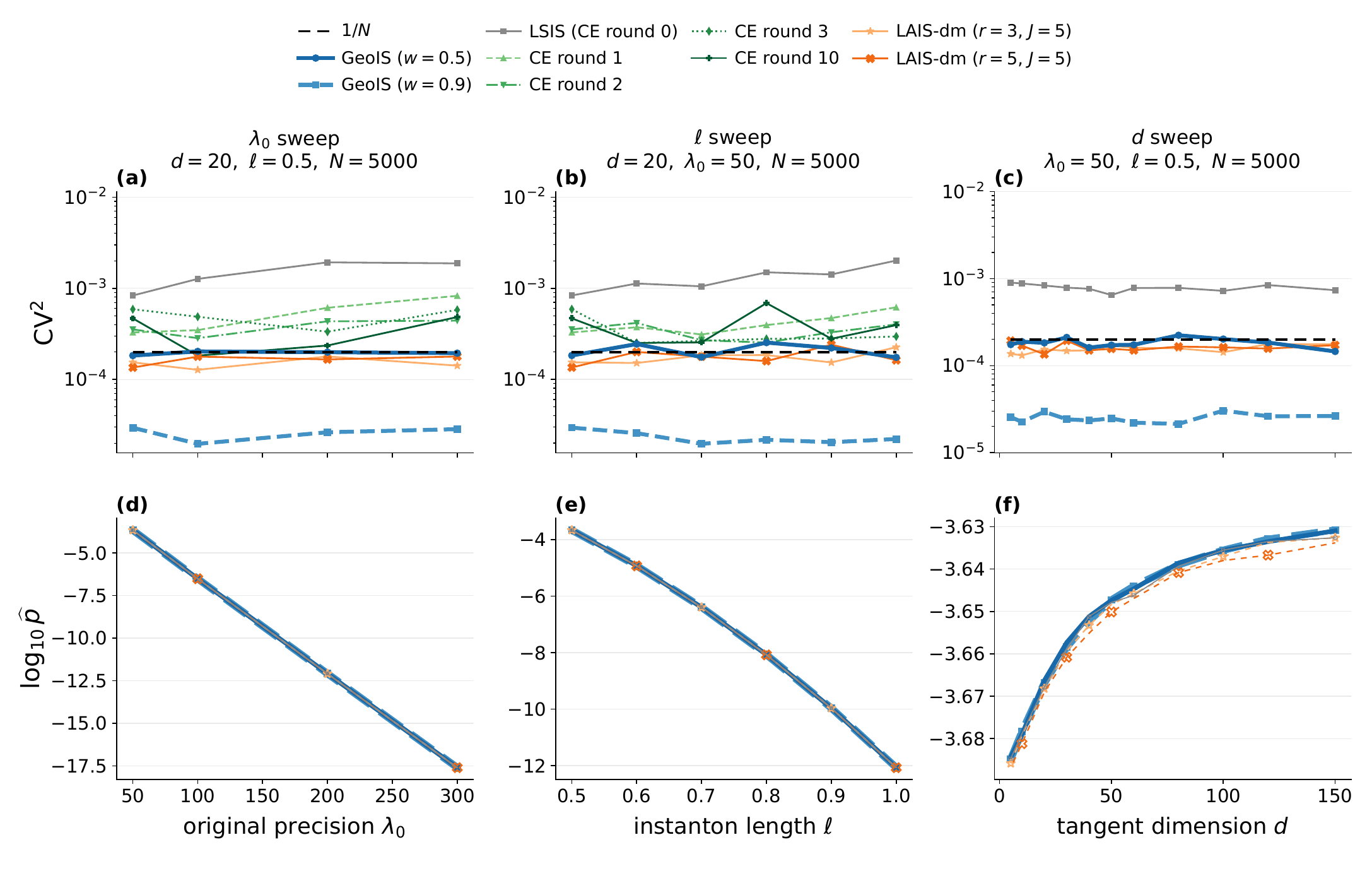}
	\caption{Comparison of performance under different dimensions and rarity parameters for the diffusion PDE example with \(N=5000\). Panels (a)--(c) show the estimator $\mathrm{CV}^2$ under the \(\la_0\), \(\ell\), and \(d\) sweeps, respectively, while panels (d)--(f) show the corresponding \(\log_{10}\bar p\). For the \(\la_0\)-sweep, \(d=20\), \(\ell=0.5\), and \(z=0.506570\). For the \(\ell\)-sweep, \(d=20\) and \(\la_0=50\), with \(z\) recalibrated for each \(\ell\). For the \(d\)-sweep, \(\ell=0.5\) and \(\la_0=50\), with \(z\) recalibrated for each \(d\).
    }
	\label{fig:diffusion_sweeps}
\end{figure}
We find the instanton $u^*$ with length $\ell$ through fixed-point iteration of $u^{(k+1)} = \ell \nabla F (u^{(k)}) /\|\nabla F (u^{(k)})\|$ and set \(z=F(u^*)\). For each parameter setting, all methods and repetitions use the same threshold.

In \Cref{fig:diffusion_N}, we first compare methods in a low-dimensional regime \(d=10\) (where \(d^2/\la=8\)) and a high-dimensional regime \(d=149\) (where \(d^2/\la=1776.08\)).
In both cases, we set \(\la_0=50\) and \(\ell=0.5\), giving \(\la=12.5\).
In both regimes, GeoIS with \(w=0.5\) closely follows the \(1/N\) reference line and performs comparably to both LAIS-dm proposals. GeoIS with \(w=0.9\) achieves the lowest empirical \(\CV^2\), roughly an order of magnitude below that of LAIS-dm.
At \(d=149\), all four reported adaptive CE rounds break down, so we do not display them.

We next examine the dependence on \(\la_0\), \(\ell\), and \(d\) in \Cref{fig:diffusion_sweeps}.
Across all three sweeps, GeoIS with \(w=0.9\) achieves the lowest empirical \(\CV^2\), about an order of magnitude below the values attained by GeoIS with \(w=0.5\) and both LAIS-dm proposals.
LSIS has higher empirical \(\CV^2\) than GeoIS and LAIS-dm throughout. As \(d\) increases, GeoIS and LAIS-dm remain stable.
The probability estimates agree closely across all methods in panels~(d) and~(e). In panel~(f), the GeoIS, LSIS, and LAIS-dm estimates change little with \(d\), consistent with the intrinsic low-dimensional structure of this diffusion problem identified in~\cite{tong2023large}.

\section{Proof of Proposition~\ref{prop:kappa}}\label{sec:proof-kappa}
We impose the hypotheses of Proposition~\ref{prop:kappa} throughout this section. We set up the proof and introduce some notation in Section~\ref{proof:set-up}, prove that $\hat p_N=\hat p_\loc$ with high probability in Section~\ref{subsec:loc}, and bound the relative MSE of $\hat p_\loc$ in Section~\ref{subsec:MSE}.
\subsection{Set-up}\label{proof:set-up}
\begin{figure}[h!]
    \centering
     \includegraphics[width=0.8\textwidth]{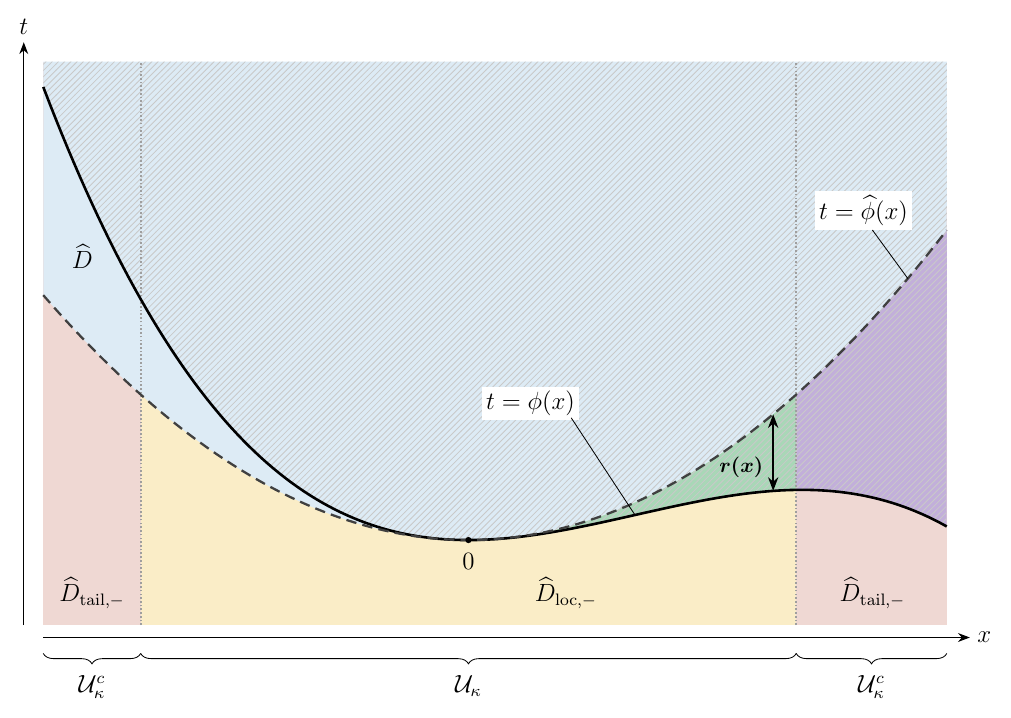}
    \caption{Regions used in the proof of Proposition~\ref{prop:kappa}. The green triangular region is $D\cap\hat D_{\loc,-}$, and $r(x)=\hat\phi(x)-\phi(x)\geq0$ denotes the height of the strip.}
    \label{fig:region}
\end{figure}

To set up the proof, we introduce several important regions, discuss the form of $\hat p_N$ and $\hat p_\loc$, and formulate Assumption~\ref{assume:bdy} in $(x,t)$ coordinates. Throughout, set 
\begin{equation}\label{r-mu}
r(x)=\hat\phi(x)-\phi(x), \qquad \mu=\mu_0/\ell.
\end{equation}
The regions are shown in Figure~\ref{fig:region} and defined below. Each set is specified through the restrictions it imposes on $(x,t)\in\R^d\times\R$.
\begin{equation}\label{D-defs}
\hat D = \{t > \hat\phi(x)\},\quad \hat D_{\loc, -} = \{x\in\us_\kappa,\,t\leq\hat\phi(x)\}, \quad \hat D_{\tail, -} = \{ x\in\us_\kappa^c,\,t\leq\hat\phi(x)\}.
\end{equation}
Note that $\hat\pi_+$ is supported in $\hat D$, the light-blue region in the figure lying above $\hat\phi$, the dotted parabola. Meanwhile, $\hat\pi_-$ is supported in $\hat D^c$, the complement of $\hat D$ in $\R^{d+1}$. This set is the disjoint union of $\hat D_{\loc, -}$ and $\hat D_{\tail, -}$. Note that $\hat D_{\loc,-}$ contains both the green and yellow regions in the figure; these are its intersections with $D$ and $D^c$, respectively.

Next, consider the formula~\eqref{hatpgeo} for $\hat p_N$. As explained below that formula, it coincides with the original form~\eqref{IS} of the IS estimator. In our proof, we return to this original formulation of $\hat p_N$, which we reproduce below for convenience:
\begin{equation}\label{hatp2}
    \hat p_N= \frac1N\sum_{i=1}^N\frac{\pi(U_i)}{\hat\pi(U_i)}\1\{U_i\in D\},\qquad U_i=(X_i,T_i)\iid\hat\pi.
\end{equation}
Furthermore, recall from~\eqref{hatploc1} that $\hat p_\loc$ is obtained from $\hat p_N$ by multiplying the $i$th summand of $\hat p_N$ by the indicator $\1\{J_i=1\text{\,or\,}X_i\in\us_\kappa\}$. The complement of this indicator is $\1\{U_i\notin\hat D_{\tail,-}\}$, since $J_i=0$ and $X_i\in\us_\kappa^c$ corresponds to sampling below $\hat\phi$ and outside $\us_\kappa$. Thus
\begin{equation}\label{loc2}
    \hat p_\loc= \frac1N\sum_{i=1}^N\frac{\pi(U_i)}{\hat\pi(U_i)}\1\{U_i\in D\}\1\{U_i\notin\hat D_{\tail,-}\}.
\end{equation} Here, the random vectors $U_i=(X_i, T_i)$ are equal to those used in~\eqref{hatp2}, so that $\hat p_N$ and $\hat p_\loc$ are coupled. Next, we state implications of Assumption~\ref{assume:bdy} in terms of $D$ and the $(x,t)$ coordinates. Namely, the assumption implies that
\begin{align}
\{(x,t)\,:\, x\in V_0/\ell,\;\; \phi(x)\leq t\leq\phi(x)+\mu\}&\subseteq D, \label{strip2}\\
\{(x,t)\,:\, x\in  V_0/\ell,\;\; t<\phi(x)\}&\subseteq D^c.\label{one-side2}
\end{align}

\subsection{High-probability localization}\label{subsec:loc}
 Comparing~\eqref{hatp2} to~\eqref{loc2}, we see that $X_i\in\us_\kappa$ for all $i=1,\dots, N$ suffices to ensure $U_i=(X_i,T_i)\notin\hat D_{\tail,-}$ for all $i$, and therefore that $\hat p_N=\hat p_\loc$.
Since $X_i\sim\N(0_d,(\la H)^{-1})$, a union bound and Lemma~\ref{lma:chi} give
\begin{align}
\Pp(\hat p_N=\hat p_\loc)&\geq\Pp(X_i\in\us_\kappa\text{ for every }i=1,\dots,N)\notag\\
&\geq 1-N\Pp(X_1\in\us_\kappa^c)=1-N\Pp(\chi_d^2\geq\la\kappa^2)\notag\\
&\geq1-Ne^{-\la\kappa^2/4}.
\label{exp-kappa}
\end{align} For the last inequality, admissibility and $\la\geq2(d+1)$ give $\la\kappa^2/d\geq36\log(\la/d)>9$, as required by Lemma~\ref{lma:chi}.
This proves the claimed lower bound~\eqref{eq:kappa-prob-clean} on the probability that $\hat p_N=\hat p_\loc$.

\subsection{Relative MSE of localized estimator}\label{subsec:MSE} This section is devoted to proving~\eqref{MSE-decomp}-\eqref{Eexpbd}. To bound $\E[(\hat p_\loc/p-1)^2]$, we first use~\eqref{loc2} to write $\hat p_\loc/p$ as follows:
\begin{equation}\label{loc3}\begin{split}
    \frac{\hat p_\loc}{p} &= \frac1N\sum_{i=1}^N\frac{\pi_D(U_i)}{\hat\pi(U_i)}\1\{U_i\notin\hat D_{\tail,-}\}.
\end{split}
\end{equation}
Here, we used that $\pi(u)\1\{u\in D\}/p$ is precisely $\pi_D(u)$. In the following lemma, we claim the decomposition~\eqref{MSE-decomp} holds with $\epsilon_\pm$ defined as follows:
\begin{equation}\label{def:epsilon-plusminus}
  \epsilon_+=  \int_{D\cap\hat D}\frac{\pi_D(u)^2}{\hat\pi_+(u)}du - 1, \qquad \epsilon_-= \int_{D\cap \hat D_{\loc,-}}\frac{\pi_D(u)^2}{\hat\pi_-(u)}du.
\end{equation}
Here, $\hat D,\hat D_{\loc,-}$ are as in~\eqref{D-defs}. The intersection $D\cap\hat D_{\loc,-}$ is the green region in Figure~\ref{fig:region}.
\begin{lemma}[Localized MSE decomposition]\label{lma:loc-MSE}
The decomposition~\eqref{MSE-decomp} holds with $\epsilon_\pm$ as in~\eqref{def:epsilon-plusminus} and
\begin{equation}\label{eps-V}
 0\leq \mathcal E_{\exp}\leq \pi_D(\hat D_{\tail,-})^2+2\pi_D(\hat D_{\tail,-})/N.
\end{equation}
\end{lemma} See Appendix~\ref{app:proof} for the short proof. Thus it remains to bound $\epsilon_+$, $\epsilon_-$, and $\pi_D(\hat D_{\tail,-})$, with a bound on the latter implying a bound on $\mathcal E_{\exp}$. 

\paragraph{Gaussian--exponential factorization.} Bounding $\epsilon_\pm$ requires controlling the ratios $\pi_D/\hat\pi_\pm$ over the relevant regions. The starting point for analyzing this ratio is the key identity~\eqref{key-identity}. For intuition, we now derive the identity. Write the density of $\pi=\mathcal N((0_d,-1),\la^{-1}I_{d+1})$ as $\pi(x,t)\propto e^{-\la z(x,t)}$, and decompose $z$ as follows:
\begin{equation}\label{zxt}
z(x,t)=\frac12\|x\|^2+\frac12(t+1)^2=\frac12+t+\frac12\|x\|^2+\frac12t^2.
\end{equation}
Substituting $t=\hat\phi(x)+y$ and using $\hat\phi(x)+\|x\|^2/2=x^\top Hx/2$ gives
$$
z(x,\hat\phi(x)+y)=\frac12+\left(y+\frac12x^\top Hx\right)+\frac12(\hat\phi(x)+y)^2.
$$
Multiplying by $-\lambda$ and exponentiating, we recognize the kernel of the Gaussian--exponential density $\varrho(x)(\la e^{-\la y})$ for $y\geq0$ on the righthand side, up to its normalizing constant. As discussed below~\eqref{key-approx}, this density has characteristic scales $x=\mathcal O(\la^{-1/2})$ and $y=\mathcal O(\la^{-1})$ for fixed $d,D$. On these scales, the terms in parentheses contribute order one to the exponent $\la z$, while the final squared term contributes order $\la^{-1}$. This informally justifies treating the latter as a remainder.

Upon introducing the needed normalizing constants on both sides, we obtain the exact identity
\begin{equation}\label{key-identity-again}
\pi(x,\hat\phi(x)+y)=\psorm\,\varrho(x)\la e^{-\la y}e^{-\la(\hat\phi(x)+y)^2/2},\qquad (x,y)\in\R^d\times\R.
\end{equation}
Importantly, this derivation shows that $\psorm$ is simply the constant that allows us to massage the elementary identity~\eqref{zxt} into the form~\eqref{key-identity-again}. The fact that $\psorm$ approximates $p=\pi(D)$ is in turn a \emph{consequence} of combining~\eqref{key-identity-again} with error bounds. Indeed, integrating~\eqref{key-identity-again} over the approximate event $\hat D=\{(x,t):t>\hat\phi(x)\}$ gives
$$
\frac{\pi(\hat D)}{\psorm}=\E\left[e^{-\la(\hat\phi(X)+Y)^2/2}\right].
$$
If the expectation is close to one and replacing $D$ by $\hat D$ changes little probability mass relative to $\psorm$, then $p/\psorm\approx 1$. 

\paragraph{Preliminary conditional density bound.}
For the upper bounds on $\pi_D/\hat\pi_\pm$ needed for our analysis, a one-sided estimate on $p/\psorm$ suffices. Indeed, divide~\eqref{key-identity-again} through by $p$ to get $\pi_D$ on the lefthand side, and $\psorm/p$ on the right. Conveniently, the correction factor $e^{-\la(\hat\phi(x)+y)^2/2}$ is at most one, so a lower bound on $p/\psorm$ gives an upper bound on $\pi_D$ in terms of $\varrho(x)\la e^{-\la y}$, and hence bounds on $\pi_D/\hat\pi_\pm$.

To obtain the lower bound on $p/\psorm$, we use the strip condition~\eqref{strip2}: the event $D$ contains the region
$$
S:=\{(x,\phi(x)+s):x\in\us_\kappa,\ 0\leq s\leq\mu\}.
$$
Integrating~\eqref{key-identity-again} over $S$ gives an expression for $\pi(S)$ (and hence a lower bound on $p=\pi(D)$) in terms of $\psorm$ times a Gaussian--exponential expectation. The following lemma records this lower bound and the implied upper bound on $\pi_D$.%
\begin{lemma}
We have
\begin{align}
&p\geq\psorm a, \qquad a:=\E\left[e^{-\frac\la2(\phi(X)+Y)^2+\la r(X)}\1_{\us_\kappa}(X)\1_{[0,\mu]}(Y)\right],\label{main1}\\
&\pi_D(x,\hat\phi(x)+y)\leq\frac1a\varrho(x)\la e^{-\la y},\qquad\forall (x,y)\in\R^d\times\R.\label{main2} \end{align}
\end{lemma} 
\begin{proof}
Since $S\subseteq D$, integrating~\eqref{key-identity-again} with $y=s-r(x)$ gives
$$
\frac{p}{\psorm}\geq\frac{\pi(S)}{\psorm}
=\int_{\us_\kappa}\int_0^\mu \varrho(x)\la e^{-\la s}e^{\la r(x)-\la(\phi(x)+s)^2/2}\,ds\,dx=a.
$$
Dividing~\eqref{key-identity-again} by $p$, restricting to $D$, and using $e^{-\la(\hat\phi(x)+y)^2/2}\leq1$ proves~\eqref{main2}.
\end{proof}
Provided we can show the exponential in the definition of $a$ is close to 1, we conclude that $p/\psorm \geq a \approx \Pp(X\in\us_\kappa, 0\leq Y\leq \mu)$. This explains the purpose of the strip condition: to ensure enough Gaussian--exponential mass is contained within $D$. 

\paragraph{Preliminary error bounds.} With~\eqref{main2} in hand, we can now bound $\epsilon_\pm$. Recall from~\eqref{def:epsilon-plusminus} that $\epsilon_-$ is an integral over $D\cap \hat D_{\loc,-}$. The key geometric observation is that by the one-sidedness condition~\eqref{one-side2}, this region is contained in a strip whose width is the positive part of the quadratic boundary approximation error $r(x)=(\hat\phi-\phi)(x)$:
\begin{equation}\label{hatDlocD}
D\cap\hat D_{\loc,-}\subseteq\{(x,\hat\phi(x)-s):x\in\us_\kappa,\ 0\leq s\leq r(x)_+\}.
\end{equation} See the green region in Figure~\ref{fig:region}. This allows us to control $\epsilon_-$ in terms of the strip width $r(x)_+$. To see this, note that for $u=(x,\hat\phi(x)-s)$ with $s\geq0$, we have $\hat\pi_-(u) = \varrho(x)\la e^{-\la s}$. Therefore, substituting this $u$ into the bound~\eqref{main2} gives $\pi_D(u)/\hat\pi_-(u)\leq a^{-1}e^{2\la s}$. Squaring this ratio and integrating against $\hat\pi_-(u)=\varrho(x)\la e^{-\la s}$ over the strip yields
\begin{equation}\label{eps-arg}
\epsilon_-\leq\frac1{a^2}\int_{\us_\kappa}\varrho(x)\int_0^{r(x)_+}\la e^{3\la s}\,ds\,dx
=\frac1{3a^2}\E\left[\left(e^{3\la r(X)_+}-1\right)\1_{\us_\kappa}(X)\right].
\end{equation}
The strip parameterization therefore reduces $\epsilon_-$ to a Gaussian expectation involving only the boundary error $r(X)$. This motivates the quantity $G$ defined in~\eqref{def:G} below. 

For $\epsilon_+$, we can use~\eqref{main2} to get $\pi_D/\hat\pi_+\leq a^{-1}$ over the entire set $\hat D$. This leads to the upper bound $\epsilon_+ \leq a^{-1}-1$. One can also show the lower bound $\epsilon_+\geq -2\pi_D(\hat D^c)$, and note that $\pi_D(\hat D^c)=\pi_D(D\cap \hat D_{\loc,-})+\pi_D(D\cap\hat D_{\tail,-})$. We can control $\pi_D(D\cap \hat D_{\loc,-})$ using the same strip characterization~\eqref{hatDlocD}, leading to a bound in terms of $G$, as for $\epsilon_-$. The tail probability is the same one needed for the bound on $\mathcal E_{\exp}$, recalling~\eqref{eps-V}. We handle it separately in Lemma~\ref{lma:analytic-controls} below. We summarize these arguments in the following lemma.
\begin{lemma}[Preliminary bounds on $\epsilon_\pm$]\label{lma:risk-reduction}
Let $X\sim\mathcal N(0_d,(\la H)^{-1})$ and $Y\sim\Exp(\la)$ be independent, let $a$ be as in~\eqref{main1}, and define
\begin{align}
G&:=\E\left[\left(e^{3\la r(X)_+}-1\right)\1_{\us_\kappa}(X)\right].\label{def:G}
\end{align}
Then
\begin{align}
&0\leq\epsilon_-\leq\frac{G}{3a^2},\label{eq:epsilon-minus-abstract}\\
&|\epsilon_+|\leq 2\pi_D(\hat D_{\tail,-}) + a^{-1}G +a^{-1}-1.\label{eq:epsilon-plus-abstract}
\end{align}
\end{lemma} See Appendix~\ref{app:proof1} for the proof of~\eqref{eq:epsilon-plus-abstract}.

\paragraph{Quantitative estimates.} Finally, we move on to quantitative estimates of $a,G$, and the tail probability. 
\begin{lemma}\label{lma:analytic-controls}
Let $\mathcal R_a=\epsilon_2^2+\epsilon_4^2+(1+\epsilon_3^2+\epsilon_4^4)/\la$ and $\mathcal R_G = \epsilon_4^2 + \epsilon_3^2 + \epsilon_4^4/d$. Then there is an absolute constant $C_\rho$ depending only on $\rho$ such that
\begin{align}
a^{-1}-1&\lesssim e^{-\la\kappa^2/4}+\mathcal R_a e^{6\mathcal R_a},
\label{eq:a-simple}\\
G&\lesssim (\epsilon_3+\epsilon_4^2)e^{C_\rho \mathcal R_G}
\label{eq:G-simple}\\
\pi_D(\hat D_{\tail,-})&\lesssim a^{-1}\la^{-1/4}e^{-\la\kappa^2/48}\lesssim a^{-1}\frac1\la.
\label{eq:beta-simple}
\end{align} 
\end{lemma}
See Appendix~\ref{app:proof2} for the proof. We give intuition for the two key mechanisms behind these estimates.

\paragraph{Mechanism 1: Taylor remainders and Gaussian moments.}
The bounds on $a$ and $G$ quantify the corrections arising from the remainder $\la(\hat\phi(x)+y)^2/2$ in the density decomposition~\eqref{key-identity-again}, and from the boundary Taylor approximation error $r=\hat\phi-\phi$. To lower bound $a$, we use the conditional Jensen inequality, in the form $a=\E[e^{-A}\1_E] \geq \Pp(E)e^{-\E[A\1_E]/\Pp(E)}$. This reduces the problem to bounding the probability of leaving the strip, the quadratic moment $\la\E[(\phi(X)+Y)^2\1_{\us_\kappa}(X)]$, and the expectation $\la\E[r(X)\1_{\us_\kappa}(X)]$. Writing $\phi=\hat\phi-r$ then reduces the quadratic moment to $\lambda$-scaled moments of $\hat\phi(X)+Y$ and $r(X)$, which are controlled by Gaussian moments and Taylor estimates. To see why the $\la$-scaling does not make these quantities blow up, recall $X\sim\la^{-1/2}$ and $Y\sim\la^{-1}$. Since $r$ is approximately cubic in $X$, we see that $\la r(X)$ scales with $\la$ as $\la\times\la^{-3/2} = \la^{-1/2}$. The term $\la(\hat\phi(X)+Y)^2$ is even smaller, of order $\la^{-1}$.

For the bound on $G$, we use $e^v-1\leq ve^v$ for $v\geq0$ and Cauchy--Schwarz to get
$$
G\leq3\left(\E[(\la r(X))^2\1_{\us_\kappa}(X)]\right)^{1/2}\left(\E[e^{6\la r(X)}\1_{\us_\kappa}(X)]\right)^{1/2}.
$$
Taylor expansion, Gaussian moment estimates, and Gaussian concentration of Lipschitz functions, show the first factor is bounded by a constant times $\epsilon_3+\epsilon_4^2$, and the second factor is bounded by a constant.

\paragraph{Mechanism 2: exponential tail separation.}
First, note that $\pi_D(\hat D_{\tail,-})=\pi(D\cap\hat D_{\tail,-})/\pi(D)$, a ratio of two Gaussian probabilities. Heuristically, the probability of a set under $\pi$ is exponentially dominated by $\exp(-\la m^2/2)$, where $m$ is the shortest distance from the set to the Gaussian center $(0_d,-1)$. For $D$, this distance is $1$. This is reflected in $\psorm$, which is within a factor $a$ of $p=\pi(D)$ and has dominant exponential scaling $e^{-\la/2}$. For $D\cap\hat D_{\tail,-}$, we show in Lemma~\ref{lma:geometric-separation} that its shortest distance from $(0_d,-1)$ is $1+\kappa^2/3$. This is by the no-return condition of Assumption~\ref{assume:no-return}. Since $\la\|U-(0_d,-1)\|^2$ has law $\chi_{d+1}^2$ under $U\sim \pi$, and $p\geq a\psorm$, it follows that
$$
\pi_D(\hat D_{\tail,-})\leq\frac{\Pp(\chi_{d+1}^2\geq\la(1+\kappa^2/3))}{a\psorm}.
$$
The numerator has leading exponential factor $e^{-\la(1+\kappa^2/3)/2}$, whereas $\psorm$ contains $e^{-\la/2}$. Their ratio therefore decays exponentially in $\la\kappa^2$. There are additional dimension-dependent terms in this ratio, stemming from the chi-square bound and from $\det H$ in the definition of $\psorm$. However, the admissibility conditions ensure the exponential decay dominates these dimension-dependent terms, giving~\eqref{eq:beta-simple}.

\paragraph{Completing the proof of Proposition~\ref{prop:kappa}.}
We now combine the bounds to prove~\eqref{MSE-err+},~\eqref{MSE-err-}, and~\eqref{Eexpbd}. First note that~\eqref{eq:a-simple} gives the simple bound $a^{-1} \lesssim e^{C\mathcal R_a}$ for an absolute constant $C$. Second, since $\la\geq2(d+1)$, we have $\mathcal R_a+\mathcal R_G\lesssim\mathcal R+1/\la$, where $\mathcal R$ is as in~\eqref{def:epsilon-234}. Thus $e^{C\mathcal R_a+C_\rho\mathcal R_G}\lesssim_\rho e^{C_\rho\mathcal R}$ after enlarging $C_\rho$, with $e^{C_\rho/\la}$ absorbed into the implicit constant. Combining these bounds with~\eqref{eq:epsilon-minus-abstract} and~\eqref{eq:G-simple} gives
\begin{equation}\label{eq:epsilon-minus-final}
    0\leq \epsilon_-\lesssim_\rho (\epsilon_3 + \epsilon_4^2)e^{C_\rho \mathcal R_G + C\mathcal R_a} \lesssim_\rho (\epsilon_3 + \epsilon_4^2)e^{C_\rho\mathcal R},
\end{equation} proving~\eqref{MSE-err-}. Similarly,~\eqref{eq:epsilon-minus-abstract}, $a^{-1} \lesssim e^{C\mathcal R_a}$,~\eqref{eq:beta-simple},~\eqref{eq:a-simple}, and~\eqref{eq:G-simple} give
\begin{equation}\label{pre-final}
\begin{split}
|\epsilon_+| &\lesssim a^{-1}/\la + a^{-1}G + a^{-1} - 1 \\
&\lesssim e^{C\mathcal R_a+C_\rho \mathcal R_G}(\epsilon_3 + \epsilon_4^2 + e^{-\la\kappa^2/4}  + \mathcal R_a  + 1/\la)\\
&\lesssim_\rho e^{C_\rho\mathcal R}(\epsilon_3 + \epsilon_2^2+\epsilon_4^2 + 1/\la + \epsilon_3^2/\la + \epsilon_4^4/\la).
\end{split} \end{equation} Since $\epsilon_3^2+\epsilon_4^4\leq\mathcal R+\mathcal R^2\lesssim e^{\mathcal R}$, the last two terms can be absorbed into $e^{C_\rho\mathcal R}/\la$ by enlarging $C_\rho$. This proves~\eqref{MSE-err+}. Finally, combining~\eqref{eps-V},~\eqref{eq:beta-simple}, $a^{-1}\lesssim e^{C\mathcal R_a}$, and bounding $\la^{-1/4}$ by 1, gives
\begin{align}
0\leq\mathcal E_{\exp}&\lesssim \pi_D(\hat D_{\tail,-})^2+\frac1N\pi_D(\hat D_{\tail,-})\lesssim e^{C\mathcal R_a}\left(e^{-\la\kappa^2/24}+\frac1N e^{-\la\kappa^2/48}\right)
\end{align}
Using $\mathcal R_a\lesssim\mathcal R+1/\la$ and absorbing $e^{C/\la}$ into the implicit constant proves~\eqref{Eexpbd} and completes the proof of Proposition~\ref{prop:kappa}.
\begin{remark}[Formulation in~\eqref{intro:CV}]
The exponentially small remainder $\mathcal E_{\exp}$ in~\eqref{intro:CV} is not identical to that in~\eqref{MSE-decomp}. Indeed, since $\hat p_N=\hat p_{\loc}$ on $A_N$, the former differs from the latter by $-\E[(\hat p_{\loc}/p-1)^2\1_{A_N^c}]$. This additional term can easily be controlled using the same localization estimates as in the proof of Proposition~\ref{prop:kappa}. The final result is that $|\mathcal E_{\exp}|\lesssim_{\eta,\rho} e^{-\la\kappa^2/48}$ for the version of $\mathcal E_{\exp}$ used in the introduction, with $\eta$ as in Proposition~\ref{prop:kappa}. %
\end{remark}

\section{Conclusion and discussions}\label{sec:conclude}
We propose GeoIS, an importance sampling method for rare-event probabilities when the underlying uncertain distribution is Gaussian, that combines a Gaussian–exponential proposal based on local boundary geometry at the instanton with a lower mixture component to preserve unbiasedness. We provide theoretical guarantees on the error bounds for GeoIS, which do not depend on the dimension, rarity, or geometry of the event set under mild assumptions, while numerical experiments demonstrate substantial reductions in relative variance and robustness with respect to changes in dimension, rarity, and geometry across the tested examples, ranging from analytical examples to PDE-involved examples.

There are several extensions for future exploration. A natural extension is to develop and rigorously justify an intrinsic low-dimensional version of GeoIS using only dominant curvature directions to reduce the cost of constructing the proposal, motivated by the intrinsic low-dimensional structure in the diffusion PDE example. This will make the method applicable to even infinite-dimensional problems, for example, systems with SDEs or SPDEs where the uncertainty is in the forcing terms, such as Brownian motion or Gaussian random fields. In addition, GeoIS can be combined with control variates to further reduce estimator variance, with a similar effect to choosing an optimal mixture weight $w$.

\section*{Acknowledgment}
ST is partially supported by National Science Foundation under Grant No.~DMS-2529292
and thanks Timo Schorlepp and Georg Stadler for earlier related discussions in 2022.

\appendix

\section{Quantitative verification of the geometric assumptions}\label{app:asymp}
\subsection{Choosing the localization radius}\label{app:radii}
We make note of several general simplifications for checking our assumptions. First, a Taylor expansion of $\phi$ gives
$\phi(x)-\frac12x^\top \nabla^2\phi(0_d)x
\geq-\frac16r_1\tau_3(r_1)x^\top Hx$ for all $x\in\us_{r_1}$ with $\ell\us_{r_1}\subseteq V_0$. Therefore, 
\begin{equation}\label{localcap-suff}
   r_1\tau_3(r_1)\leq2 \implies \text{local cap}.
\end{equation}  This condition is convenient but not necessary: it restricts both upward and downward deviations of $\phi$ from its quadratic approximation, whereas the local cap condition only penalizes downward deviations.

Next, suppose there is some $R>0$ satisfying $\ell\us_R\subseteq V_0$ for which we have derivative control in $\us_R$; namely, $\tau_k(R)\leq T_k$, $k=3,4$ for some $T_3,T_4\geq0$. Then a reasonable choice of $r_1$ to satisfy $r_1\tau_3(r_1)\leq 2$ is the following largest possible one given knowledge that $\tau_3(R)\leq T_3$, namely, 
\begin{equation}\label{r1-choice}
r_1= \min\{R,2/T_3\},
\end{equation}
where $2/T_3=+\infty$ if $T_3=0$. Indeed, $r_1\leq R$ gives $\tau_3(r_1)\leq\tau_3(R)\leq T_3$, and hence $r_1\tau_3(r_1)\leq r_1T_3\leq2$. The reason to take $r_1$ large is that it weakens the no-return condition in Assumption~\ref{assume:no-return} and it increases the range of admissible $\kappa$'s in~\eqref{kap:admit}. Note that since this $r_1\leq R$ and $\ell \us_R\subseteq V_0$, we have $\ell\us_{r_1}\subseteq V_0$ as well, which is required in Assumption~\ref{assume:bdy}.

Next, suppose the no-return condition is indeed satisfied with this choice of $r_1$ and some $\gamma>0$, and assume the strip condition holds with some height $\mu_0>0$. Then a natural choice of $\rho$-admissible $\kappa$ is
\begin{equation}\label{kap-choice}
\kappa^2 = \min\left(\left(\frac{\rho r_1}{2}\right)^2, 3\gamma, \frac{4\mu_0}{\ell}\right) 
\end{equation} for any $r_1$ satisfying $r_1\tau_3(r_1)\leq 2$, e.g. as in~\eqref{r1-choice}, provided this $\kappa$ satisfies
\begin{equation}\label{kap-choice-need}
\frac{12}{\lambda}
\left(3d\log\frac{\lambda}{d}
+2(1\vee\log\det H)\right)\leq\kappa^2.
\end{equation} 
Note that for this $\kappa$ we have, $\kappa\leq r_1$ and $\kappa\tau_3(\kappa)\leq (\rho r_1/2)\tau_3(r_1)\leq\rho$, so the two upper bounds in~\eqref{kap:admit} are indeed satisfied and only the lower bound~\eqref{kap-choice-need} remains to establish $\rho$-admissibility.

The choice of $\kappa$ in~\eqref{kap-choice} is close to the largest it can be under~\eqref{kap:admit}. The reason for choosing a large $\kappa$ is first, that doing so decreases the exponentially small term $\mathcal E_{\exp}$ from~\eqref{Eexpbd} and increases the localization probability~\eqref{eq:kappa-prob-clean}. Second, a smaller $\kappa$ could only help to decrease $\tau_k(\kappa)$, $k=3,4$ in the bounds~\eqref{MSE-err+} and~\eqref{MSE-err-} on $\epsilon_\pm$. Thus if we assume knowledge of the derivative bounds only in the larger region of radius $R$, then there is no use in decreasing $\kappa$. 

\subsection{Explicit boundary}\label{app:explicit}
Suppose $D_0$ is the global epigraph
\begin{equation}\label{D0-epi}
D_0=\{u^*+Qv+sn:v\in\R^d,\ s\geq\phi_0(v)\},
\end{equation}
where $\phi_0\in C^4(\R^d)$, $u^*=\ell n$ is the unique instanton, and $Q,n$ are as in Section~\ref{sec:setup}. In particular, $\phi_0(0_d)=0$ and $\nabla\phi_0(0_d)=0_d$, with $\phi_0\in C^4(\R^d)$.

The global epigraph representation then gives Assumption~\ref{assume:bdy} with $V_0=\R^d$ and arbitrary finite $\mu_0>0$. It remains to verify the local cap and no-return conditions, find an admissible $\kappa>0$, and control the derivatives of $\phi$ to compute $\epsilon_j$, $j=2,3,4$.

\begin{lemma}[Explicit boundary: derivative bounds and local cap]\label{lma:explicit-local}
Suppose $D_0$ is as in~\eqref{D0-epi} and $H=I_d+\ell\nabla^2\phi_0(0_d)\succ0$. Let $h$ be a lower bound on $H$, i.e. $hI_d\preceq H$ and suppose that 
\begin{equation}\label{phi0-deriv}
    \|\nabla^2\phi_0(0_d)\|\leq C_2,\qquad \sup_{\|v\|\leq R_0}\|\nabla^j\phi_0(v)\|\leq C_j,\qquad j=3,4,
\end{equation} for some $R_0>0$ and $C_2,C_3,C_4\geq0$. 
Then
\begin{equation}\label{eq:explicit-tau}
\tau_2(0)\leq\ell C_2/h=:T_2,\qquad
\tau_j(R)\leq\ell^{j-1}C_j/h^{j/2}=:T_j,\quad j=3,4,
\end{equation}
for $R:=R_0\sqrt{h}/\ell$. Assumption~\ref{assume:bdy} holds with $V_0=\mathbb R^d$ and any finite $\mu_0>0$, and the local cap condition in Assumption~\ref{assume:no-return} holds with $r_1=\min(R, 2/T_3)$.
\end{lemma}
The derivative bounds follow from~\eqref{eq:weighted-rescaling} and $\|T\|_H\leq h^{-k/2}\|T\|$ for a $k$th order derivative tensor $T$. For $\|H^{1/2}x\|\leq R=R_0\sqrt{h}/\ell$, we have $\|v=\ell x\|\leq R_0$, so~\eqref{phi0-deriv} applies. The choice $r_1=\min(R, 2/T_3)$ gives the local cap by~\eqref{localcap-suff}.

Verifying the no-return condition in Assumption~\ref{assume:no-return} requires global information about $\phi_0$. In the below lemma, we consider one natural case, in which $\phi_0$ is strongly convex. In addition to checking the no-return condition, we also simplify some of the quantities in Lemma~\ref{lma:explicit-local} using an explicit lower bound on $h$.
\begin{lemma}[Full assumption verification for explicit boundary, strongly convex $\phi_0$]\label{lma:cvx}
Suppose $D_0$ is as in~\eqref{D0-epi}, $\phi_0$ is globally $m$-strongly convex for some $m>0$, and $\phi_0$ satisfies~\eqref{phi0-deriv}. Then Assumptions~\ref{assume:bdy} and~\ref{assume:no-return} hold with
\begin{equation}\label{eq:explicit-convex-radii}
V_0=\mathbb R^d,\qquad a=\min\{R_0,2m/C_3\},\qquad r_1=\frac{a\sqrt{1+m\ell}}{\ell},\qquad \gamma=\frac{1+m\ell}{1+C_2\ell}r_1^2,
\end{equation}
and any finite $\mu_0>0$. The derivative bounds~\eqref{eq:explicit-tau} hold with $R=R_0\sqrt h/\ell$ and $h=1+m\ell$, and we have $r_1\leq R$.
\end{lemma}
\begin{proof}
We have $\nabla^2\phi_0(0_d)\succeq mI_d$ by assumption, so $h=1+m\ell$ is a valid lower bound on $H$. Also, $\|H\|\leq1+C_2\ell$. For $a,r_1$ from~\eqref{eq:explicit-convex-radii},  $R=R_0\sqrt h/\ell$ and $T_3=\ell^2C_3/h^{3/2}$ as in Lemma~\ref{lma:explicit-local}, these bounds give
\begin{align*}
 r_1=\frac{a\sqrt h}{\ell}&\leq\frac{R_0\sqrt h}{\ell}=R,\\
 r_1\tau_3(r_1)&\leq r_1\tau_3(R) \leq r_1T_3 \\
 &= \frac{aC_3\ell}{ h}=\frac{aC_3\ell}{1+m\ell}\leq\frac{aC_3}{m}\leq2.   
\end{align*}
Thus~\eqref{localcap-suff} gives the local cap condition. It remains to prove no return. Strong convexity and the normalization at zero give
$$
\phi_0(v)\geq\frac{m}{2}\|v\|^2,\qquad
\phi(x)\geq\frac{m\ell}{2}\|x\|^2.
$$
For $(x,t)\in D$, the global epigraph representation implies $t\geq\phi(x)$. Using this, the fact that $\|H\| \leq 1+ C_2\ell$, and that $x^\top Hx\geq r_1^2$ when $x\notin\us_{r_1}$, gives
\begin{align*}
\|x\|^2+(1+t)^2 &=1+\|x\|^2+2t+t^2\geq1+(1+m\ell)\|x\|^2\\
&\geq1+\frac{1+m\ell}{1+C_2\ell}x^\top Hx \geq 1+\frac{1+m\ell}{1+C_2\ell}r_1^2.
\end{align*}
Thus no return is satisfied with $\gamma=\frac{1+m\ell}{1+C_2\ell}r_1^2$.
\end{proof}
Finally, we discuss the remaining checks needed to apply Proposition~\ref{prop:kappa}. Fix $0<\rho<1/2$ and set
\begin{equation}\label{kap-cvx}
\kappa=r_1\min\left\{\frac{\rho}{2},\sqrt{\frac{3(1+m\ell)}{1+C_2\ell}}\right\},
\end{equation}
where $r_1$ is given by~\eqref{eq:explicit-convex-radii}. This is precisely as in~\eqref{kap-choice}, provided $\mu_0$ is chosen sufficiently large. Thus $\kappa$ is $\rho$-admissible provided~\eqref{kap-choice-need} holds, and the only remaining hypotheses of Proposition~\ref{prop:kappa} is $\la\ge 2(d+1)$.

We now apply this argument to the large $\ell$ regime to prove~\eqref{P-CV-asymp-ell}. 
\begin{proof}[Proof of~\eqref{P-CV-asymp-ell}]
Throughout this proof, implicit constants may depend on $m,R_0,C_2,C_3,C_4$, and the fixed $\lambda_0$, but not on $d,\ell,w$, or $N$. Choose $\kappa$ as in~\eqref{kap-cvx} with $\rho=1/4$. Lemma~\ref{lma:cvx} and~\eqref{eq:explicit-tau} give
$$
\kappa^2\asymp\ell^{-1},\qquad T_2=\mathcal O(1),\qquad T_3=\mathcal O(\sqrt\ell),\qquad T_4=\mathcal O(\ell).
$$
Since $\|H\|=O(\ell)$, the left-hand side of~\eqref{kap-choice-need} is $O(d(\log\ell)/\ell^2)=o(\ell^{-1})$ when $d(\log\ell)/\ell\to0$. Thus $\kappa^2\asymp\ell^{-1}$ is indeed $\rho$-admissible for $\rho=1/4$ and all $\ell$ large enough. Furthermore, $\la=\ell^2\lambda_0 \geq 2(d+1)$ also holds under $d(\log\ell)/\ell\to0$. Proposition~\ref{prop:kappa} therefore applies. Finally, we compute $\epsilon_j$, $j=2,3,4$. Substituting $\tau_j(0)\leq\tau_j(R)\leq T_j$, $j=1,2,3$ into $\epsilon_j$ gives
$$
\epsilon_2 \lesssim d/\ell, \qquad \epsilon_3 \lesssim d/\sqrt\ell, \qquad \epsilon_4^2 \lesssim d^2/\ell.
$$
Substituting these bounds and $\la\kappa^2 \asymp \ell$ into~\eqref{MSE-err+}-\eqref{Eexpbd} gives $|\epsilon_+|+\epsilon_-=\mathcal O(e^{Cd^2/\ell}\,d/\sqrt\ell)$ and $\mathcal E_{\exp}=\mathcal O(e^{-c\ell})$. This uses that, in the exponent $\mathcal R$ defined in~\eqref{def:epsilon-234}, we have $\epsilon_4^2 + \epsilon_4^4/d \lesssim \epsilon_4^2(1 + d/\ell) \leq Cd^2/\ell$ by the assumption $d(\log\ell)/\ell\to0$. Substituting into~\eqref{eq:kappa-prob-clean} and~\eqref{MSE-decomp} proves~\eqref{P-CV-asymp-ell}.
\end{proof}

\subsection{Implicit boundary}\label{app:implicit}
 Next, we consider the case in which the boundary of $\Dorig$ is specified implicitly through a function. This is a common rare-event setting in which a response function $F:\R^{d+1}\to\R$ maps the random parameter to an observable, and $\Dorig$ includes all events that the observable exceeds a given threshold $z$. Namely, suppose
$$
\Dorig:=\{u\in\R^{d+1}:F(u)\geq z\},
$$
where $F\in C^4(\mathcal O)$ on an open neighborhood $\mathcal O$ of the instanton $u^*$, and assume that $\nabla F(u^*)\neq0$. %
The first-order optimality condition gives a multiplier $\eta>0$ such that $u^*=\eta\nabla F(u^*)$. Consequently, $n=\nabla F(u^*)/\|\nabla F(u^*)\|$. The implicit function theorem can be applied to $G(v,s) = F(u^*+Qv+sn)$, since $F$ is sufficiently regular in a neighborhood of $u^*$ and $G$ satisfies $\partial_sG(0_d,0)=\nabla F(u^*)^\top  n>0$. We conclude there is a neighborhood $V_0$ of $0_d$ and a unique function $\phi_0\in C^4(V_0)$ such that 
\begin{equation}\label{F-implicit}
F(u^*+Qv + \phi_0(v)n)=z, \qquad \phi_0(0_d)=0,\qquad \nabla\phi_0(0_d)=0_d.
\end{equation}
Differentiating twice yields $\nabla^2\phi_0(0_d)=-\frac{1}{\|\nabla F(u^*)\|}Q^\top\nabla^2F(u^*)Q.$ This gives~\eqref{eq:implicit-BH}.

Recall that $H$ must be positive definite. A sufficient condition is 
\begin{equation}\label{HFsuff}
\frac{\ell}{\|\nabla F(u^*)\|}\lambda_{\max}(\nabla^2F(u^*))<1,
\end{equation}  where $\lambda_{\max}(\nabla^2F(u^*))$ is the largest eigenvalue of $\nabla^2F(u^*)$. If $\nabla^2F(u^*)\preceq 0$ so that the largest eigenvalue is nonpositive, then the condition~\eqref{HFsuff} is always satisfied.

We now proceed to check the assumptions in Proposition~\ref{prop:kappa}. First, we give quantitative guarantees on the width of $V_0$, the height of the strip inside $D_0$, and the derivative quantities $\tau_2,\tau_3,\tau_4$.

\begin{lemma}[Implicit boundary: strip, derivative bounds, local cap]\label{lma:quantitative-implicit}
Let $\gradnorm=\|\nabla F(u^*)\|>0$ and $h=\lambda_{\min}(H)>0$. Suppose $F\in C^4$ on an open set containing $\{\|u-u^*\|\leq R_0\}$ and, for constants $M_2>0$ and $M_3,M_4\geq0$,
\begin{equation}\label{Ms}
    \sup_{\|u-u^*\|\leq R_0}\|\nabla^jF(u)\|\leq M_j,\qquad j=2,3,4.
\end{equation}
Define
\begin{equation}\label{eq:implicit-neighborhood}
R_* = \min\{R_0,\gradnorm/(2M_2)\},\qquad V_0=\{v\in\R^d:\|v\|<R_*/2\},\qquad \mu_0=R_*/4.
\end{equation}
Then there is a unique $\phi_0\in C^4(V_0)$ satisfying~\eqref{F-implicit}, and the strip condition of Assumption~\ref{assume:bdy} is satisfied with the above $V_0,\mu_0$. Furthermore,
\begin{equation}\label{eq:implicit-tau}
\tau_2(0)\leq\frac{\ell C_2}{h}=:T_2,\qquad \tau_j(R)\leq A_j\frac{\ell^{j-1}C_j}{h^{j/2}}=:T_j,\quad j=3,4,
\end{equation}
for $R:=R_*\sqrt{h}/(4\ell)$. Here $A_3,A_4$ are absolute constants, and
\begin{equation}\label{eq:implicit-derivative-constants}\begin{gathered}
C_2 = M_2/\gradnorm,\qquad C_3=M_3/\gradnorm+(M_2/\gradnorm)^2,\\
C_4=M_4/\gradnorm+(M_2/\gradnorm)(M_3/\gradnorm)+(M_2/\gradnorm)^3.
\end{gathered}
\end{equation}
With this $R$ and $T_3$ from~\eqref{eq:implicit-tau}, the local cap condition of Assumption~\ref{assume:no-return} is satisfied with $r_1=\min(R,2/T_3)$, as is the condition $\ell \us_{r_1}\subseteq V_0$.
\end{lemma}
\begin{proof}%
Let $G(v,s)=F(u^*+Qv+sn)$. Orthogonality of $Q$ and $n$ gives $\|Qv+sn\|^2=\|v\|^2+s^2$. Thus the cylinder $\|v\|\leq R_*/2$, $|s|\leq R_*/2$ lies in $\overline B(u^*,R_*)$. The Hessian bound and $\nabla F(u^*)=\gradnorm n$ imply throughout this ball that
\begin{equation}\label{eq:implicit-gradient-variation}
\|\nabla F(u)-\gradnorm n\|\leq M_2\|u-u^*\|\leq M_2R_*\leq \gradnorm/2.
\end{equation}
Since $\|n\|=1$, equation~\eqref{eq:implicit-gradient-variation} implies throughout the same ball that
\begin{equation}\label{eq:implicit-normal-lower}
n^\top\nabla F(u)=\gradnorm+n^\top(\nabla F(u)-\gradnorm n)\geq\gradnorm/2.
\end{equation}
In particular, $\partial_sG\geq \gradnorm/2$ on the cylinder. Since $G(0_d,0)=z$ and $\nabla_vG(0_d,0)=0_d$, Taylor's theorem also gives
$$
|G(v,0)-z|\leq\frac{M_2}{2}\|v\|^2\leq\frac{M_2R_*^2}{8}\leq\frac{\gradnorm R_*}{16}.
$$
Integrating the lower bound on $\partial_sG$ from $s=0$ to $s=\pm R_*/2$ therefore yields
$$
G(v,-R_*/2)\leq z-\frac{3\gradnorm R_*}{16},\qquad G(v,R_*/2)\geq z+\frac{3\gradnorm R_*}{16}.
$$
Strict monotonicity and the intermediate value theorem give a unique root $s=\phi_0(v)$ in this interval. The implicit function theorem makes these roots a $C^4$ function on $V_0$. At $v=0_d$, uniqueness gives $\phi_0(0_d)=0$, and implicit differentiation gives $\nabla\phi_0(0_d)=0_d$. Comparing $G(v,\phi_0(v))=z$ with $G(v,0)$ gives
\begin{equation}\label{eq:implicit-height-comparison}
\frac{\gradnorm}{2}|\phi_0(v)|\leq|G(v,0)-z|\leq\frac{M_2}{2}\|v\|^2.
\end{equation}
At $k(v):=u^*+Qv+\phi_0(v)n$, implicit differentiation also gives
\begin{equation}\label{eq:implicit-gradient-identity}
\nabla\phi_0(v)=-\frac{Q^\top\nabla F(k(v))}{n^\top\nabla F(k(v))}.
\end{equation}
The point $k(v)$ lies in the cylinder and hence in $\overline B(u^*,R_*)$. Since $Q^\top n=0_d$ and $\|Q^\top\|=1$, equation~\eqref{eq:implicit-gradient-variation} gives
\begin{align*}
\|Q^\top\nabla F(k(v))\|&=\|Q^\top(\nabla F(k(v))-\gradnorm n)\|\leq\|\nabla F(k(v))-\gradnorm n\|,\\
&\leq M_2\|k(v)-u^*\|\leq M_2R_*.
\end{align*}
Substituting this numerator bound and the denominator bound~\eqref{eq:implicit-normal-lower} at $u=k(v)$ into~\eqref{eq:implicit-gradient-identity} yields $\|\nabla\phi_0(v)\|\leq2M_2R_*/\gradnorm$. Separately, dividing~\eqref{eq:implicit-height-comparison} by $\gradnorm/2$ gives $|\phi_0(v)|\leq M_2\|v\|^2/\gradnorm$. Using $\|v\|<R_*/2$ and $M_2R_*\leq\gradnorm/2$ from~\eqref{eq:implicit-neighborhood}, these estimates become
\begin{equation}\label{eq:implicit-graph-control}
|\phi_0(v)|\leq\frac{M_2}{\gradnorm}\|v\|^2\leq\frac{R_*}{8},\qquad \|\nabla\phi_0(v)\|\leq\frac{2M_2R_*}{\gradnorm}\leq1.
\end{equation}
Strict monotonicity further gives, for $v\in V_0$ and $|s|\leq R_*/2$,
\begin{equation}\label{eq:implicit-local-epigraph}
F(u^*+Qv+sn)\geq z\quad\Longleftrightarrow\quad s\geq\phi_0(v).
\end{equation}
Since $|\phi_0(v)|+\mu_0\leq3R_*/8<R_*/2$, the claimed strip lies in this cylinder and hence in $D_0$ after mapping to the original coordinates.

We next bound $\tau_2(0)$. Differentiating the boundary identity twice at $v=0_d$, where $\nabla\phi_0(0_d)=0_d$, gives
$$
\nabla^2\phi_0(0_d)=-\frac{1}{\gradnorm}Q^\top\nabla^2F(u^*)Q,\qquad \|\nabla^2\phi(0_d)\|=\ell\|\nabla^2\phi_0(0_d)\|\leq\frac{\ell M_2}{\gradnorm}.
$$
Therefore, 
\begin{align*}
\tau_2(0)=\|\nabla^2\phi(0_d)\|_H\leq\|H^{-1}\|\|\nabla^2\phi(0_d)\|\leq\frac{\ell M_2}{\gradnorm h}.
\end{align*}
This proves the $\tau_2$ bound in~\eqref{eq:implicit-tau}. For the higher derivatives, fix $v\in V_0$ and a unit vector $a$, and write
$$
p_k=\langle\nabla^k\phi_0(v),a^{\otimes k}\rangle,\qquad
w=Qa+p_1n,\qquad A=n^\top\nabla F(k(v)).
$$
Differentiating $F(q(v+ta))=z$ repeatedly gives
\begin{align*}
Ap_2&=-\langle\nabla^2F,w\otimes w\rangle,\\
Ap_3&=-\langle\nabla^3F,w\otimes w\otimes w\rangle-3p_2\langle\nabla^2F,w\otimes n\rangle,\\
Ap_4&=-\langle\nabla^4F,w\otimes w\otimes w\otimes w\rangle-6p_2\langle\nabla^3F,w\otimes w\otimes n\rangle\\
&\quad-3p_2^2\langle\nabla^2F,n\otimes n\rangle-4p_3\langle\nabla^2F,w\otimes n\rangle,
\end{align*}
where all derivatives of $F$ are evaluated at $k(v)$. Since $k(v)$ belongs to the ball on which the derivative bounds hold, $A\geq \gradnorm/2$ and $\|w\|^2=1+p_1^2\leq2$ imply successively
\begin{align*}
|p_2|&\leq4\frac{M_2}{\gradnorm}=4C_2,\\
|p_3|&\leq4\sqrt2\frac{M_3}{\gradnorm}+24\sqrt2\left(\frac{M_2}{\gradnorm}\right)^2\leq24\sqrt2 C_3,\\
|p_4|&\leq8\frac{M_4}{\gradnorm}+160\frac{M_2M_3}{\gradnorm^2}+480\left(\frac{M_2}{\gradnorm}\right)^3\leq480C_4,
\end{align*}
with $C_2, C_3,C_4$ as in~\eqref{eq:implicit-derivative-constants}. Taking the supremum over unit vectors $a$ gives
\begin{equation}\label{AjCj}
 \sup_{v\in V_0}\|\nabla^j\phi_0(v)\|\leq A_jC_j,\qquad j=3,4,   
\end{equation}
where we may take $A_3=24\sqrt2$ and $A_4=480$. If $x\in\us_{R}$ then $\|\ell x\|\leq \ell\|H^{1/2}x\|/\sqrt h \leq \ell R/\sqrt h \leq R_*/4$, and hence $\ell x\in V_0$. Applying~\eqref{AjCj},~\eqref{eq:weighted-rescaling} and $\|T\|_H\leq h^{-j/2}\|T\|$ therefore proves~\eqref{eq:implicit-tau}. In particular, $\ell\us_R\subset V_0$ and $\tau_3(R)\leq T_3$, so $r_1=\min(R,2/T_3)$ gives $r_1\tau_3(r_1)\leq2$ and the claimed local cap condition by~\eqref{localcap-suff}.
\end{proof}
To verify one-sidedness and the no-return condition, we need global assumptions on $F$. One natural assumption is concavity.
\begin{lemma}[Full assumption verification for implicit boundary, concave $F$]\label{lma:concave-level-set}
Suppose $F$ is globally concave and satisfies the conditions of Lemma~\ref{lma:quantitative-implicit}. Then $H\succeq I_d$, and Assumptions~\ref{assume:bdy} and~\ref{assume:no-return} hold with $V_0,\mu_0,r_1$ from Lemma~\ref{lma:quantitative-implicit} and $\gamma=r_1^2/\|H\|$. The derivative bounds of that lemma continue to apply.\end{lemma}
\begin{proof}
Concavity gives $\nabla^2F(u^*)\preceq0$, so~\eqref{eq:implicit-BH} implies $H\succeq I_d$. To prove the one-sidedness part of Assumption~\ref{assume:bdy}, fix $v\in V_0$ and $s<\phi_0(v)$, for $V_0$ as in Lemma~\ref{lma:quantitative-implicit}. Let $u=u^*+Qv+\phi_0(v)n$. By the proof of Lemma~\ref{lma:quantitative-implicit}, $u$ satisfies~\eqref{eq:implicit-normal-lower}. This, negativity of $s-\phi_0(v)$, and concavity of $F$ at $u$, give
$$
F(u^*+Qv+sn)\leq F(u)+(s-\phi_0(v))n^\top\nabla F(u) < F(u)=z.
$$ Therefore, $u^*+Qv+sn\in D_0^c$, proving the one-sidedness part of Assumption~\ref{assume:bdy}. To prove the no-return part of Assumption~\ref{assume:no-return}, note that
$$
z\leq F(u)\leq F(u^*)+\nabla F(u^*)^\top(u-u^*)=z+\gradnorm\,n^\top(u-u^*).
$$ for all $u\in D_0$, by concavity of $F$ at $u^*$. This implies that ever $(x,t)\in D$ satisfies $t\geq0$. If also $x\notin\us_{r_1}$, then
$$
\|x\|^2+(1+t)^2\geq1+\|x\|^2\geq1+\frac{x^\top Hx}{\|H\|}>1+\frac{r_1^2}{\|H\|}.
$$
This proves no return with the stated $\gamma$. The other parts of the assumptions have already been checked in Lemma~\ref{lma:quantitative-implicit}.
\end{proof}
To apply Proposition~\ref{prop:kappa}, assume $\la\geq 2(d+1)$, fix $0<\rho<1/2$ and set
$$
\kappa=\min\left\{\frac{\rho r_1}{2},\frac{\sqrt3\,r_1}{\sqrt{\|H\|}},\sqrt{\frac{R_*}{\ell}}\right\},
$$
where $R_*$ and $r_1$ are given by Lemma~\ref{lma:quantitative-implicit}. This is the choice~\eqref{kap-choice} with $\gamma=r_1^2/\|H\|$ and $\mu_0=R_*/4$, so this $\kappa$ is $\rho$-admissible provided~\eqref{kap-choice-need} holds.

\section{Details for the examples}\label{app:examples}
\subsection{Baseline importance-sampling methods}
\label{app:baselines}
For comparison, we use the LDT-based shifted importance sampling (LSIS) method in \cite{tong2021extreme}. In the original coordinates, the LSIS proposal is \(q_{\mathrm{LSIS}}^{u}=\mathcal N\!\left(u^\star,\lambda_0^{-1}I_{d+1}\right)\). Under the coordinate transformation in Section~\ref{sec:setup}, this becomes \(q_{\mathrm{LSIS}}^{xt}=\mathcal N\!\left((0_d,0),\lambda^{-1}I_{d+1}\right)\), with \(\lambda=\ell^2\lambda_0\). For samples \(U_i\) and \((X_i,T_i)\) related by this transformation, the LSIS estimator has the equivalent representations
\begin{align*}
\hat p_{\mathrm{LSIS}}
&=\frac1N\sum_{i=1}^N\1_{\Dorig}(U_i)\frac{\piorig(U_i)}{q_{\mathrm{LSIS}}^{u}(U_i)},\qquad U_i\overset{\mathrm{i.i.d.}}{\sim}q_{\mathrm{LSIS}}^{u},\\
&=\frac1N\sum_{i=1}^N\1_{\Dnew}(X_i,T_i)\frac{\pinew(X_i,T_i)}{q_{\mathrm{LSIS}}^{xt}(X_i,T_i)},\qquad (X_i,T_i)\overset{\mathrm{i.i.d.}}{\sim}q_{\mathrm{LSIS}}^{xt}.
\end{align*}
The Jacobian factor \(\ell^{d+1}\) cancels between the transformed target and proposal densities, so the importance weights are unchanged by the coordinate transformation. 

For CE, we follow \cite{tong2023large} and initialize the Gaussian proposal with \(q_0=q_{\mathrm{LSIS}}^{xt}\). We perform all updates of the mean and full covariance in normalized coordinates. Let \(q_j=\mathcal N(m_j,\Sigma_j)\) denote the proposal obtained after \(j\) CE updates, so that round \(0\) corresponds to LSIS, and let \(Z_i^{(j)}=(X_i^{(j)},T_i^{(j)})\overset{\mathrm{i.i.d.}}{\sim}q_j\). Define the importance weights \(\omega_i^{(j)}=\1_{\Dnew}(Z_i^{(j)})\pinew(Z_i^{(j)})/q_j(Z_i^{(j)})\). For a Gaussian proposal, the CE updates of the mean and covariance have the closed forms
\[
m_{j+1}=\frac{\sum_{i=1}^N\omega_i^{(j)}Z_i^{(j)}}{\sum_{i=1}^N\omega_i^{(j)}},\qquad
\Sigma_{j+1}=\frac{\sum_{i=1}^N\omega_i^{(j)}\bigl(Z_i^{(j)}-m_{j+1}\bigr)\bigl(Z_i^{(j)}-m_{j+1}\bigr)^\top}{\sum_{i=1}^N\omega_i^{(j)}}.
\]
For a given \(q_j\), the corresponding estimator is \(\hat p_{\mathrm{CE},j}=N^{-1}\sum_{i=1}^N\1_{\Dnew}(Z_i^{(j)})\pinew(Z_i^{(j)})/q_j(Z_i^{(j)})\). 

For the diffusion PDE example, we additionally compare with LAIS using deterministic-mixture weights (LAIS-dm) \cite{tong2023large}. Recall that, for \(\Dorig=\{u:F(u)\ge z\}\), \(n=\nabla F(u^*)/\|\nabla F(u^*)\|\) is the normal direction at the instanton. Let \(P_\perp=I_{d+1}-nn^\top\). LAIS performs CE adaptation in the \(r\)-dimensional LDT-informed subspace spanned by \(n\) and the \(r-1\) eigenvectors of \(H_{\mathrm{LDT}}=P_\perp\nabla^2F(u^*)P_\perp\) corresponding to the largest eigenvalues in absolute value. In our experiments, we compare \(r=3\) and \(r=5\). LAIS-dm reuses samples from all iterations using deterministic-mixture MIS weights.

\subsection{Exact quadratic example}\label{app:quadratic-details}

\begin{proof}[Proof of Proposition~\ref{prop:quadratic}]
We prove the result directly without reference to Proposition~\ref{prop:kappa}. Since $\phi=\hat\phi$ globally, the upper proposal covers the entire event and every lower-component draw has zero weight almost surely. Thus we see from~\eqref{tildWi-components} that $$\hat p_N = \left(\frac1N\sum_{i=1}^N\frac{J_iA_i}{w}\right)\psorm,$$ where $J_i\sim\operatorname{Bernoulli}(w)$ are i.i.d. and independent of $A_i=e^{-\la T_i^2/2}$. Here, the $T_i$ are i.i.d. with $T_i=\phi(X_i)+Y_i$ and $X_i,Y_i$ have the usual Gaussian and exponential distributions. Also, $\phi(x) = \frac12\sum_{i=1}^d(\ell a_i)x_i^2$. The density identity~\eqref{key-identity-again}, integrated over the global epigraph, gives $p=\psorm \E[A_1]$. Therefore,
$$\CV^2(\hat p_N) = \Var(\hat p_N/p) = \frac1N\Var\left((J_1/w)(A_1/\E[A_1])\right).$$ Write $U=J_1/w$ and $V=A_1/\E[A_1]$. Then $\E[U]=\E[V]=1$ and $U$ and $V$ are independent. Also, $\E[U^2] = 1/w$. Therefore,
\begin{equation}\label{JA}
\begin{split}
N\CV^2(\hat p_N) &=\Var(UV)=\E[U^2]\E[V^2]-1 \\
&= \E[U^2]-1 + \E[U^2](\E[V^2]-1)\\
&= w^{-1} - 1 + \frac{\epsilon_+}{w}, \qquad \epsilon_+=\Var(V)=\E[V^2]-1.
\end{split}\end{equation}
for every $w\in(0,1]$. Since $\epsilon_+$ is a variance, it is nonnegative. Furthermore, since $A_1\leq1$, we have $\E[V^2] =\E[A_1^2]/\E[A_1]^2 \leq \E[A_1]/\E[A_1]^2 = 1/\E[A_1]$. Thus $0\leq \epsilon_+ \leq 1/\E[A_1] - 1$. Finally, Jensen's inequality gives 
\begin{equation}\label{jensengives}
\begin{split}
\log(1/\E[A_1]) &\leq \la\E[T_1^2]/2 \leq \la\E[\phi(X)^2]+\la\E[Y^2] \\
&\leq \frac{3\tau_2(0)^2d^2}{4\la} + \frac{2}{\la}.
\end{split}
\end{equation} To get the second line, we used that $X_i^2\stackrel{d}{=}\la^{-1}\frac{1}{1+\ell a_i}Z_i^2$ for i.i.d. $Z_i\sim\mathcal N(0,1)$. Hence
\begin{equation}
|\phi(X)|\leq\frac12\sum_{i=1}^d\frac{|\ell a_i|}{\la(1+\ell a_i)}Z_i^2 \leq \frac{\tau_2(0)}{2\la}\|Z\|^2,
\end{equation} from which $\la\E[\phi(X)^2]\leq 3\tau_2(0)^2d^2/(4\la)$ follows.
Exponentiating~\eqref{jensengives} proves the desired bound~\eqref{eq:parabolic-cv2} on $\epsilon_+$.\end{proof}

\subsection{Ellipsoid level set}\label{app:ellipsoid-details}

\begin{proof}[Proof of Proposition~\ref{prop:ellipsoid}]
All implicit constants are absolute. For the ellipsoidal event, $u^*=(0_d,\ell)$ and we may take $Q=(e_1,\dots,e_d)$. Then~\eqref{eq:implicit-BH} gives
\[
  \nabla^2\phi(0_d)=-\ell\operatorname{diag}(a_1,\ldots,a_d),\qquad H=I_d+\nabla^2\phi(0_d),\qquad \la=\ell^2\la_0.
\]
To verify the no-return part of Assumption~\ref{assume:no-return}, substitute $u=u^*+\ell(Qx+tn)$ into $F(u)\geq z$ to get that for all $(x,t)\in D$ it holds
\[
  0\leq 2t+\ell t^2 +\ell\sum_{i=1}^d a_i x_i^2 =2t+\ell t^2-x^\top \nabla^2\phi(0_d)x \leq (1+t)^2-1-x^\top \nabla^2\phi(0_d)x.
\]
Consequently,
\[
  1+x^\top Hx\leq (1+t)^2+\|x\|^2,
\]
which implies no return with $\gamma=r_1^2$ for any local radius $r_1$. Next, we apply Lemma~\ref{lma:quantitative-implicit} for any arbitrary $R_0$. Note that $\nabla F(u^*)=(0_d,1)$, $\nabla^2F(u^*)=\mathrm{diag}(a_1,\dots,a_d,1)$, and $\nabla^k F(u)=0$ for all $k>2$. Therefore,
\[
  m_1=\|\nabla F(u^*)\|=1,\qquad M_2=\max\{1,a\},\qquad M_3=M_4=0
\]
satisfy~\eqref{Ms}. The assumptions $\ell<1$ and $\ell a<1$ imply $m_1/(2M_2)\geq\ell/2$. Thus, setting the arbitrary $R_0$ to be $R_0=\ell/2$, we get $R_*=\ell/2$, and the strip condition of Assumption~\ref{assume:bdy} is satisfied with $V_0=\{\|v\|<\ell/4\}$ and $\mu_0=\ell/8$. Lemma~\ref{lma:quantitative-implicit} also gives the following derivative bounds on a neighborhood of radius $R=\sqrt h/8$:
\[
  \tau_2(0)\leq h^{-1},\qquad \tau_3(R)\lesssim h^{-3/2},\qquad \tau_4(R)\lesssim h^{-2}.
\]
Choosing $r_1=\min(R,2/T_3)$ as in the lemma (with $T_3=C_3h^{-3/2}$ an upper bound on $\tau_3(R)$) therefore gives $r_1\gtrsim h^{3/2}$ and verifies the local cap condition from Assumption~\ref{assume:no-return}, as well as the condition $\ell\us_{r_1}\subseteq V_0$. Finally, the one-sidedness condition of Assumption~\ref{assume:bdy} follows because we may write $\Dorig$ explicitly as the epigraph $\Dorig=\{(v,s+\ell):v\in\R^d,\ s\geq\phi_0(v)\}$ where
\[
  \phi_0(v_1,\dots,v_d) =\sqrt{\left(1-\sum_{i=1}^d a_i v_i^2\right)_+}-1.
\]
Thus $\Dorig$ contains no points below this boundary. Next, we choose a $(1/4)$-admissible $\kappa$. The above choices of $r_1,\gamma,\mu_0$, as well as the fact that $H\preceq I_d$ imply that~\eqref{kap:admit} with $\rho=1/4$ is ensured by
\begin{equation}\label{kap2}
  \frac{12}{\la}\left(3d\log\frac{\la}{d}+2\right) \leq\kappa^2\leq\min(r_1^2,1/2),\qquad
  \kappa\tau_3(R)\leq 1/4.
\end{equation}
Since $r_1\gtrsim h^{3/2}$ and $\tau_3(R)\lesssim h^{-3/2}$, the choice $\kappa=k h^{3/2}$ satisfies the two upper bounds in~\eqref{kap2} for a sufficiently small absolute $k>0$. Its lower bound follows from the assumed condition~\eqref{h-ell-la-cond}. Indeed, the condition gives $\la\geq\eta(\frac{d}{h^3}\log\frac eh)\geq4$ when $\eta\geq e$. Thus $1/\la\leq(d/\la)\log(\la/d)$. Since $t\mapsto(\log t)/t$
is decreasing for $t\geq e$, we obtain
\[
  \frac{12}{\la}\left(3d\log\frac{\la}{d}+2\right)
  \leq 60\frac d\la\log\frac{\la}{d}
  \leq \frac{60h^3}{\eta}
       \frac{\log\eta+3\log(1/h)+\log\log(e/h)}{\log(e/h)}
  \leq 60h^3\frac{\log\eta+4}{\eta},
\]
where the last inequality uses $\log(e/h)\geq1$ and
$\log\log(e/h)\leq\log(e/h)$. Choosing $\eta$ sufficiently large
makes the last expression at most $k^2h^3=\kappa^2$. Furthermore, $\la\geq 2(d+1)$ for $\eta$ sufficiently large, so Proposition~\ref{prop:kappa} applies. We now compute the $\epsilon_j$, $j=2,3,4$. 
The boundary $\phi$ is even near $0_d$, so $\nabla^3\phi(0_d)=0$ and
$\tau_3(0)=0$. The derivative bounds therefore give
\begin{equation}
    \begin{gathered}
        \epsilon_3=0,\qquad \epsilon_2^2+\epsilon_4^2+\la^{-1} \lesssim\frac{d^2}{\la h^2},\\
        \mathcal R=(\epsilon_2\vee\epsilon_4)^2+\frac{\epsilon_4^4}{d}
  \lesssim\frac{d^2}{\la h^2}+\frac{d^3}{\la^2 h^4}\lesssim\frac{d^2}{\la h^2},
    \end{gathered}
\end{equation}
where the last inequality uses $d/\la\le h^3$. Substituting these bounds into~\eqref{MSE-err+}-\eqref{Eexpbd} gives
\[
  |\epsilon_+|+\epsilon_-\lesssim\frac{d^2}{\la h^2}\exp\left(C\frac{d^2}{\la h^2}\right),\qquad
 \mathcal E_{\exp}\lesssim e^{Cd^2/(\la h^2)-\la\kappa^2/48}.
\]
for some absolute $C>0$. Since $d/\la\leq h^3/\eta$, we have $Cd^2/(\la h^2) = C\la (d/\la)^2/h^2 \leq C\la h^4/\eta^2 \leq C\la h^3/\eta^2$. Thus
\[
 C\frac{d^2}{\la h^2}-\la\kappa^2/48 \leq \la h^3[C/\eta^2 - k^2/48] \leq -c\la h^3
\] for some constant $c$, increasing $\eta$ further if necessary. Substituting these bounds into~\eqref{MSE-decomp} and using $\la=\la_0\ell^2$ proves~\eqref{eq:ellipsoid-cv2}. Similarly, using $\la \kappa^2 = k^2\la_0\ell^2 h^3$ in~\eqref{eq:kappa-prob-clean} proves~\eqref{eq:ellipsoid-cv2}.
\end{proof}
\subsection{Portfolio example}\label{app:portfolio-details}

\begin{proof}[Proof of Proposition~\ref{prop:portfolio}]
Constants with subscript $a_0$ depend only on $a_0$ and may change between occurrences. Fix $\rho=1/4$ and let $a=|\log z|$. Recall that we assume $a\geq a_0$ for some fixed $a_0$. We have $u^*=(\log z)\mathbf 1_{d+1}$ and therefore $\ell =\sqrt{d+1}a$. Using~\eqref{eq:implicit-BH} and $d+1\asymp d$, we have
$$
\ell\asymp \sqrt {d}a,\qquad H=(1+a)I_d,\qquad \la\asymp da^2\la_0.
$$
Apply Lemmas~\ref{lma:quantitative-implicit} and~\ref{lma:concave-level-set} with $R_0=1$. Note that $F(u)=-(d+1)^{-1}\sum_{i=1}^{d+1} e^{u_i}$ is globally concave and we may take
$$
m_1=\frac z{\sqrt {d+1}},\qquad M_j=\frac{ze}{d+1},\qquad j=2,3,4,
$$ to satisfy~\eqref{Ms}. Thus these lemmas verify the geometric assumptions. Since $M_j/m_1\asymp 1/\sqrt d$ for $j=2,3,4$, we get $C_j \asymp 1/\sqrt d$. Also, $h=\lambda_{\min}(H) = 1+a\asymp_{a_0}a$. Substituting the above quantities into the prescribed $\mu_0, r_1,\gamma$ from these lemmas, and into the derivative bounds~\eqref{eq:implicit-tau}, gives
\begin{equation}\label{eps-portfolio}\begin{gathered}
\mu_0\asymp1,\qquad r_1\asymp_{a_0}\frac{1}{\sqrt {d\,a}},\qquad \gamma\asymp\frac1{da^2},\\
\tau_2(0)\lesssim 1,\qquad \tau_3(r_1)\lesssim\sqrt{da},\qquad \tau_4(r_1)\lesssim da.
\end{gathered}\end{equation}
Since $\log\det H=d\log(1+a)$, substituting~\eqref{eps-portfolio}
into~\eqref{kap:admit} gives the following sufficient conditions for
$\kappa$ to be $(1/4)$-admissible, with $C>0$ sufficiently large and
$c_{a_0}>0$ sufficiently small:
\begin{equation}\label{kap-portfolio}
  C\frac{1+\log(a^2\la_0)+\log(1+a)}{a^2\la_0}
  \leq\kappa^2\leq\frac{c_{a_0}}{da^2}.
\end{equation}
Indeed, $r_1^2\asymp_{a_0}1/(da)$, $\gamma\asymp1/(da^2)$, and
$\mu_0/\ell\asymp1/(\sqrt d\,a)$ give the upper bounds. They also
ensure $\kappa\leq r_1$ and
\[
  \kappa\tau_3(\kappa)\leq\kappa\tau_3(r_1)
  \lesssim\sqrt{c_{a_0}/a}\leq\sqrt{c_{a_0}/a_0},
\] so $\kappa\tau_3(\kappa)\leq\rho=1/4$
holds for $c_{a_0}$ sufficiently small. Choose $k_{a_0}>0$ sufficiently small and set
\begin{equation}\label{eq:portfolio-kappa-explicit}
  \kappa=\frac{k_{a_0}}{a\sqrt d},\qquad
  \la\kappa^2=\frac{d+1}{d}k_{a_0}^2\la_0.
\end{equation}
Then the upper bound in~\eqref{kap-portfolio} holds. For the lower
bound, we use that $\la_0/d\geq\eta_{a_0}(1+\log d+\log(1+a))$ by the assumption~\eqref{port-assume}. Since $\log a\leq\log(1+a)$, we have
\begin{equation}\label{eq:portfolio-admissibility-explicit}\begin{split}
 1+\log(a^2\la_0)+\log(1+a)
  &\leq 1 + 3\log(1+a) +\log d + \log (\la_0/d) \\
  &\leq 3(\la_0/d)/\eta_{a_0} + \log (\la_0/d).
  \end{split}
\end{equation} Therefore,
\begin{equation}\label{kap-portfolio-2}
  C\frac{1+\log(a^2\la_0)+\log(1+a)}{a^2\la_0}
  \leq C\frac{\la_0/(d\eta_{a_0}) + \log(\la_0/d)}{a^2\la_0} = \frac{C}{da^2}\left(\eta_{a_0}^{-1} + \frac{\log(\la_0/d)}{\la_0/d}\right)
\end{equation}
The right hand side can be made smaller than $\kappa^2 = k_{a_0}^2/(da^2)$ by choosing $\eta_{a_0}$ sufficiently large, which ensures $\log(\la_0/d)/(\la_0/d)$ is sufficiently small. Finally, $a^2\la_0\geq a_0^2\eta_{a_0}d\geq2$ for $\eta_{a_0}$
sufficiently large, so $\la=(d+1)a^2\la_0\geq2(d+1)$.
Thus Proposition~\ref{prop:kappa} applies.

To quantify the resulting error bound, set $q=d^2/(a\la_0)$. By~\eqref{eps-portfolio} and $\la\asymp da^2\la_0$,
\begin{equation}\label{eq:portfolio-epsilon-explicit}
\la^{-1}\lesssim_{a_0}q, \qquad \epsilon_2^2\lesssim\frac d{a^2\la_0}=\frac q{da}\lesssim_{a_0}q,\qquad \epsilon_3\lesssim\frac d{\sqrt{a\la_0}}=\sqrt q,\qquad \epsilon_4^2\lesssim\frac{d^2}{a\la_0}=q.
\end{equation}
The condition~\eqref{port-assume} gives $d/\la_0\leq\eta_{a_0}$, so $\mathcal R\lesssim_{a_0}q+\frac{q^2}{d}\lesssim_{a_0}q$. Applying~\eqref{MSE-err+}--\eqref{MSE-err-} gives
$|\epsilon_+|+\epsilon_-\lesssim_{a_0}\sqrt q\,e^{C_{a_0}q}.$
For the localization remainder,~\eqref{Eexpbd} and~\eqref{eq:portfolio-kappa-explicit} give
$$
\mathcal E_{\exp}\lesssim e^{C_{a_0}q-k_{a_0}^2\la_0/48}
$$
Since $q/\la_0=(d/\la_0)^2/a\leq\eta_{a_0}^2/a_0$, we may choose $\eta_{a_0}$ sufficiently small that $C_{a_0}q\leq k_{a_0}^2\la_0/96$. Therefore, for $N\geq1$,
$$
\mathcal E_{\exp}\lesssim e^{-k_{a_0}^2\la_0/96}.
$$
Combining these estimates with~\eqref{eq:kappa-prob-clean}
and~\eqref{MSE-decomp}, and recalling $a=|\log z|$,
we obtain~\eqref{eq:portfolio-localization}
and~\eqref{eq:portfolio-cv2}.
\end{proof}

\section{Proofs deferred from Section~\ref{sec:proof-kappa}}\label{app:proof}
Throughout, we impose the assumptions of Proposition~\ref{prop:kappa}. Recall that $r=\hat\phi-\phi$ and $\mu=\mu_0/\ell$ as in~\eqref{r-mu}, $\epsilon_\pm$ are as in~\eqref{def:epsilon-plusminus}, and $a,G$ are as in~\eqref{main1} and~\eqref{def:G}. Whenever expectations are taken, $X\sim\N(0_d,(\la H)^{-1})$ and $Y\sim\Exp(\la)$ are independent, and $\varrho$ denotes the density of $X$. We break up the appendix into two parts. Appendix~\ref{app:proof1} proves preliminary bounds, namely Lemma~\ref{lma:loc-MSE} and~\ref{lma:risk-reduction}. Appendix~\ref{app:proof2} proves the quantitative bounds of Lemma~\ref{lma:analytic-controls}.

\subsection{Preliminary bounds}\label{app:proof1}
\begin{proof}[Proof of Lemma~\ref{lma:loc-MSE}]
Write $b:=\pi_D(\hat D_{\tail,-})$ and $R_i=(\pi_D/\hat\pi)(U_i)\1\{U_i\notin\hat D_{\tail,-}\}$, so that $\hat p_\loc/p=N^{-1}\sum_{i=1}^N R_i$. Then $\E[R_1]=1-b$. Since $\hat\pi=w\hat\pi_++(1-w)\hat\pi_-$ and the two components have disjoint supports up to a null set,
$$
\E[R_1^2]=\int_{D\setminus\hat D_{\tail,-}}\frac{\pi_D(u)^2}{\hat\pi(u)}\,du=\frac{1+\epsilon_+}{w}+\frac{\epsilon_-}{1-w}.
$$
Therefore,
\begin{align*}
\E\left[\left(\frac{\hat p_\loc}{p}-1\right)^2\right]&=\frac1N\Var(R_1)+b^2\\
&=\frac1N\left(\frac1w-1+\frac{\epsilon_+}{w}+\frac{\epsilon_-}{1-w}\right)+b^2(1-1/N)+2b/N.
\end{align*}
This proves~\eqref{MSE-decomp} with $\mathcal E_{\exp}=b^2(1-1/N)+2b/N$. Since $N\geq1$, this quantity lies between zero and $b^2+2b/N$, proving~\eqref{eps-V}.
\end{proof}
\begin{proof}[Proof of Lemma~\ref{lma:risk-reduction}]The bound on $\epsilon_-$ has already been given in the argument leading up to~\eqref{eps-arg}. To bound $\epsilon_+$, we use the two representations 
$$1+\epsilon_+  = \int_{\hat D}(\pi_D(u)/\hat\pi_+(u))\pi_D(u)du= \int_{\hat D}(\pi_D(u)/\hat\pi_+(u))^2\hat\pi_+(u)du.$$ The first of these representations and the upper bound~\eqref{main2} immediately imply $1+\epsilon_+ \leq a^{-1}\pi_D(\hat D)\leq a^{-1}$. The second representation and Cauchy-Schwarz give $1+\epsilon_+ \geq \pi_D(\hat D)^2$. Combining the upper and lower bound, and using that $1-\pi_D(\hat D)^2\leq2\pi_D(\hat D^c)$, it follows that
\begin{equation}\label{eps+-prelim}
|\epsilon_+| \leq\max\left\{2\pi_D(\hat D^c),\frac1a-1\right\}.%
\end{equation} Next, write $\pi_D(\hat D^c) =\pi_D(\hat D_{\loc,-}) + \pi_D(\hat D_{\tail,-})$. To bound $\pi_D(\hat D_{\loc,-}) = \pi_D(D\cap\hat D_{\loc,-})$, we again use~\eqref{hatDlocD} and the pointwise density bound~\eqref{main2}. We obtain 
\begin{equation}\label{eps+-prelim2}
\pi_D(\hat D_{\mathrm{loc},-})
\leq\frac1a\int_{\us_\kappa}\int_{-r(x)_+}^0\varrho(x)\la e^{-\la y}dydx = \frac1a\E\left[\left(e^{\lambda r(X)_+}-1\right)\1_{\us_\kappa}(X)\right]
\leq\frac{G}{3a}.
\end{equation}
The last inequality uses $e^v-1\leq(e^{3v}-1)/3$ for $v\geq0$. Combining~\eqref{eps+-prelim} and~\eqref{eps+-prelim2} gives $|\epsilon_+| \leq\max\{\pi_D(\hat D_{\tail,-}) + 2G/3a, 1/a-1\}$, which is further bounded by~\eqref{eq:epsilon-plus-abstract}.
\end{proof}

\subsection{Quantitative bounds: proof of Lemma~\ref{lma:analytic-controls}}\label{app:proof2}
We first use admissibility to control the sizes of two terms.
\begin{lemma}Let
\begin{equation}\label{def:Psi}
\Psi = \frac{\kappa^2}{3}\left(1-\frac{d+1}{\la}\right)-\frac{d+1}{\la}\left(1+\log\frac\la{d+1}\right).
\end{equation} Then
\begin{align}
\la\kappa^2&\geq36\log\la+24(1\vee\log\det H),\label{eq:admissibility-absorption}\\
\la\kappa^2&\geq12(d+1)\left(1+\log\frac{\la}{d+1}\right),\label{eq:admissibility-tail-threshold}\\
\Psi &\geq \kappa^2/12.
\end{align}
\end{lemma}
\begin{proof}
For $d\geq1$ and $\la\geq2(d+1)$, the elementary inequalities
$$
d\log\frac{\la}{d}\geq\log\la,\qquad 3d\log\frac{\la}{d}\geq(d+1)\left(1+\log\frac{\la}{d+1}\right)
$$
and~\eqref{kap:admit} prove~\eqref{eq:admissibility-absorption} and~\eqref{eq:admissibility-tail-threshold}. Next, the definition of $\Psi$ gives
\begin{equation}\begin{split}
\Psi&=\frac {\kappa^2}{3}\left(1-\frac{d+1}{\la}\right)-\frac{d+1}{\la}\left(1+\log\frac{\la}{d+1}\right)\\
&\geq \frac{\kappa^2}{6} - \frac{\kappa^2}{12} =  \frac{\kappa^2}{12}.
\label{eq:Psi-expanded}
\end{split}
\end{equation} The second line uses $\la\geq2(d+1)$ and~\eqref{eq:admissibility-tail-threshold}.
\end{proof}
We now prove the three inequalities~\eqref{eq:a-simple},~\eqref{eq:G-simple}, and~\eqref{eq:beta-simple} to finish the proof of Lemma~\ref{lma:analytic-controls}.

\begin{proof}[Proof of~\eqref{eq:a-simple}]
Set
$$
E:=\{X\in\us_\kappa,\ Y\leq\mu\},\qquad q:=\Pp(E),\qquad V:=-\frac\la2(\phi(X)+Y)^2+\la r(X).
$$
Then $a=\E[e^V\1_E]$ by~\eqref{main1}. By $\rho$-admissibility, $\la\kappa^2/d\geq9$ and $\mu\geq\kappa^2/4$. Lemma~\ref{lma:chi} therefore gives
\begin{equation}\label{eq:a-probability-bound}
q\geq1-\Pp(X\notin\us_\kappa)-\Pp(Y>\mu)\geq1-2e^{-\la\kappa^2/4}\geq\frac12.
\end{equation}
For the last inequality, use $\la\kappa^2\geq9d\geq9$.

Since
$$
\frac12(\phi(X)+Y)^2=\frac12(\hat\phi(X)+Y-r(X))^2\leq(\hat\phi(X)+Y)^2+r(X)^2,
$$
and $X$ and $Y$ are independent, we have
\begin{align}
\E[V\1_E]&\geq-\la\E[(\hat\phi(X)+Y)^2]-\la\E[r(X)^2\1\{X\in\us_\kappa\}]\notag\\
&\qquad-\la\left|\E[r(X)\1\{X\in\us_\kappa\}]\right|.
\label{eq:a-master-bound}
\end{align}
Lemmas~\ref{lma:quadratic-exp-moments} and~\ref{lma:r-moments} give
\begin{align}
\la\E[(\hat\phi(X)+Y)^2]&\leq\epsilon_2^2+\frac3\la,\label{eq:a-quadratic-bound}\\
\la\E[r(X)^2\1\{X\in\us_\kappa\}]&\leq\frac1\la\left(\epsilon_3^2+\frac12\epsilon_4^4\right),\label{eq:a-square-bound}\\
\la\left|\E[r(X)\1\{X\in\us_\kappa\}]\right|&\leq\frac18\epsilon_4^2.\label{eq:a-mean-bound}
\end{align}
Consequently,
$$
\E[V\1_E]\geq-\left(\epsilon_2^2+\frac{\epsilon_4^2}{8}+\frac{3+\epsilon_3^2+\epsilon_4^4/2}{\la}\right)\geq-3\mathcal R_a.
$$
Applying Jensen's inequality conditionally on $E$ yields
\begin{equation}\label{eq:a-jensen-bound}
a=q\E[e^V\mid E]\geq q\exp\left(\frac{\E[V\1_E]}q\right)\geq q e^{-3\mathcal R_a/q}.
\end{equation}
Using $e^v-1\leq ve^v$ for $v\geq0$, followed by~\eqref{eq:a-probability-bound}, we obtain
\begin{align}
a^{-1}-1&\leq q^{-1}e^{3\mathcal R_a/q}-1=\frac{1-q}{q}+\frac{e^{3\mathcal R_a/q}-1}{q}\notag\\
&\leq\frac{1-q}{q}+\frac{3\mathcal R_a}{q^2}e^{3\mathcal R_a/q}\notag\\
&\lesssim 1-q + \mathcal R_ae^{6\mathcal R_a} \leq e^{-\la\kappa^2/4}+\mathcal R_a e^{6\mathcal R_a}.
\end{align}
This proves~\eqref{eq:a-simple}.
\end{proof}

\begin{proof}[Proof of~\eqref{eq:G-simple}]
For $X\in\us_\kappa$, set
$$
A:=3\la r(X).
$$
On $\{r(X)\geq0\}$ we have $A\geq0$ and $e^A-1\leq Ae^A$, so Cauchy-Schwarz gives
\begin{align}
G&\leq\E[A e^A\1\{r(X)\geq0,\ X\in\us_\kappa\}]\notag\\
&\leq\left(\E[A^2\1\{X\in\us_\kappa\}]\right)^{1/2}\left(\E[e^{2A}\1\{X\in\us_\kappa\}]\right)^{1/2}\notag\\
&=3\left(\la^2\E[r(X)^2\1\{X\in\us_\kappa\}]\right)^{1/2}\left(\E[e^{6\la r(X)}\1\{X\in\us_\kappa\}]\right)^{1/2}.
\label{eq:G-cauchy-schwarz}
\end{align}
Lemmas~\ref{lma:r-moments} and~\ref{lma:exp-r} now give
\begin{align}
G\lesssim(\epsilon_3+\epsilon_4^2)e^{C_\rho(\epsilon_4^2 + \epsilon_3^2 + \epsilon_4^4/d)},
\end{align}
as desired.
\end{proof}
To prove the tail probability bound, we start with the following lemma.
\begin{lemma}[No-return starting at $\kappa$]\label{lma:geometric-separation}
For every $v=(x,t)\in D$ satisfying $\|H^{1/2}x\|\geq\kappa$,
\begin{equation}
\|v-(0_d,-1)\|^2\geq1+\kappa^2/3.
\label{eq:geometric-separation}
\end{equation}
\end{lemma}
\begin{proof}
First suppose $\kappa\leq\|H^{1/2}x\|\leq r_1$. Assumption~\ref{assume:bdy} gives $t\geq\phi(x)$ and Assumption~\ref{assume:no-return} gives $\phi(x)\geq \frac12x^\top \nabla^2\phi(0_d)x - \frac13x^\top Hx$. Hence
\begin{equation}\label{v-below}
\begin{split}
    \|v-(0_d,-1)\|^2&=\|x\|^2+(t+1)^2=\|x\|^2+t^2+2t+1\\
&\geq \|x\|^2+2\phi(x)+1\\
&\geq1+x^\top \nabla^2\phi(0_d)x-\frac23x^\top Hx+\|x\|^2\\
&=1+\frac13x^\top Hx\geq1+\kappa^2/3.
\end{split}
\end{equation}
Now suppose $\|H^{1/2}x\|>r_1$. The no-return part of Assumption~\ref{assume:no-return} gives $\|v-(0_d,-1)\|^2\geq1+\gamma$, and $\rho$-admissibility gives $\gamma\geq\kappa^2/3$, proving~\eqref{eq:geometric-separation}.
\end{proof}

\begin{proof}[Proof of~\eqref{eq:beta-simple}] Lemma~\ref{lma:geometric-separation},~\eqref{main1}, and the definition of $\psorm$ imply
\begin{equation}\label{tail1}
\pi_D(\hat D_{\tail,-})\leq\frac{\Pp(\chi_{d+1}^2\geq\la(1+\kappa^2/3))}{a\psorm}\lesssim a^{-1}\sqrt{\la\det H}e^{\la/2}\Pp(\chi_{d+1}^2\geq\la(1+\kappa^2/3)).
\end{equation}
Since $\la>d+1$, Lemma~\ref{lma:chi} and $\log(1+u)\leq u$ give
\begin{align}
\Pp(\chi_{d+1}^2\geq\la(1+\kappa^2/3))&\leq\exp\left\{-\frac\la2-\frac{\la\kappa^2}{6}+\frac{d+1}{2}\left(1+\log\frac\la{d+1}+\log(1+\kappa^2/3)\right)\right\}\notag\\
&\leq\exp\left\{-\frac\la2-\frac{\la\kappa^2}{6}+\frac{d+1}{2}\left(1+\log\frac\la{d+1}+\frac{\kappa^2}{3}\right)\right\}\notag\\
&=e^{-\la/2}e^{-\la\Psi/2},
\label{eq:unnormalized-tail-final}
\end{align} recalling~\eqref{def:Psi}. Substituting this bound into~\eqref{tail1} gives
$$
\pi_D(\hat D_{\tail,-}) \lesssim a^{-1}\exp\left(-\frac12\left[\la\Psi - \log\det H - \log\la\right]\right) \leq  a^{-1}\exp\left(-\frac12\left[\la\kappa^2/12 - \log\det H - \log\la\right]\right).
$$
By~\eqref{eq:admissibility-absorption},
$$
\log\det H+\log\la\leq\frac{\la\kappa^2}{24}-\frac12\log\la.
$$
Consequently,
$$
\pi_D(\hat D_{\tail,-})\lesssim a^{-1}\exp\left(-\frac12\left[\frac{\la\kappa^2}{24}+\frac12\log\la\right]\right)=a^{-1}\la^{-1/4}e^{-\la\kappa^2/48}.
$$
Finally,~\eqref{eq:admissibility-absorption} also gives $\la\kappa^2\geq36\log\la$, so $\la^{-1/4}e^{-\la\kappa^2/48}\leq\la^{-1}$. This proves~\eqref{eq:beta-simple}.
\end{proof}

\section{Taylor and Gaussian estimates}\label{app:taylor-gaussian}

This appendix contains the Taylor expansions, Gaussian expectations, and Gaussian tail bounds used in Section~\ref{sec:proof-kappa} and Appendix~\ref{app:proof}. We impose the assumptions of Proposition~\ref{prop:kappa}. Recall that $r=\hat\phi-\phi$, and the definitions~\eqref{def:epsilon-234} of $\epsilon_2,\epsilon_3,\epsilon_4$. Throughout, $X\sim\N(0_d,(\la H)^{-1})$.

\subsection{Weighted Taylor estimates}

\begin{lemma}[Weighted Taylor estimates]\label{lma:weighted-taylor}
Recall $r(x)=\hat\phi(x)-\phi(x)$. For every $x\in\us_\kappa$,
\begin{align}
r(x)&=-\frac16\left\langle\nabla^3\phi(\xi_3x),x^{\otimes3}\right\rangle
\label{eq:r-third-order}\\
r(x)&=-\frac16\left\langle\nabla^3\phi(0),x^{\otimes3}\right\rangle-\frac1{24}\left\langle\nabla^4\phi(\xi_4x),x^{\otimes4}\right\rangle,
\label{eq:r-fourth-order}
\end{align}
for some $\xi_3,\xi_4\in[0,1]$ that may depend on $x$. Consequently,
\begin{align}
|r(x)|&\leq\frac{\tau_3(\kappa)}6\|H^{1/2}x\|^3,
\label{eq:r-pointwise}\\
\left|r(x)+\frac16\left\langle\nabla^3\phi(0),x^{\otimes3}\right\rangle\right|&\leq\frac{\tau_4(\kappa)}{24}\|H^{1/2}x\|^4,
\label{eq:r-fourth-remainder}\\
\|H^{-1/2}\nabla r(x)\|&\leq\frac{\tau_3(0)}2\|H^{1/2}x\|^2+\frac{\tau_4(\kappa)}6\|H^{1/2}x\|^3.
\label{eq:r-gradient}
\end{align}
\end{lemma}

\begin{proof}
Apply the one-dimensional Taylor theorem to $s\mapsto\phi(sx)$, using $\phi(0)=0$ and $\nabla\phi(0)=0$.
Since $r(x)=\frac12x^\top\nabla^2\phi(0_d)x-\phi(x)$, negating the Taylor remainder gives~\eqref{eq:r-third-order} and~\eqref{eq:r-fourth-order}. The weighted tensor norm then gives~\eqref{eq:r-pointwise} and~\eqref{eq:r-fourth-remainder}.

To justify the mixed derivative bound with the diagonal norm in Definition~\ref{def:norms}, use Banach's theorem for symmetric multilinear forms on a real Hilbert space; see~\cite[equation (2)]{carando2019symmetric}. Applied after the change of variables $v\mapsto H^{1/2}v$, it gives
\begin{equation}\label{eq:mixed-derivative-bound}
\left|\left\langle\nabla^j\phi(z),v_1\otimes\cdots\otimes v_j\right\rangle\right|\leq\|\nabla^j\phi(z)\|_H\prod_{i=1}^j\|H^{1/2}v_i\|,\qquad j=3,4.
\end{equation}
For every $v\in\R^d$, Taylor's theorem applied to $s\mapsto\langle\nabla\phi(sx),v\rangle$, with the quadratic term evaluated at $s=0$ and the cubic remainder controlled on $[0,1]$, gives
$$
|\langle\nabla r(x),v\rangle|\leq\left(\frac{\tau_3(0)}2\|H^{1/2}x\|^2+\frac{\tau_4(\kappa)}6\|H^{1/2}x\|^3\right)\|H^{1/2}v\|.
$$
Taking $v=H^{-1/2}u$ and then the supremum over $\|u\|=1$ proves~\eqref{eq:r-gradient}.
\end{proof}

\subsection{Gaussian polynomial moments}

\begin{lemma}[Gaussian radial moments]\label{lma:gaussian-radial-moments}
If $Z\sim\N(0,I_d)$, then
\begin{align}
\E\|Z\|^4&=d(d+2),
\label{eq:gaussian-fourth-radial}\\
\E\|Z\|^8&=d(d+2)(d+4)(d+6)\leq105d^4.
\label{eq:gaussian-eighth-radial}
\end{align}
\end{lemma}
Both identities follow from the moments of the chi-square distribution.

\begin{lemma}[Quadratic and exponential moments]\label{lma:quadratic-exp-moments}
Let $X\sim\N(0_d,(\la H)^{-1})$ and $Y\sim\Exp(\la)$ be independent. Then
\begin{equation}
\la\E[(\hat\phi(X)+Y)^2]\leq\epsilon_2^2+\frac3\la.
\label{eq:quadratic-exp-bound}
\end{equation}
\end{lemma}

\begin{proof}
Let
$$
C:=H^{-1/2}\nabla^2\phi(0_d)H^{-1/2},\qquad u_1:=\tr (C),\qquad u_2:=\tr(C^2).
$$
If $Z\sim\N(0,I_d)$, then
$$
\hat\phi(X)\stackrel{d}{=}\frac1{2\la}Z^\top CZ.
$$
The standard Gaussian quadratic-form identities give
$$
\E[\hat\phi(X)]=\frac{u_1}{2\la},\qquad \E[\hat\phi(X)^2]=\frac{u_1^2+2u_2}{4\la^2}.
$$
Since $\E[Y]=1/\la$ and $\E[Y^2]=2/\la^2$,
\begin{align}
\la\E[(\hat\phi(X)+Y)^2]&=\frac1{4\la}\left(2u_2+(u_1+2)^2+4\right)\notag\\
&\leq\frac1{4\la}\left(2d\tau_2(0)^2+(d\tau_2(0)+2)^2+4\right)\notag\\
&\leq\tau_2(0)^2\frac{d^2}{\la}+\frac3\la   =\epsilon_2^2+\frac3\la.
\end{align}
For the penultimate inequality, set $b=d\tau_2(0)$ and use $2d\tau_2(0)^2\leq2b^2$ and $b\leq b^2/4+1$.
\end{proof}

\begin{lemma}[Gaussian cubic contraction]\label{lma:gaussian-cubic}
Let $Z\sim\N(0,I_d)$. Then
\begin{equation}
\E\left[\left\langle\nabla^3\phi(0),(H^{-1/2}Z)^{\otimes3}\right\rangle^2\right]\leq15\tau_3(0)^2d^2.
\label{eq:cubic-contraction}
\end{equation}
\end{lemma} 
\begin{proof}
Define the symmetric trilinear form
$$
T[u,v,w]:=\left\langle\nabla^3\phi(0),H^{-1/2}u\otimes H^{-1/2}v\otimes H^{-1/2}w\right\rangle.
$$
Let $T_{ijk}:=T[e_i,e_j,e_k]$. Wick's formula gives
\begin{equation}
\E[T[Z,Z,Z]^2]=6\sum_{i,j,k=1}^dT_{ijk}^2+9\sum_{k=1}^d\left(\sum_{i=1}^dT_{iik}\right)^2.
\label{eq:wick-cubic}
\end{equation}
By~\eqref{eq:mixed-derivative-bound}, $|T[u,v,w]|\leq\tau_3(0)\|u\|\|v\|\|w\|$. Hence
\begin{align}
\sum_{i,j,k=1}^dT_{ijk}^2&=\sum_{i,j=1}^d\|T[e_i,e_j,\cdot]\|^2\leq d^2\tau_3(0)^2,
\label{eq:cubic-frobenius}\\
\sum_{k=1}^d\left(\sum_{i=1}^dT_{iik}\right)^2&=\left\|\sum_{i=1}^dT[e_i,e_i,\cdot]\right\|^2\leq d^2\tau_3(0)^2.
\label{eq:cubic-trace}
\end{align}
Substituting~\eqref{eq:cubic-frobenius} and~\eqref{eq:cubic-trace} into~\eqref{eq:wick-cubic} proves~\eqref{eq:cubic-contraction}.
\end{proof}

\subsection{Moment bounds for the Taylor remainder}

\begin{lemma}[Moments of the Taylor remainder]\label{lma:r-moments}
Let $X\sim\N(0_d,(\la H)^{-1})$. Then
\begin{align}
\la\left|\E[r(X)\1\{X\in\us_\kappa\}]\right|&\leq\frac18\epsilon_4^2,
\label{eq:r-mean-bound}\\
\la^2\E[r(X)^2\1\{X\in\us_\kappa\}]&\leq\epsilon_3^2+\frac12\epsilon_4^4.
\label{eq:r-square-bound}
\end{align}
\end{lemma}

\begin{proof}
Write $X=(\la H)^{-1/2}Z$, where $Z\sim\N(0,I_d)$. On $\{\|Z\|\leq\kappa\sqrt\la\}$, equation~\eqref{eq:r-fourth-order} gives
\begin{equation}
r((\la H)^{-1/2}Z)=-\frac1{6\la^{3/2}}\left\langle\nabla^3\phi(0),(H^{-1/2}Z)^{\otimes3}\right\rangle+R_4(Z),
\label{eq:r-scaled-expansion}
\end{equation}
where
\begin{equation}
|R_4(Z)|\leq\frac{\tau_4(\kappa)}{24\la^2}\|Z\|^4\qquad\text{on }\{\|Z\|\leq\kappa\sqrt\la\}.
\label{eq:R4-bound}
\end{equation}
The cubic term in~\eqref{eq:r-scaled-expansion} has zero expectation over $\{\|Z\|\leq R\}$, by symmetry. Hence
\begin{align}
\la\left|\E[r(X)\1\{X\in\us_\kappa\}]\right|&\leq\frac{\tau_4(\kappa)}{24\la}\E\|Z\|^4=\frac{\tau_4(\kappa)d(d+2)}{24\la}\leq\frac{\tau_4(\kappa)d^2}{8\la}=\epsilon_4^2/8.
\end{align}
This proves~\eqref{eq:r-mean-bound}. Using $(u+v)^2\leq2u^2+2v^2$, Lemma~\ref{lma:gaussian-cubic}, and~\eqref{eq:gaussian-eighth-radial}, we obtain
\begin{align}
\la^2\E[r(X)^2\1\{X\in\us_\kappa\}]&\leq\frac1{18\la}\E\left[\left\langle\nabla^3\phi(0),(H^{-1/2}Z)^{\otimes3}\right\rangle^2\right]+\frac{\tau_4(\kappa)^2}{288\la^2}\E\|Z\|^8\notag\\
&\leq\frac{5\tau_3(0)^2d^2}{6\la}+\frac{35\tau_4(\kappa)^2d^4}{96\la^2}\leq\epsilon_3^2+\frac12\epsilon_4^4.
\end{align}
This proves~\eqref{eq:r-square-bound}.
\end{proof}

\subsection{Gaussian tail estimate}

\begin{lemma}[$\chi^2$ tail bound]\label{lma:chi}
Let $x\geq d$. Then
\begin{equation}
\Pp(\chi_d^2\geq x)\leq\exp\left\{-\frac x2\left(1-\frac dx\left(1+\log\frac xd\right)\right)\right\}.
\label{eq:chi-tail}
\end{equation}
If $x/d\geq9$, then
\begin{equation}
\Pp(\chi_d^2\geq x)\leq e^{-x/4}.
\label{eq:chi-tail-simple}
\end{equation}
\end{lemma}

\begin{proof}
If $x=d$, the claimed bound is one. For $x>d$ and $\theta\in(0,1/2)$, the Chernoff inequality gives
$$
\Pp(\chi_d^2\geq x)\leq e^{-\theta x}(1-2\theta)^{-d/2}.
$$
Optimizing at $\theta=(1-d/x)/2$ proves~\eqref{eq:chi-tail}. If $b=x/d\geq9$, then
$(1+\log b)/b\leq\frac12$, and~\eqref{eq:chi-tail-simple} follows.
\end{proof}

\subsection{Exponential Gaussian estimate}

\begin{lemma}[Corollary D.3 in~\cite{laplaceHD}]\label{lma:truncated-gaussian}
Let $Z\sim\N(0,I_d)$, let $U\subseteq\R^d$ be nonempty, open, and convex, and let $f\in C^1(\overline U)$ have bounded gradient on $U$. Then
\begin{equation}
\E[e^{f(Z)}\1\{Z\in U\}]\leq\exp\left\{\frac{\E[f(Z)\1\{Z\in U\}]}{\Pp(Z\in U)}+\frac12\sup_{z\in U}\|\nabla f(z)\|^2\right\}.
\label{eq:truncated-gaussian-exp}
\end{equation}
\end{lemma}

\begin{lemma}[Exponential remainder bound]\label{lma:exp-r}
We have
\begin{equation}
\E\left[e^{6\la r(X)}\1\{X\in\us_\kappa\}\right]\lesssim e^{C_\rho(\epsilon_4^2 + \epsilon_3^2 + \epsilon_4^4/d)}
\label{eq:exp-r}
\end{equation} for some $C_\rho$ depending only on $\rho$.
\end{lemma}

\begin{proof}
Set
$$
s:=\kappa\tau_3(\kappa)\leq\rho<1/2, \qquad q:=\frac2d\vee\frac{2s}{1-2s},\qquad c^2:=\frac{(1+\sqrt q)^2}{1-2s}.
$$ The function $h(v):=v-1-\log v$ is increasing for $v>1$, and
$$
h((1+\sqrt q)^2)>q.
$$
Since $q\geq2/d$ and $q\geq2s/(1-2s)\geq-\log(1-2s)$, this choice ensures
\begin{align}
c^2-1-\log(c^2)&\geq\frac2d,
\label{eq:c-condition-one}\\
(1-2s)c^2-1-\log(c^2)&>0.
\label{eq:c-condition-two}
\end{align}
We split $\us_\kappa$ into the ball $\|H^{1/2}X\|\leq\min(c\sqrt{d/\la},\kappa)$ and the annulus $\min(c\sqrt{d/\la},\kappa)\leq \|H^{1/2}X\|\leq \kappa$. On $\us_\kappa$,~\eqref{eq:r-pointwise} gives $|r(X)|\leq\frac{s}{6}X^\top HX$. Consequently, the expectation over the annulus, if it is nonempty, is bounded by
\begin{align}
\E\left[e^{6\la r(X)}\1\{c\sqrt{d/\la}\leq\|H^{1/2}X\|\leq\kappa\}\right]&\leq\E\left[e^{s\la X^\top HX}\1\{c\sqrt{d/\la}\leq\|H^{1/2}X\|\}\right]\notag\\
&=(1-2s)^{-d/2}\Pp(\chi_d^2\geq c^2(1-2s)d)\notag\\
&\leq\exp\left\{-\frac d2\left((1-2s)c^2-1-\log(c^2)\right)\right\}\leq1.
\label{eq:annular-exp-bound}
\end{align}
It remains to bound the expectation in the inner ball. To do this, we will apply Lemma~\ref{lma:truncated-gaussian}. Set
$$
R:=\min\{c\sqrt d,\kappa\sqrt\la\},\qquad f(z):=6\la r((\la H)^{-1/2}z),
$$ so that
\begin{equation}
   \E\left[e^{6\la r(X)}\1\left\{\|H^{1/2}X\|\leq\min(c\sqrt{d/\la},\kappa)\right\}\right]  = \E\left[e^{f(Z)}\1\{\|Z\|\leq R\}\right].
\end{equation}
Using~\eqref{eq:c-condition-one} and Lemma~\ref{lma:chi}, we have $\Pp(\|Z\|>c\sqrt d)\leq e^{-1}
$. By $\rho$-admissibility, $\la\kappa^2/d\geq9$, so Lemma~\ref{lma:chi} also gives
$\Pp(\|Z\|>\kappa\sqrt\la)\leq e^{-\la\kappa^2/4}\leq e^{-1}$. Combining the two estimates, we conclude
\begin{equation}
\Pp(\|Z\|\leq R)\geq1-e^{-1}.
\label{eq:inner-ball-probability}
\end{equation}
By a similar calculation as in the proof of~\eqref{eq:r-mean-bound}, we have
\begin{align}
\frac{|\E[f(Z)\1\{\|Z\|\leq R\}]|}{\Pp(\|Z\|\leq R)}&\leq\frac{6}{(1-e^{-1})}\frac{\epsilon_4^2}{8}\leq 2\epsilon_4^2.
\label{eq:inner-exp-mean}
\end{align}
Moreover,~\eqref{eq:r-gradient} implies
\begin{align}
\sup_{\|z\|\leq R}\|\nabla f(z)\|^2&=36\la\sup_{\|H^{1/2}x\|\leq\min\{c\sqrt{d/\la},\kappa\}}\|H^{-1/2}\nabla r(x)\|^2\notag\\
&\leq\left(3c^2\epsilon_3+\frac{c^3}{\sqrt d}\epsilon_4^2\right)^2\leq18c^4\epsilon_3^2+\frac{2c^6}{d}\epsilon_4^4.
\label{eq:inner-exp-gradient}
\end{align}
Applying Lemma~\ref{lma:truncated-gaussian} on the ball $\{\|z\|<R\}$ and using~\eqref{eq:inner-exp-mean}--\eqref{eq:inner-exp-gradient}, we obtain
\begin{align}
\E[e^{f(Z)}\1\{\|Z\|\leq R\}]&\leq\exp\left\{2\epsilon_4^2+9c^4\epsilon_3^2+\frac{c^6}{d}\epsilon_4^4\right\}\leq e^{9(1\vee c)^6(\epsilon_4^2 + \epsilon_3^2 + \epsilon_4^4/d)}.\label{eq:inner-exp-bound}
\end{align} Finally, from the definition of $s,q,c$ and the fact that $s\leq\rho<1/2$ implies we can bound $9(1\vee c)^6$ by some $C_\rho$. Making this replacement and combining~\eqref{eq:annular-exp-bound} with~\eqref{eq:inner-exp-bound} proves~\eqref{eq:exp-r}.\end{proof}

\bibliographystyle{plain}
\bibliography{sampling}
\end{document}